\documentclass[11pt]{article}          
\usepackage[english]{babel}
\usepackage{bbold}
\renewcommand{\baselinestretch}{1.5}
\makeatletter
\def\@begintheorem#1#2{\trivlist%
 \item[\hskip \labelsep{\sffamily\bfseries #2\ #1}]\itshape}
\newtheorem{teo}{Theorem}[section]
\newtheorem{defi}[teo]{Definition}

\newtheorem{lem}[teo]{Lemma}
\newtheorem{pro}[teo]{Proposition}
\newtheorem{_rem}[teo]{Remark}
\newtheorem{_eje}[teo]{Example}

\newenvironment{rem}{\def\@begintheorem##1##2{\trivlist%
 \item[\hskip\labelsep{\sffamily\bfseries ##2\ ##1}]}\begin{_rem}}{\end{_rem}}
\newenvironment{eje}{\def\@begintheorem##1##2{\trivlist%
 \item[\hskip\labelsep{\sffamily\bfseries ##2\ ##1}]}\begin{_eje}}{\end{_eje}}
\makeatother

\newenvironment{beweis}{{\em Proof:}}{\hfill $\rule{2mm}{2mm}$
\vspace{3mm}

}

\DeclareMathAlphabet{\Ma}{U}{msa}{m}{n}
\DeclareMathAlphabet{\Mb}{U}{msb}{m}{n}
\DeclareMathAlphabet{\Meuf}{U}{euf}{m}{n}

\def\got#1{\Meuf{#1}}

\DeclareSymbolFont{ASMa}{U}{msa}{m}{n}
\DeclareSymbolFont{ASMb}{U}{msb}{m}{n}
\DeclareMathSymbol{\hrist}{\mathord}{ASMa}{"16}
\DeclareMathSymbol{\varkappa}{\mathalpha}{ASMb}{"7B}
\DeclareMathSymbol{\CrPr}{\mathord}{ASMb}{"6F}

\newfont{\EinsFont}{cmr7 scaled 1070}

\def\restriction{{\mathchoice{%diplaystyle
 \mbox{\unitlength1cm\begin{picture}(.2,.4)%
  \bezier{5}(.07,.3)(.1,.27)(.13,.24)%
  \put(.07,.35){\line(0,-1){.5}}\end{picture}}}{%textstyle
 \mbox{\unitlength1cm\begin{picture}(.2,.4)%
  \bezier{5}(.07,.3)(.1,.27)(.13,.24)%
  \put(.07,.35){\line(0,-1){.5}}\end{picture}}}{%scriptstyle
  \hrist}{\hrist}}}

  \def\al #1.{{\mathcal{#1}}}
  \def\ot #1.{{\got{#1}}}
 \def\CCR{\overline{\Delta({\al S.},\,\sigma)}}
 \def\CCRX{\overline{\Delta( X,\,\sigma)}}
  \def\ccr #1,#2.{\overline{\Delta(#1,\,#2)}}

  \def\b #1.{{\bf #1}}
  \def\cross#1.{\mathrel{\mathop{\times}\limits_{#1}}}
  \def\C{\Mb{C}}
  \def\N{\Mb{N}}
  
  \def\R{\Mb{R}}
  \def\Z{\Mb{Z}}
  \def\J{\Mb{J}}
 \def\un{{\mathbb 1}}

  \def\wh{\widehat}
  \def\wt{\widetilde}

\def\slim{\mathop{\hbox{\rm s-lim}}}
  \def\cross #1.{\mathrel{\raise 3pt\hbox{$\mathop\times\limits_{#1}$}}}
  \def\ol #1.{\overline{#1}}
\def\b #1.{{\bf #1}}
\def\f #1,#2.{\mathsurround=0pt \hbox{${#1\over #2}$}\mathsurround=5pt}
\def\hlf{{\f 1,2.}}
\def\ker{{\rm Ker}\,}
\def\span{{\rm Span}\,}
\def\aut{{\rm Aut}\,}
\def\dom{{\rm Dom}\,}
\def\ran{{\rm Ran}\,}
\def\rep{{\rm Rep}\,}
\def\reg{{\rm Reg}}
\def\supp{{\rm supp}\,}
\def\rlf{{R(\lambda,f)}}

\def\set #1,#2.{\left\{\,#1\;\bigm|\;#2\,\right\}}
\def\maprightu #1;{\smash{\mathop{\longrightarrow}\limits^{#1}}}
\def\maprightd #1;{\smash{\mathop{\longrightarrow}\limits_{#1}}}
\def\maprightt #1,#2.{\mathrel{\smash{\mathop{\longrightarrow}\limits_{#1}^{#2}}}}
\def\chop{\hfill\break}
\def\s #1.{_{\smash{\lower2pt\hbox{\mathsurround=0pt $\scriptstyle #1$}}\mathsurround=5pt}}
\def\v #1.{\mathord{\raise 3pt\hbox{\mathsurround=0pt $\mathop\vee\limits^{#1}$\mathsurround=5pt}}}

\def\bx{{\bf x}}
\def\ClifS{{{\rm Cliff}\big(\al S.(\R)\big)}}
\def\ab{\allowbreak}
\def\bra #1,#2.{{\left\langle #1,\,#2\right\rangle_{\al A.}}}

\def\XP#1!{\renewcommand{\baselinestretch}{.7}\marginpar{{\footnotesize #1}\hfil}
\renewcommand{\baselinestretch}{1.5}}
\def\XB{\marginpar{
{\footnotesize\bf Change~starts-----}\lower 11pt\hbox{\mathsurround=0pt$
\!\!\displaystyle{
\Bigg\downarrow}$\mathsurround=3pt}}}
\def\XE{\marginpar{{\footnotesize\bf Change~ends-----}\raise 10pt\hbox{\mathsurround=0pt$
\!\!\displaystyle{
\Bigg\downarrow}$\mathsurround=3pt}}}

\title{\bf Algebraic Supersymmetry: A case study}

\author{
 {\sc Detlev Buchholz}            \\[1mm]
  {\footnotesize Institut f\"ur Theoretische Physik,}  \\
 {\footnotesize Universit\"at G\"ottingen, Friedrich-Hund-Platz 1,} \\
 {\footnotesize  D-37077 G\"ottingen, Germany} \\
 {\footnotesize buchholz@theorie.physik.uni-goettingen.de} \\
 {\footnotesize FAX: +49-551-399263} \\
\and 
 {\sc Hendrik Grundling}                                            \\[1mm] 
 {\footnotesize Department of Mathematics,}                         \\ 
 {\footnotesize University of New South Wales,}                      \\
 {\footnotesize Sydney, NSW 2052, Australia.}                             \\ 
 {\footnotesize hendrik@maths.unsw.edu.au}                  \\ 
 {\footnotesize FAX: +61-2-93857123}}
\date{\small Dedicated to Daniel Kastler on the occasion of his 80th birthday}
\begin{document}
\maketitle

%%%%%%%%%%%%%%%%%%%%%%%%%%%%%%%%%%%%%%%%%%%%%%%%%%%%%%%%%%%%%%%%%%%%%%%%%%%%%%
\begin{abstract}
\noindent The treatment of supersymmetry is known to cause  
difficulties in the C$^*$--algebraic framework of relativistic quantum field theory;
several no--go theorems indicate that super--derivations and super--KMS 
functionals must be quite singular objects in a C$^*$--algebraic setting.
In order to clarify the situation, 
a simple supersymmetric chiral field theory 
of a free Fermi and Bose field defined on $\R$ is analyzed. It is shown that 
a meaningful  
$C^*$--version of this model can be based on the tensor 
product of a CAR--algebra and a novel version of a  
CCR--algebra, the ``resolvent algebra''. 
The elements of this resolvent algebra serve as 
mollifiers for the super--derivation. Within this model, unbounded
(yet locally bounded) graded KMS--functionals are constructed and 
proven to be supersymmetric. From these KMS--functionals, Chern 
characters are obtained by generalizing formulae of Kastler and 
of Jaffe, Lesniewski and Osterwalder. The characters are used to 
define cyclic cocycles in 
the sense of Connes' noncommutative geometry which are  
``locally entire''. \end{abstract}

%%%%%%%%%%%%%%%%%%%%%%%%%%%%%%%%%%%%%%%%%%%%%%%%%%%%%%%%%%%%%%%%%%%%%%%%%%%%%%
\section{Introduction}

Graded (super) derivations occur in many parts of physics: supersymmetry, 
BRS-constraint reduction
and cyclic homology, to name a few. To adequately model these in a C*-algebra
setting involves notorious domain problems. Kishimoto and Nakamura~\cite{KiNa}
showed, for example, that apparently natural domain assumptions on the 
supersymmetry graded derivations lead to an empty theory. 
Similarly, supersymmetric KMS--functionals underlying the 
construction of cyclic cocycles as in \cite{Ka,JLO}
cannot exist in the case of infinitely extended systems \cite{BuLo}. 
These obstructions may explain why a general C$^*$-algebraic framework 
for supersymmetry has not yet emerged. It thus seems worthwhile 
to explore representative examples in  more detail in order 
to identify the pertinent structures.

In the present article we aim to develop tools to define 
and analyze in a C*-algebra setting a simple but, with regard to 
the mathematical problems under investigation, generic supersymmetric 
quantum field theory. It is the model of a chiral Fermi-- and 
Bose--field, defined on the light ray $\R$. As the construction of 
this model is easily accomplished in the Wightman setting of (unbounded)
quantum fields, we can concentrate here on the specific problems arising in 
the passage to a $C^*$--framework.

Although the model has formally the structure of a tensor product
of a CAR--algebra and a CCR--algebra, the adequate formulation 
of its $C^*$--version requires some care. It turns out that the 
standard Weyl algebra description of the CCR part is not 
suitable for the formulation of supersymmetry. We therefore introduce
a more viable variant of the CCR--algebra, the resolvent algebra, 
which formally may be thought of as being generated 
by the resolvents of the underlying Bose--field. These resolvents
act as mollifiers for the super--derivation and allow one to 
define it on a domain which is weakly dense in the underlying C$^*$--algebra 
in all representations of interest. The resolvents also lead to a mollified
version of the fundamental relation of supersymmetry, relating the 
square of the super--derivation and the generator of time translations.
These rather weak variants of supersymmetry turn out to be sufficient
for the further analysis.

Having clarified the C$^*$--algebraic formulation of supersymmetry, one  
has the necessary tools for 
the analysis of the supersymmetric KMS--functionals in this model. 
Again, these functionals are easily constructed in the
Wightman setting. Yet, as follows from general arguments \cite{BuLo},
they cannot be extended continuously 
to the full underlying C$^*$--algebra. In fact,
one does not have any \textit{a priori} information on their domains
of definition.

In the present model, the restrictions of 
the supersymmetric KMS--functionals to any local subalgebra 
of the underlying C$^*$--algebra turn out to be
bounded. Thus these functionals 
are densely defined, but their domain of definition does not contain any
non--trivial analytic elements with regard to the dynamics, as 
is required in the construction of cyclic cocycles given in \cite{Ka,JLO}.
Nevertheless, by relying on techniques from the theory of analytic 
functions of several complex variables, it is possible to define cyclic 
cocycles in the present model as well. The restrictions of these 
cocycles to any fixed local subalgebra of the underlying C$^*$--algebra
turn out to be entire in the sense of Connes \cite{Co}.

So the present field--theoretic model allows for a satisfactory 
C$^*$--algebraic formulation of supersymmetry and the analysis of
its consequences.
There are three observations which are of interest going beyond the 
present model: First, 
a C$^*$--algebraic formulation of supersymmetry has to rely on   
the concept of mollifiers or, complementary, of unbounded operators 
affiliated with the underlying C$^*$--algebra \cite{Georg}. 
Second, there is growing evidence that supersymmetric KMS--functionals,
although being unbounded, are locally bounded, in accordance with the
heuristic considerations in \cite{BuLo}. And third, although these 
functionals generically do not have analytic elements in their 
domain of definition, they can still be used to define local 
versions of Connes' entire cyclic cocycles by relying on techniques
from complex analysis. Based  on these insights, a proper 
C$^*$--algebraic framework for the formulation of supersymmetry 
and the analysis of its consequences in quantum field theory
seems within reach. We hope to return to this problem elsewhere.

The plan of our paper is as follows. 
We will state our results in the body of the
paper, and defer almost all the proofs to the appendix.
In Sect.~2 we present in the Wightman framework
the basic supersymmetry model which we wish to analyze;
in Sect.~3 we prepare for its analysis in a C*--setting by considering 
algebraic mollifying relations
for the quantum fields, which leads to the study of the C*--algebras 
generated by the resolvents of the fields.
In Sect.~4 we use these tools to present the C*--algebraic framework
of the model.
In Sect.~5 we define (unbounded) graded KMS--functionals on
the model and prove basic properties for them, including their
supersymmetry invariance and their local boundedness.
In Sect.~6 we use these KMS--functionals to define a Chern character
formula (generalizing the construction in ~\cite{Ka,JLO}),
from which we obtain a (locally) entire cyclic cocycle 
in the sense of Connes. 
This can then be taken as input to an index theory for supersymmetric quantum
field theories, of the type proposed by Longo~\cite{Lo}.

%%%%%%%%%%%%%%%%%%%%%%%%%%%%%%%%%%%%%%%%%%%%%%%%%%%%%%%%%%%%%%%%%%%%%%%%%%%
\section{The model}
\label{Heur}

We begin by presenting here our model in the Wightman framework, 
which we would like to model in 
a C*-algebra setting. It is the the simplest example for supersymmetry
on noncompact spacetime, in that we have one dimension, one boson and one fermion.

We assume chiral fields, so there is only one space-time dimension, $\R.$
The Fermi field is given by the Clifford operators $c(f)=c(f)^*,$ where
$f\in\al S.(\R,\,\R)$ and
\[
\big\{c(f),\,c(g)\big\}=(f,g):=\int fg\,dx\;.
\]
The boson field is $j(f)=j(f)^*,$ where $f\in\al S.(\R,\,\R)$ and
\[
\big[j(f),\,j(g)\big]=i\sigma(f,g):=i\int fg'\,dx\;.
\]
The $\Z_2\hbox{--grading}$ automorphism $\gamma$ comes from the 
Fermi field by
\[
  \gamma\big(c(f)\big)=-c(f)\,,\qquad
  \gamma\big(j(f)\big)=j(f)
\]
and defines even and odd parts of the polynomial 
field algebra by
$A_{\pm}=\big(A\pm\gamma(A)\big)\big/2\,.$
The heuristic supercharge $Q:=\int c(x)j(x)\,dx$ defines the supersymmetry
generator $\delta$ as the graded derivation:
\[
\delta(A):=[Q,\,A_+]+\{Q,\,A_-\}
\]
which satisfies $\delta(AB)=\delta(A)B+\gamma(A)\delta(B)\;.$ Note that on
the generating elements of the field algebra we have:
\begin{equation}
\label{SusyD}
\delta\big(c(f)\big)=j(f)\,,\qquad \delta\big(j(f)\big)=ic(f')\;.
\end{equation}
Time evolution is given by translation, i.e.
\[
\alpha_t(c(f)):=c(f_t)\,\qquad\alpha_t(j(f)):=j(f_t)
\]
where $f_t(x):=f(x-t),$ $x\in\R\,.$
The generator of time evolution is the derivation:
\begin{equation}
\label{TimeD}
\delta_0(c(f))=ic(f')\,,\qquad\delta_0(j(f))=ij(f')\;.
\end{equation}
The supersymmetry relation is valid on the field algebra:
\begin{equation}
\label{SuSy}
\delta^2=\delta_0\;.
\end{equation}
Our problem is to realize this structure in a C*-algebra setting.
Some problems already arise from the relation
$\delta((c(f))=j(f),$ in which $\delta$ takes
a bounded operator to an unbounded one. We will deal with this issue
in the next section. A deeper source of problems will come from the
theorems of Kishimoto and Nakamura~\cite{KiNa} which
will make it hard to realize the supersymmetry relation
(\ref{SuSy}) on a dense domain.
% state:\chop
% Let $(\al A.,\,\gamma)$ be a graded C*-algebra and $\alpha:\R\to\aut\al A.$
% a pointwise continuous action, with a generator $\delta_0$ such that
% $C^\infty(\delta_0):=\bigcap\limits_{n=1}^\infty{\rm Dom}(\delta_0^n)$
% is dense. Let $\delta$ be a closable graded derivation with 
% $C^\infty(\delta_0)\subset{\rm Dom}(\delta)$ such that
% $\delta\circ\alpha_t=\alpha_t\circ\delta$ for all $t,$ and
% $\delta^2=\delta_0$ on $C^\infty(\delta_0).$
% Then $\delta$ is bounded.

%%%%%%%%%%%%%%%%%%%%%%%%%%%%%%%%%%%%%%%%%%%%%%%%%%%%%%%%%%%%%%%%%%%%%%%%%%%%%%
%%%%%%%%%%%%%%%%%%%%%%%%%%%%%%%%%%%%%%%%%%%%%%%%%%%%%%%%%%%%%%%%%%%%%%%%%%%%%%
\section{On Mollifiers and Resolvent Algebras}
\label{Moll}

Here we develop tools to handle the unboundedness of the
range elements of $\delta\,.$
Recall that a selfadjoint operator $A$ on a Hilbert space $\al H.$
is {\it affiliated} with a C*-algebra $\al A.\subset\al B.(\al H.)$ if
the resolvent ${(i\lambda\un-A)^{-1}}\in\al A.$ for some 
$\lambda\in\R\backslash0$ (hence for all $\lambda\in\R\backslash0).$ 
This notion is used by Georgescu~\cite{Georg} e.a.
(and is weaker than the one  used by  Woronowicz~\cite{Wor1}) 
and it implies the usual one, i.e. that $A$ commutes with all unitaries
commuting with $\al A.$ (but not conversely).
% It also implies that
% $\al A.$ contains all bounded continuous functions of $A.$
Observe that 
\[
A(i\lambda\un-A)^{-1}=\ol(i\lambda\un-A)^{-1}A.=i\lambda(i\lambda\un-A)^{-1}-\un
\in\al A.\;.
\]
Thus the resolvent ${(i\lambda\un-A)^{-1}}=M$ acts as a ``mollifier''
for $A,$ i.e. $\ol MA.$ and $AM$ are bounded and in $\al A.,$ and $M$
is invertible such that $M^{-1}\ol MA.=A=AMM^{-1}.$
This suggests that as $AM$ and $\ol MA.$ in $\al A.$ carries the
information of $A$ in bounded form, 
we can ``forget'' the original representation, and study the affiliated
$A$ abstractly through these elements.

We want to apply this idea to a representation of the bosonic fields
$j(f)=j(f)^*,$ $f\in\al S.(\R)$ where
\[
\big[j(f),\,j(g)\big]=i\sigma(f,\, g):=i\int fg'\, dx
\]
on some common dense invariant core $\al D.\subset\al H.$
of the selfadjoint fields $j(f)\,.$
It seems natural to look for mollifiers in the Weyl algebra
\[
\CCR=C^*\set\exp(ij(f)),{f\in\al S.(\R)}.\,,
\]
(abstractly $\CCR$ is the C*-algebra generated by a set
of unitaries ${\set\delta_f,{f\in\al S.(\R)}.}$ such that
$\delta_f^*=\delta_{-f}$ and $\delta_f\delta_g=e^{-i\sigma(f,g)/2}
\delta_{f+g}).$
Unfortunately this is not possible because:
\begin{pro}
\label{CCRX}
The Weyl algebra $\CCR$ contains no 
nonzero element $M$ such that $j(f)M$ is bounded 
for some $f\in\al S.(\R)\backslash0\,.$
Thus $\CCR$ contains no mollifier for any nonzero
$j(f),$  and $j(f)$ is not affiliated with $\CCR\,.$
\end{pro}
\begin{beweis}
Assume that $M\in\CCR$ is nonzero such that
$j(f)M$ is bounded for some nonzero
$f\in\al S.(\R)\,.$
Let $U(t):=\exp(itj(f)),$ and denote the spectral resolution
of $j(f)$ by $j(f)={\int\lambda\,dP(\lambda)},$ then
\begin{eqnarray*}
\left\|{(U(t)-\un)M}\right\| &=&
\bigg\|\int(e^{it\lambda}-1)dP(\lambda)M\bigg\|  \\[1mm]
&=& |t|\bigg\|\int{(e^{it\lambda}-1)\over t\lambda}\,dP(\lambda)
\int\lambda'\,dP(\lambda')M\bigg\|   \\[1mm] 
&\leq& C|t|\|j(f)M\|\longrightarrow 0
\end{eqnarray*}
as $t\to0,$ where we used the bound
${|{e^{ix}-1\over x}|}<C$ for some constant $C.$
Let $\al J.\subset\CCR$ consist of all elements $M$
such that $\left\|{(U(t)-\un)M}\right\|\to 0$ as $t\to 0.$
This is clearly a  norm-closed linear space,
and by the inequality $\left\|{(U(t)-\un)MA}\right\| 
\leq\left\|{(U(t)-\un)M}\right\|\,\|A\|$ it is also 
a right ideal. To see that it is a two sided ideal note that
\[
\left\|{(U(t)-\un)e^{ij(g)}M}\right\|=
\left\|{(U(t)e^{it\sigma(f,g)}-\un)M}\right\|
\]
still converges to $0$ as $t\to 0,$
and use the fact that $\CCR$ is the norm closure of the span 
of $\set e^{ij(g)},g\in{\al S.(\R)}.\;.$
But $\CCR$ is simple, hence $\al J.\ni M$ must be zero.
\end{beweis}

Our solution is to abandon the Weyl algebra as the appropriate
C*-algebra to model the bosonic fields $j(f),$ and instead to
choose the unital C*-algebra generated by the resolvents:
\[
C^*\set{\un,\;R(\lambda,f)},\lambda\in\R\backslash 0,\;
f\in{\al S.}(\R)\backslash 0.
\]
where $R(\lambda,f):=(i\lambda\un-j(f))^{-1}\,.$ Then by construction
all $j(f)$ are
 affiliated to this C*-algebra and it contains
mollifiers $R(\lambda,f)$ for all of them.

The above discussion took place in a concrete setting, 
i.e. represented on a Hilbert space, and we would like to abstract this.
Just as the Weyl algebra can be abstractly defined by the Weyl relations,
we now want to abstractly define the C*-algebra of resolvents
(of the $j(f))$ by generators and relations.
\begin{defi}
\label{ResAlg}
Given a symplectic space $(X,\,\sigma),$ we define %{\bf Resolvent
% Algebra} $\al R.(X,\,\sigma)$ 
$\al R._0$ to be the universal unital 
*-algebra generated by
the set $\set{R(\lambda,f)},\lambda\in\R\backslash 0,\;f\in X\backslash 0.$
and the relations
\begin{eqnarray}
\label{Rinvol}
R(\lambda,f)^*&=&R(-\lambda,f) \\[1mm]
\label{Rhomog}
\rlf &=& {1\over\lambda}\,R(1,\,{1\over\lambda}f)  \\[1mm]
\label{Resolv}
\rlf - R(\mu,f) &=& i(\mu-\lambda)\rlf R(\mu,f)  \\[1mm]
\label{Rccr}
\big[\rlf,\,R(\mu,g)\big] &=&
i\sigma(f,g)\,\rlf\,R(\mu,g)^2\rlf \\[1mm]
\label{Rsum}
\rlf R(\mu,g)&=& R(\lambda+\mu,\,f+g)[\rlf+R(\mu,g)
+i\sigma(f,g)\rlf^2R(\mu,g)]
\end{eqnarray}
where $\lambda,\;\mu\in\R\backslash 0$ and $f,\,g\in X\backslash 0\,,$
and for (\ref{Rsum}) we require $\lambda+\mu\not=0$ and
$f+g\not=0\,.$ 
That is, start with the free unital *-algebra generated by
 $\set{R(\lambda,f)},\lambda\in\R\backslash 0,\;f\in X\backslash 0.$
 and factor out by the ideal generated by the relations
(\ref{Rinvol}) to (\ref{Rsum}) to obtain the *-algebra
 $\al R._0\,.$ % Define the enveloping C*--seminorm on  $\al R._0$
 % by 
% \[
% \|A\|_0:=\sup\set\|\pi(A)\|,\pi\in{\rm Rep}\,{\al R._0}.
% \]
% where ${\rm Rep}\,{\al R._0}$ is the bounded Hilbert space representations
% of $\al R._0\,.$ Then $\al R.(X,\,\sigma)$ is the closure of
% $\al R._0\big/\ker\|\cdot\|_0$ w.r.t.  the norm $\|\cdot\|_0\;.$
\end{defi}
\begin{rem}
\begin{itemize}
\item[(i)]
The *-algebra $\al R._0$ is nontrivial, because it has nontrivial
representations. For instance, in a Fock representation of the
CCRs over $(X,\,\sigma)$ we have the CCR-fields $\varphi(f)$
from which we can define $\pi(\rlf)=(i\lambda\un-\varphi(f))^{-1}$
to obtain a representation of $\al R._0.$
\item[(ii)] Obviously (\ref{Rinvol}) encodes the selfadjointness of 
$j(f),$ (\ref{Rhomog}) encodes $j(\lambda f)=\lambda j(f),$
(\ref{Resolv}) encodes that $\rlf$ is a resolvent,
(\ref{Rccr}) encodes the canonical commutation relations
and (\ref{Rsum}) encodes additivity
% Note that we did not require full linearity;- in  fact if we also
$j(f+g)=j(f)+j(g)\,.$ % we will have to add the condition
% to those in Definition~\ref{ResAlg}.
Moreover, the identity was added explicitly, we do not have that
$R(1,0)=-i\un\,,$ in fact $R(1,0)$ is undefined.
% \item[(iii)] Let $\mu=-\lambda$ in Equation~(\ref{Resolv}) to get that
% $\rlf-\rlf^*=-2i\lambda\rlf\rlf^*\,.$
\end{itemize}
\end{rem}
To define our resolvent C*-algebra, we need to decide on which 
C*-seminorm to define on $\al R._0.$
The obvious choice is the enveloping
C*-norm, however for the purpose of our model, it is more convenient 
to use a different norm, which we now define.
We will say that a state $\omega$ on the Weyl algebra $\CCRX$
is {\bf strongly regular} if the functions
\[
\R^n\ni(\lambda_1,\ldots,\lambda_n)\to\omega\big(\delta_{\lambda_1f_1}
\cdots\delta_{\lambda_nf_n}\big)
\]
are smooth for all $f_1,\ldots,\,f_n\in X$ and all 
$n\in\N\,.$ Of special importance is that the GNS-representation 
of a strongly regular state has a common dense invariant domain
for all the generators $j(f)$ of the one parameter groups
$\lambda\to\pi_\omega(\delta_{\lambda f})$
(this domain is obtained by applying the polynomial algebra of the Weyl operators
$\set\pi_\omega(\delta_f),f\in X.$ to the cyclic GNS-vector).
Some important classes of states, e.g. quasi-free states are
strongly regular. Denote by $\pi_S$ the direct sum of the GNS-representations
of all strongly regular states, then as the resolvents of the fields
are in $\pi_S\big(\CCRX\big)'',$ 
we can extend $\pi_S$ to a representation
of $\al R._0$ by the Laplace transform:
\begin{equation}
\label{Laplace1}
\pi_S(R(\lambda,f)):=-i\int_0^\infty e^{-\lambda t}\pi_S(\delta\s -tf.)\,dt\;,
\qquad\lambda>0\,.
\end{equation}
We define our {\bf resolvent algebra} $\al R.(X,\,\sigma)$ as the  abstract
C*-algebra generated by $\pi_S(\al R._0),$ i.e. we factor $\al R._0$ by
$\ker\pi_S$ and complete w.r.t. the operator norm of $\pi_S\,.$

We state some elementary properties of $\al R.(X,\,\sigma)\,.$
\begin{teo}
\label{Relemen}
Let $(X,\,\sigma)$ be a given nondegenerate symplectic space, and define
$\al R.(X,\,\sigma)$ as above. Then for all $\lambda,\,\mu\in\R\backslash 0$
and $f,\, g\in X\backslash 0$ we have:
\begin{itemize}
\item[(i)] $[\rlf,\,R(\mu,f)]=0\,.$ 
 Substitute $\mu=-\lambda$ to see that
$\rlf$ is normal.
\item[(ii)] $\big\|\rlf\big\|=|\lambda|^{-1}\,.$
\item[(iii)] $\rlf$ is analytic in $\lambda\,.$ Explicitly, the series 
expansion:
\[
\rlf=\sum_{n=0}^\infty(\lambda_0-\lambda)^n\,R(\lambda_0,f)^{n+1}i^n,
\qquad\lambda,\;\lambda_0\not=0\qquad\qquad
\hbox{(Von Neumann series)}
\]
converges in norm whenever $|\lambda_0-\lambda|<|\lambda_0|\;.$
\item[(iv)] $R(\lambda,tf)$ is norm continuous in $t\in\R\backslash 0\,.$
\item[(v)] $\rlf R(\mu,g)^2\rlf= R(\mu,g)\rlf^2 R(\mu,g)\,.$
\item[(vi)] Let $T\in{\rm Sp}(X,\sigma)$ be a symplectic transformation.
then $\alpha\big(\rlf\big):=R(\lambda,Tf)$ defines  a unique
automorphism $\alpha\in\aut\al R.(X,\,\sigma)\,.$
\end{itemize}

\end{teo}
Note that the von Neumann series for $\rlf$ converges for any
$z\in\C$ with $|z-\lambda_0|<|\lambda_0|,$ i.e. on a disk
which stays off the real line. Using different $\lambda_0's$ 
we can thus define $R(z,f)$ for any complex $z$ not on the real line
and deduce the properties in the definition for these from
the series. Thus we obtain also resolvents $R(z,f)$ for complex
$z$ in $\al R.(X,\,\sigma)\,.$

Any operator family $R_{\lambda}$ satisfying the resolvent equation
(\ref{Resolv}) is called by Hille a pseudo-resolvent (cf. p215
in~\cite{Yos}), and for such a family we know (cf. Theorem~1 p216
in~\cite{Yos}) that:
\begin{itemize}
\item{} All $R_\lambda$ have a common range and a common null space.
\item{} A pseudo resolvent $R_{\lambda}$ is the resolvent for an operator
$B$ iff $\ker R_\lambda=\{0\}\,,$ and in this case
$\dom B=\ran R_\lambda$ for all $\lambda\,.$
\end{itemize}
Thus we define:
\begin{defi}
A {\bf regular representation} $\pi\in\rep\al R.(X,\,\sigma)$
is a Hilbert space representation such that 
\[
\ker\pi\big(R(1,f)\big)=\{0\}\qquad\forall\;f\in
\al S.(\R,\R)\backslash 0\;.
\]
We denote the collection of regular representations by $\reg\,.$
\end{defi}
Obviously many regular representations are known, e.g. $\pi_S$ and the
Fock representation.
Given a $\pi\in\rep\al R.(X,\,\sigma)$ with 
$\ker\pi\big(R(1,f)\big)=\{0\},$ we can define a field 
operator by 
\[
j_\pi(f):=i\un-\pi\big(R(1,f)\big)^{-1}\;
\]
with domain $\dom j_\pi(f)=\ran\pi\big(R(1,f)\big)\,.$
Thus for $\pi\in\reg,$ all the field operators
$j_\pi(f),$ $f\in
\al S.(\R,\R)$ are defined, and we have the resolvents 
$\pi(\rlf)=(i\lambda\un-j_\pi(f))^{-1}\,.$
\begin{teo} 
\label{RegThm}
Let $\al R.(X,\,\sigma)$ be as above, and let $\pi\in\rep\al R.(X,\,\sigma)$
satisfy $\ker\pi\big(R(1,f)\big)=\{0\}= \ker\pi\big(R(1,h)\big)$ for given
$f,\; h\in X.$ Then
\begin{itemize}
\item[(i)]  $j_\pi(f)$ is selfadjoint, and $\pi(\rlf)\dom j_\pi(h)
\subseteq\dom j_\pi(h)\,.$
\item[(ii)]
$\lim\limits_{\lambda\to\infty}i\lambda\pi(\rlf)\psi=\psi$ for all
$\psi\in\al H._\pi,$
\item[(iii)]
$\lim\limits_{s\to 0}i\pi(R(1,sf))\psi=\psi$ for all
$\psi\in\al H._\pi.$
\item[(iv)]
The space $\al D.:={\pi\big(R(1,f)R(1,h)\big)\al H._\pi}$
is a joint dense domain 
for $j_\pi(f)$ and $j_\pi(h)$ 
and we have:
$[j_\pi(f),\,j_\pi(h)]=i\sigma(f,h)$ on $\al D.,$
\item[(v)] $j_\pi(\lambda f+h)=\lambda j_\pi(f)+j_\pi(h)$ for all
$\lambda\in\R$ on $\al D.,$
\item[(vi)]
$j_\pi(f)\pi(\rlf)=\pi(\rlf)j_\pi(f)=i\lambda\pi(\rlf)-\un$ on
$\dom j_\pi(f),$
\item[(vii)]
$\big[j_\pi(f),\pi(R(\lambda,h))\big]=i\sigma(h,f)\pi(R(\lambda,h)^2)$
on $\dom j_\pi(f),$
\item[(viii)]
Denote $W(f):=\exp(ij_\pi(f))\,,$ then
\begin{eqnarray*}
W(f)W(h) &=& e^{i\sigma(f,h)}W(h)W(f)  \\[1mm]
W(f)\pi\big(R(\lambda,h)\big)W(f)^* &=&
\pi\big(R(\lambda+i\sigma(f,h),\,h)\big)\,.
\end{eqnarray*}
Moreover  $W(f)\al D.\subseteq\al D.\supseteq W(h)\al D.,$ hence
 $\al D.:={\pi\big(R(1,f)R(1,h)\big)\al H._\pi}$
is a common core
for $j_\pi(f)$ and $j_\pi(h)\,.$
\end{itemize}
\end{teo}
A distinguished regular representation of $\al R.(X,\,\sigma)$ is of course
the defining strongly regular representation $\pi_S.$ 
By definition
$\al R.(X,\,\sigma)$ is faithfully represented in it, and moreover, there
is a common dense invariant domain $\al D._0$ for all the field operators
$j\s\pi_S.(f),$ $f\in X.$ 
This domain can be enlarged to a dense invariant domain $\al D._T$
for both the resolvents and the fields simply by applying
 all polynomials
in $j\s\pi_S.(f)$ and $\pi_S(R(\lambda,f))$ to $\al D._0,$ 
which makes sense, because from (i) above all resolvents
preserve the joint domain $\mathop{\bigcap}\set
\dom j\s\pi_S.(f),{f\in X}..$ 
Thus we can form the *-algebra
 of (unbounded) operators  
\[
\al E._0:=*\hbox{--alg}\set{
j\s\pi_S.(f),\;
\pi_S(R(\lambda,f))},f\in X,\;\lambda\in\R\backslash 0.
\]
on  $\al D._T\,.$
Then $\al E._0$ contains of course the *-algebra 
$\pi_S(\al R._0)$ generated by resolvents alone, which is dense in
$\al R.(X,\,\sigma).$ 
We will need these *-algebras $\al E._0\supset\pi_S(\al R._0)$
below, and will generally not indicate the faithful representation
$\pi_S$ w.r.t. which they are defined.
Note that for any strongly regular state $\omega,$ its cyclic GNS-vector
is in the domain of all $j\s\pi_\omega.(f)\,,$ hence $\omega$
extends to define a functional on $\al E._0\,.$ Thus 
we give a meaning to all
expressions of the form 
\begin{eqnarray*}
&&\!\!
\omega\big(j(f_1)\cdots j(f_n)R(\lambda_1,g_1)\cdots R(\lambda_k,g_k)\big)
\\[1mm]
&&\qquad
:=\left(\Omega\s\omega.,\,j\s\pi_\omega.(f_1)\cdots j\s\pi_\omega.(f_n)
\pi_\omega\big(R(\lambda_1,g_1)\cdots)R(\lambda_k,g_k)\big)\Omega\s\omega.
\right)
\end{eqnarray*} 
as above.
A very important class of states on $\CCRX$ are the quasifree states,
which we will need below.
They are given by 
\[
\omega(\delta_f) = \exp\big(-\hlf\langle f | f \rangle_\omega\big) , 
\quad f \in X,
\]
where $\langle \, \cdot \, | \, \cdot \, \rangle_\omega $ 
is a (possibly semi--definite) scalar product on the complex 
linear space $X + i X$ satisfying 
$$ \langle f | g \rangle_\omega- 
\langle g | f \rangle_\omega = i \sigma(f,g), 
\quad f,g \in X. $$
Any quasifree state is 
also regular in the strong sense.
By a routine computation one can represent the expectation
values of products of Weyl operators in a quasifree state 
in the form
$$ \omega(\delta_{f_1} \cdots \delta_{f_n}) = 
\exp\Big(- \sum_{k<l}  \langle f_k | f_l \rangle_\omega 
-\hlf \sum_l   \langle f_l | f_l \rangle_\omega \Big). $$
Making use of the Laplace transform~(\ref{Laplace1}) for the GNS--represntation
of the resolvents, we 
 have for $\lambda_1, \dots \lambda_n > 0$
\begin{eqnarray}
& & \omega(R(\lambda_1,f_1) \cdots R(\lambda_n,f_n))  \nonumber\\[1mm]
& & = (-i)^n 
\int_0^\infty \! dt_1 \dots  \int_0^\infty \! dt_n \,
e^{-{\sum}_k t_k \lambda_k} \omega(\delta\s t_1 f_1. \cdots
\delta\s t_n f_n.) \nonumber\\[1mm]
\label{QFresolvent}
& & =  (-i)^n  
\int_0^\infty \! dt_1 \dots  \int_0^\infty \! dt_n \,
\exp\Big(-\sum_k t_k \lambda_k
- \sum_{k<l} \, t_k t_l \langle f_k | f_l \rangle_\omega 
-\hlf \sum_l \,  t_l^2 \langle f_l | f_l \rangle_\omega\Big). 
\end{eqnarray}

\noindent {\bf Remark:} One can replace anywhere in this
equation $f_k$ by $-f_k$, thus it does not impose any restriction of
generality to assume that $\lambda_1, \dots \lambda_n > 0$. The
 relation (\ref{QFresolvent}) should be regarded as the definition 
of quasifree states on the resolvent algebra.

In our calculations below, we will frequently need the following
differentiablility of quasifree states:
\begin{pro}
\label{MasterLemma}
Let $\omega$ be a  quasifree state as above, and let
$x_k\in\R \mapsto f_k(x_k) \in X$, $k = 1, \ldots n$
be paths in $X$ for which the functions 
$ x_k,x_l \longmapsto \langle  f_k( x_k) | f_l( x_l)
\rangle_\omega, 
\quad k,l = 1, \dots n,$ 
are smooth. Then
\begin{eqnarray*}
&&\!\!\!\!\! \frac{\partial}{\partial x_r} \,
\omega\Big(R(\lambda_1, f_1(x_1)) \cdots R(\lambda_n, f_1(x_n))\Big) 
\\[1mm]
&=& -  \sum_{k=1}^{r-1}
  \frac{\partial }{\partial x_r}\Big(\big\langle f_k(x_k) | 
f_r(x_r) \big\rangle_\omega  \Big)
\, \frac{\partial^2}{\partial \lambda_r\partial\lambda_k}
% \frac{\partial}{\partial \lambda_k}
\, \omega\Big(R(\lambda_1, f_1(x_1)) \cdots R(\lambda_n, f_1(x_n))\Big)\\[1mm]
&&\qquad-\hlf\frac{\partial }{\partial x_r}\Big(\big\langle f_r(x_r) | 
f_r(x_r) \big\rangle_\omega  \Big)
\, \frac{\partial^2}{\partial \lambda_r^2} % \frac{\partial}{\partial \lambda_r}
\, \omega\Big(R(\lambda_1, f_1(x_1)) \cdots R(\lambda_n, f_1(x_n))\Big)\\[1mm] 
&&\qquad -  \sum_{k=r+1}^n  
  \frac{\partial }{\partial x_r}\Big(
\big\langle f_r(x_r) | 
  f_k(x_k)
\big\rangle_\omega  \Big)
\, \frac{\partial^2}{\partial \lambda_r\partial \lambda_k} 
% \frac{\partial}{\partial \lambda_k}
\, \omega\Big(R(\lambda_1, f_1(x_1)) \cdots R(\lambda_n, f_1(x_n))\Big)
\end{eqnarray*}
% \begin{eqnarray*}
% &&\!\!\!\!\! \frac{\partial}{\partial x_r} \,
% \omega\Big(R(\mu_1, f_1(x_1)) \cdots R(\mu_n, f_1(x_n))\Big) 
% \\[1mm]
% &=& -  \sum_{k=1}^n  
%   \frac{\partial }{\partial x_r}\Big(\big\langle f_r(x_r) | 
% f_k(x_k) \big\rangle_\omega  +
% \big\langle f_k(x_k) | 
%   f_r(x_r)
% \big\rangle_\omega \Big)
% \, \frac{\partial}{\partial \mu_r} \frac{\partial}{\partial \mu_k}
% \, \omega\Big(R(\mu_1, f_1(x_1)) \cdots R(\mu_n, f_1(x_n))\Big).
% \end{eqnarray*}
and all the partial derivatives involved in this formula exist.
\end{pro}

\section{C*-algebra formulation of Supersymmetry.}

Here we want to write our model of Section~\ref{Heur} in a 
C*-algebra framework.
However, to motivate our choices made below, let us recall
a theorem of Kishimoto and Nakamura~\cite{KiNa}:
\begin{teo}
Let $\al A.$ be a C*-algebra with $\Z_2\hbox{--grading}$ $\gamma,$
let $\alpha:\R\to\aut\al A.$ be a pointwise continuous action with
generator $\delta_0$ having a smooth domain $C^\infty(\delta_0):=
\bigcap\limits_{n=1}^\infty\dom(\delta_0^n)\,.$
Let $\delta$ be a closable graded derivation with 
$\dom(\delta)\supset C^\infty(\delta_0),$ $\delta\circ\alpha_t
=\alpha_t\circ\delta$ for all $t,$ and $\delta^2=\delta_0$
on $ C^\infty(\delta_0).$ Then $\delta$ is bounded.
\end{teo}
Thus it will be hard to obtain the supersymmetry relation on 
natural dense domains.

For the fermion field, let $\al H.=L^2(\R)$ and define ${\rm CAR}(
\al H.)$ in Araki's self-dual form (cf.~\cite{Ar}) as follows.
On $\al K.:=\al H.\oplus\al H.$ define an antiunitary involution
$\Gamma$ by $\Gamma(h_1\oplus h_2):=\ol h._2\oplus\ol h._1\;.$
Then ${\rm CAR}(\al H.)$ is the unique simple C*--algebra with
generators $\set\Phi(k),k\in{\al K.}.$ such that
$k\to\Phi(k)$ is antilinear, $\Phi(k)^*=\Phi(\Gamma k)\,,$
and
\[
\left\{\Phi(k_1),\,\Phi(k_2)^*\right\}=(k_1,\,k_2)\un\;,\qquad
k_i\in\al K.\;.
\]
The correspondence with the heuristic creators and annihilators
of fermions is given by $\Phi(h_1\oplus h_2)=a(h_1)+a^*(\ol h._2)\;,$
where
\[
a(h)=\int a(x)\,\ol h(x).\,dx\;,\qquad
a^*(h)=\int a^*(x)\, h(x)\, dx\;.
\]
To obtain the Clifford operators $c(f)=c(f)^*,$ 
$f\in\al S.(\R,\,\R)$ we take $c(f):=\Phi(f\oplus f)/\sqrt{2}\,,$
in which case we have  $c(f)=c(f)^*$
and $\{c(f),\,c(g)\}=(f,g)=\int fg\,dx\,.$
Let $\ClifS:=C^*\set c(f),{f\in\al S.(\R)}.$ 
and notice that $\wt{c}(f):=i\Phi(f\oplus -f)/\sqrt{2}$
also satisfies the Clifford relations, hence generates another 
copy of $\ClifS$ in ${\rm CAR}(\al H.)\,,$
and together these two Clifford algebras generate all
of ${\rm CAR}(\al H.)\,.$ In fact, since the $c(f)$ and 
$\wt c(g)$ anticommute, we have that ${\rm CAR}(\al H.)
\cong{{\rm Cliff}\big(\al S.(\R)\oplus\al S.(\R)\big)}\,.$
 Conversely, if we are given a real pre-Hilbert space $X$
 with complexified completion $Y$ and a projection $P$
 and antiunitary involution $\Gamma$ such that
 $\Gamma P\Gamma=\un-P$ and these preserve $X,$ then 
 we have an isomorphism ${\rm Cliff}(X)\cong{\rm CAR}(Y)$
 given by $\Phi(x)=\big(c(Px)-ic(\Gamma Px)
+c(P\Gamma x)+ic((\un-P)x)\big)/\sqrt{2}\,.$
%%
% [Can't see why the C*-alg generated by $\{c(f);f\in\al S.\}$ is
% ${\rm CAR}(\al H.)$- not true in one dim, when $a=\sigma_x+i\sigma_y$
% and $c=(a+a^*)/\sqrt{2}=\sigma_x\sqrt{2}$]\chop

For the bosonic part we take the resolvent algebra $\al R.(\al S.(\R),\sigma)$
where $\sigma(f,g):=\int fg'\,dx\,,$ and so the full C*-algebra in which we want to
define our model is
\[
\al A.:=\ClifS\otimes\al R.(\al S.(\R),\sigma)
\]
where the tensor norm is unique because the CAR-algebra is nuclear.
The grading automorphism $\gamma$ is the identity on $\al R.(\al S.(\R),\sigma)\,,$
and $\gamma(\Phi(k))=-\Phi(k)$ for all $k$ on the CAR-part.

Next, we want to define on some suitable domain in $\al A.$ the supersymmetry
graded derivation $\delta$ corresponding to the relations~(\ref{SusyD}).
First, considering
$\delta\big(j(f)\big)=ic(f'),$ since $\delta$ is a derivation on the 
bosonic part, it is natural to define
\begin{equation}
\label{DerR}
\delta(\rlf):=ic(f')\,\rlf^2\quad\in\al A.\,.
\end{equation}
However due to the unbounded rhs of
$\delta\big(c(f)\big)=j(f)$ we cannot define $\delta$ directly on the
$c(f),$ so we need to multiply by mollifiers. Define
\begin{eqnarray}
\label{DefZ}
\zeta(f)&:=&c(f)R(1,f)\,, \\[1mm]
\label{DerZ}
\hbox{then}\quad\qquad
\delta(\zeta(f))&:=&i R(1
,f)-\un+ic(f)c(f')R(1,f)^2\quad\in\al A.
\end{eqnarray}
where we made use of the graded derivation property, the
relations~(\ref{SusyD}) and $j(f)R(\lambda,f)=i\lambda R(\lambda,f)-\un\,.$
Next, we would like to extend $\delta$ as a graded derivation to the 
*-algebra generated by these basic objects:
\[
\al D._S:=\hbox{*--alg}
\set{\un,\;\rlf,\,\;\zeta(f)},{\lambda\in\R\backslash 0,\;
f\in\al S.(\R)\backslash 0}.\subset\al A.\;.
\]
Observe that $\al D._S$ is not norm-dense in $\al A.,$ however
due to~Theorem~\ref{RegThm}(ii) applied to  $R(\lambda,f)c(f)=
\zeta(f/\lambda),$
it will be strong operator dense
in $\al A.$ in any regular representation.
Note that $\delta$ does not preserve $\al D._S,$ it takes its
image in the norm dense *-algebra
\[
\al A._0:=\hbox{*--alg}
\set{\un,\;\rlf,\,\;c(f)},{\lambda\in\R\backslash 0,\;
f\in\al S.(\R)\backslash 0}.\subset\al A.\;.
\]
To see that $\delta$ extends as a graded derivation to $\al D._S,$
we proceed as follows.
 Let $\pi_0$ be any representation of 
% ${\rm CAR}(\al H.),$ 
$\ClifS$ then $\pi_0\otimes\pi_S$ is a faithful representation of
$\al A.$ and there is a common dense invariant domain $\al D.:=\al H.\s\pi_0.
\otimes\al D._T$ for all $\pi_0(c(f))\otimes\un,$
$\un\otimes j\s\pi_S.(f)$ and $\un\otimes \pi_S(\rlf)$ where 
$\al D._T$ denotes the domain 
of $\al E._0$ defined at the end of Section~\ref{Moll}. (Henceforth we will not
indicate tensoring by $\un$ nor the representations $\pi_0,\;\pi_S$
when the context makes clear what is meant).
Let 
\[
\al E.:=*\hbox{--algebra}\set{c(f),\;j(f),\;\rlf},{f\in\al S.(\R),\;
\lambda\in\R\backslash 0}.\supset\al A._0
\]
so we have the *-algebras $\al R._0\subset\al E._0\subset\al E.$
on $\al D.\,.$
Define on the generating elements of $\al E.$
 a map $\overline{\delta}$, setting
\begin{eqnarray*}
 \overline{\delta}(j(f)) &=& i c(f^\prime), \\[1mm]
 \overline{\delta}(R(\lambda,f)) &=& i c(f^\prime) R(\lambda,f)^2, \\[1mm]
 \overline{\delta}(c(f)) &=& j(f).
\end{eqnarray*}
We will see that this map extends to a graded derivation
on $\al E.$. For the proof it suffices to show that $\overline{\delta}$
is linear and satisfies the graded Leibniz rule on any finite
polynomial involving operators $j(f)$, $R(\lambda,f)$ and $c(f)$, 
i.e. in each instance only a finite number of test functions $f$ and real
parameters $\lambda$ are involved. We will take advantage of this
fact as follows.

Let $X_s \subset \al S.(\R)$ be any \textit{finite}--dimensional subspace
and consider the subalgebra $\al E.(X_s) \subset \al E.$
generated by the elements $j(f)$, $R(\lambda,f)$ and $c(f)$ with
$\lambda \in \R\backslash 0$, $f \in X_s$. We extend $X_s$ to a space $
{X_s}^\prime \subset\al S.(\R) $
by adding to the elements of $X_s$ also their \textit{first}
derivatives. Picking in  ${X_s}^\prime$ some (finite) 
orthonormal basis $\{h_n\}$ with regard to the
scalar product $( \cdot  , \, \cdot ),$ 
we have for any $f \in X_s$ the ``completeness
relations''
$ \sum_n \, ( h_n , f)  \, h_n = f, \
\sum_n \, ( h_n , f^\prime ) \, h_n = f^\prime $.

Next, we define an operator $Q_s \in \al E.$, setting
$$ Q_s = {\sum}_n c(h_n) j(h_n). $$
As $Q_s$ is of fermionic (odd) type, we can consistently define with the 
help of it a graded derivation $\overline{\delta}_s$ on $\al E.$, setting 
for even and odd elements $E_\pm \in \al E.$, respectively, 
$$
\overline{\delta}_s(E_+) = [Q_s, E_+] =  Q_s E_+ - E_+ Q_s, \quad 
\overline{\delta}_s(E_-) = \{Q_s,E_-\} = Q_s E_- + E_- Q_s.
$$
Computing the action of $\overline{\delta}_s$ on the
even elements $j(f)$, $R(\lambda,f)$ and odd elements $c(f)$, where 
$\lambda \in \R\backslash 0$, $f \in X_s$, we obtain from the basic relations in 
$\al E.$ by some elementary algebraic manipulations.
\begin{eqnarray*}
 \overline{\delta}_s(j(f)) &=& 
i {\sum}_n \, c(h_n)\, ( h_n , f^\prime ) = i c(f^\prime)\,, \\[1mm]
 \overline{\delta}_s(R(\lambda,f)) &=&  
i {\sum}_n \, c(h_n) \,( h_n , f^\prime) \,  R(\lambda,f)^2 
= i c(f^\prime) R(\lambda,f)^2, \\[1mm]
 \overline{\delta}_s(c(f)) 
&=&  {\sum}_n \, ( h_n,\,  f)\, j(h_n) = j(f).
\end{eqnarray*}

Thus we conclude that the action of $\overline{\delta}$ on the
generating elements of $\al E.(X_s)$ coincides with the action
of the graded derivation $\overline{\delta}_s$. As the choice
of the subspace $X_s$ was arbitrary, it follows that
$\overline{\delta}$ extends to a graded derivation on the 
whole polynomial algebra $\al E.$.

The final step consists in showing that the action of 
$\delta$ on the generating elements 
$R(\lambda,f)$, $\zeta(f)$ of $\al D._S$  coincides
with the action of the graded derivation 
$\overline{\delta}$. But this follows 
immediately from the relations given above. 
Thus $\delta$ extends to a graded derivation
with domain $\al D._S\,$  and range in $\al A._0$.
Uniqueness is clear from the graded derivation property, so we
have proven:

\begin{teo}
\label{DwellDf}
There is a unique graded derivation $\delta:\al D._S\to\al A.$
satisfying relations (\ref{DerR}) and (\ref{DerZ}).
\end{teo}

Next we need to define the time evolution derivation $\delta_0$ in
this C*-setting. From the  equations (\ref{TimeD})
this suggest that we define on $\al E.$ a *-derivation
$\overline{\delta}_0$ satisfying:
\begin{eqnarray*}
% \label{TimeE}
 \overline{\delta}_0(j(f)) &=& i j(f^\prime), \\[1mm]
 \overline{\delta}_0(R(\lambda,f)) &=& 
i R(\lambda,f) j(f^\prime) R(\lambda,f), \\[1mm]
 \overline{\delta}_0(c(f)) &=& i \, c(f^\prime)
\end{eqnarray*}
and then  proceed to the corresponding mollified relations in $\al A..$
For the proof that $\overline{\delta}_0$ extends to a % (skew symmetric)
*-derivation on  $\al E.,$
we proceed as in the discussion of the superderivation: We pick
any finite dimensional subspace $X_s \subset \al S.(\R),$ 
consider the corresponding subalgebra $\al E.(X_s) \subset \al E.$
and  choose in the extended space ${X_s}^\prime \subset\al S.(\R)$, 
containing the elements of $X_s$ and their first
derivatives, some orthonormal basis $\{h_n\}$. In addition 
to the completeness relations mentioned above we will also make use of  
$ \sum_n \, (h_n,  f ) \, h_n^\prime = f^\prime$
for $f \in X_s$.

We consider now the symmetric operator in $\al E.$
$$ H_s = \frac{1}{2} \, {\sum}_n \big\{ i c({h_n}^\prime) c(h_n) +
j(h_n) j(h_n) \big\}. $$     
Putting 
$$  \overline{\delta}_{0 \, s}(\,\cdot \,) =  [H_s, \, \cdot \,], $$
it induces a *--derivation on $\al E.$. Its action on the generating elements
of  $\al E.(X_s)$ can easily be computed:
\begin{eqnarray*}
 \overline{\delta}_{0 \, s}(j(f)) &=& 
 {\sum}_n \, j(h_n) \, i (h_n , f^\prime ) = i \,  
 j(f^\prime), \\[1mm]
 \overline{\delta}_{0 \, s}(R(\lambda,f)) &=&  
{\sum}_n \, 
i (h_n , f^\prime )\,  R(\lambda,f) j(h_n)  R(\lambda,f) 
= i \, R(\lambda,f) j(f^\prime)  R(\lambda,f), \\[1mm]
 \overline{\delta}_{0 \, s}(c(f)) 
&=& \frac{1}{2} \, {\sum}_n \, i \big\{ ( h_n , f )\, c({h_n}^\prime) 
+ (h_n , f^\prime )\, c(h_n) \big\} = i \, c(f^\prime).
\end{eqnarray*}
Thus we conclude as in the preceding discussion
that the action of $\overline{\delta}_0$ on the
generating elements of $\al E.(X_s)$ coincides with the action
of the derivation $\overline{\delta}_{0 \, s}$. 
As $X_s$ was arbitrary, it follows that
$\overline{\delta}_0$ extends to a derivation on the 
whole polynomial algebra $\al E.$.

The supersymmetry relation  $\overline{\delta}^2 =
\overline{\delta}_0$ can now be verified on the generating elements 
of $\al E.$ and thus holds on the whole algebra $\al E..$

The question now is how one should define the time evolution
$\delta_0$ and the square $\delta^2$ on the C*-algebra $\al A.$
from the unbounded versions in $\al E..$
Since $\delta:\al D._S\to\al A._0,$ its square
$\delta^2$ does not make sense on $\al D._S.$ 
Note however that for every $A\in\al A._0$ there is a monomial $M\in\al D._S$
of resolvents $\rlf$
 such that 
\[
AM\in\al D._S\ni MA\,.
\]
(By Theorem~\ref{RegThm}(ii) we know that in regular representations we can let
these mollifiers $M$ go to $\un$ in the strong operator topology.)
\begin{defi}
\label{deltasquared}
For each $A\in \al D._S$ let $M_A\in \al D._S$ be a monomial of resolvents
$\rlf$ such that $M_A\delta(A)\in \al D._S.$ Define
\[
M_A\delta^2(A):=\delta\big(M_A\delta(A)\big)-\delta(M_A)\delta(A)\in\al A._0\;.
\]
\end{defi}
Note that this definition coincides with $M_A\overline\delta^2(A)$
in $\al E.,$ however the definition above involves only bounded
quantities, so it can be defined independently in the C*-setting
on $\al D._S\,.$
Of course we then have the mollified SUSY--relations
$M_A\delta^2(A)=M_A\overline{\delta}_0(A)$ for all
$A\in\al D._S$ from the unbounded SUSY relation in $\al E.\,.$
This is not however acceptable for a bounded SUSY--relation until we have
demonstrated the connection of $M_A\overline{\delta}_0(A)$
with the time evolution.
The time evolution $\alpha:\R\to\aut\al A.$
is just translation, as this is a chiral theory
\[
\alpha_t(c(f)):=c(f_t)\,,\quad\alpha_t\left(\rlf\right)
=R(\lambda,f_t)\;.
\]
The desired connection 
\begin{equation}
\label{timeconn}
M_A\overline{\delta}_0(A)=-i{d\over dt}
M_A\,\alpha_t(A)\Big|_0
\end{equation}
exists only in specific regular representations
on suitable domains,
and for these one will then have
supersymmetry. 
In many applications, one only needs the supersymmetry weakly, i.e.
\[
\omega(BM_A\overline{\delta}_0(A)C)=-i{d\over dt}
\omega(BM_A\,\alpha_t(A)C)\Big|_0=\omega(BM_A\delta^2(A)C)
\]
for $A,B,C$ in a suitable domain and $\omega$ a
distinguished functional.
We will verify this relation explicitly below for the
functionals used in our constructions.

\section{ Graded KMS--functionals.}

Graded KMS--functionals are used in supersymmetric theories to
calculate cyclic cocycles~\cite{JLO, Ka}, and here we want to 
develop this theory in the current context for our
simple supersymmetric model as a first application of it.

\begin{defi}
\label{KMSfunDef}
Let $\al A.$ be a unital C*-algebra with a grading automorphism
$\gamma\in\aut\al A.,$ $\gamma^2=\iota,$ and a (pointwise continuous)
action $\alpha:\R\to \aut\al A.$ such that $\alpha_t\circ\gamma
=\gamma\circ\alpha_t$ for all $t\,.$ Then a {\bf graded KMS--functional}
is a (possibly unbounded) functional $\varphi$ on $\al A.$
such that 
\begin{itemize}
\item[(i)] $\dom\varphi$ is a unital dense *--subalgebra of $\al A.$
such that
\[
\gamma(\dom\varphi)\subseteq\dom\varphi\supseteq\alpha_t(\dom\varphi)
\quad\forall\,t,
\]
\item[(ii)]
For all $A,\,B\in\dom\varphi$ there is a continuous complex function
$F\s{A,B}.:S\to\C$ on the strip $S:=\R+i{[0,1]}$ which is analytic on the
interior of $S$ and satisfying on the boundary:
\begin{eqnarray*}
F\s{A,B}.(t)&=& \varphi\left(A\,\alpha_t(B)\right)\quad\forall\,t  \\[1mm]
F\s{A,B}.(t+i)&=&\varphi\left(\alpha_t(B)\gamma(A)\right)\quad\forall\,t\in\R\,.
\end{eqnarray*}
\item[(iii)]
For $A,\,B\in\dom\varphi$ we have 
\[
\left|F\s{A,B}.(t+is)\right|<C(1+|t|)^N\quad\forall\,t\in\R,\;
s\in(0,1)
\]
and some $C\in\R_+$ and $N\in\N$ depending on $A$ and $B.$
\end{itemize}
\end{defi}
\begin{eje}
Below for our model, we will define on 
\[
\al A.=
\ClifS\otimes\al R.(\al S.(\R),\sigma)
\]
a functional $\varphi=\psi\otimes\omega,$ with $\dom\varphi=
\al A._0$ where $\psi$ and $\omega$ are quasi-free with
two-point functions
\begin{eqnarray*}
\omega(j^2(f))&=&\int{p\over 1-e^{-p}}\left|\widehat{f}(p)\right|^2dp\;, \\[1mm]
\psi(c(f)c(g))&=&\lim_{\varepsilon\to 0^+}\int{p\over 1-e^{-p}}\,
{p\over p^2+\varepsilon^2}\,{\widehat{f}(p)}\,\ol\hat{g}(p).\,dp
\end{eqnarray*}
and we will verify that  $\varphi=\psi\otimes\omega$
is a graded KMS--functional. Note that $\omega$ is a state on
$\al R.(\al S.(\R),\sigma),$ but $\psi$ is unbounded and
nonpositive. It does however satisfy supersymmetry, in that 
$\varphi\circ\delta=0,$ and equation~(\ref{timeconn}) holds weakly.
\end{eje}

The motivation for using graded KMS--functionals come from several sources:
\begin{itemize}
\item{} Physicists used graded KMS--functionals to construct supersymmetric
field theories in a thermal background \cite{Fu,Ho}.
\item{} Jaffe e.a.~\cite{JLO} and Kastler~\cite{Ka}
 used graded KMS--functionals to construct cyclic cocycles in
 Connes' cyclic cohomology.
\end{itemize}
The reasons why one has to use nonpositive unbounded  KMS--functionals
for field theories on noncompact spacetime are as follows. First,
there is the theorem of Buchholz and Ojima~\cite{BuOj}
that supersymmetry breaks down in spatially homogeneous KMS--states,
and second there is the theorem of Buchholz and Longo~\cite{BuLo}
that if $\varphi$ is a bounded graded KMS--functional
of $\al A.$ with time evolution $\alpha:\R\to\aut\al A.,$
and if there are $\beta_n\in\aut\al A.$ such that
\[
\lim_{n\to\infty}\varphi\left(C[A,\,\beta_n(B)]\right)=0
\quad\forall\; A,\,B,\,C\in\al A.
\]
then $\alpha=\iota\,.$
On noncompact spaces, the translations will produce the 
$\beta_n$ in a local field theory. 
Thus in local field theories on noncompact spaces,
we are inevitably led to unbounded graded KMS--functionals
for supersymmetry.

First, we would like to establish a few general properties of
graded KMS--functionals.

\begin{pro}
\label{KMSbasics}
Given a graded KMS-functional $\varphi$
defined w.r.t. the data ${(\al A.,\,\gamma,\,\alpha),}$ then
\begin{itemize}
\item[(i)] $\varphi$ is $\alpha\hbox{--invariant}.$
\item[(ii)] $\varphi$ is $\gamma\hbox{--invariant}.$
\end{itemize}
Moreover if a functional $\varphi$ satisfies the graded KMS--property 
on a subset $Y\subset\dom\varphi\subset\al A.,$ then it also
satisfies the graded KMS--property on $\span Y\,.$
\end{pro}
\begin{pro}
\label{KMStensor}
Let $\al A.=\al C.\otimes\al B.$ where $\al C.$ and $\al B.$ are
unital C*-algebras with $\al C.$ nuclear. Let
$\sigma:\R\to\aut\al C.$ and $\beta:\R\to\aut\al B.$ be
dynamical systems, and let $\gamma\in\aut\al C.$ be a grading
automorphism, $\gamma^2=\iota.$
Let $\omega\in\ot S.(\al B.)$ be a KMS--state on $\al B.$
w.r.t. $\beta,$ and let $\psi$ be a graded KMS--functional on
$\al C.$ w.r.t. $\sigma.$ Define a functional
$\varphi:=\psi\otimes\omega$ with $\dom\varphi:={\span
\set C\otimes B,C\in\dom\psi,\;B\in{\al B.}.}$
by $\varphi(C\otimes B):=\psi(C)\omega(B)\,.$
Then $\varphi$ is a graded KMS--functional w.r.t. the
grading $\gamma\otimes\iota,$ and the C*-dynamical system
$\sigma\otimes\beta:\R\to\aut(\al C.\otimes\al B.)\,.$
\end{pro}
Thus for our model, as 
$\al A.=\ClifS\otimes\al R.(\al S.(\R),\sigma),$
it suffices to define a graded KMS--functional $\psi$
on $\ClifS$ and a KMS--state $\omega$ on
$\al R.(\al S.(\R),\sigma)$ from which we can then construct
the graded KMS--functional $\varphi:=\psi\otimes\omega.$
We start by defining
the KMS--state $\omega$ on $\al R.(\al S.(\R),\sigma)\,.$
\begin{teo}
\label{KMSresolv}
\begin{itemize}
\item[(i)] There is a quasi--free state on $\CCR$ defined by
$\omega(\delta_f):=\exp[-s(f,f)/2]\,,$ $f\in\al S.(\R,\R)$ where
\begin{equation}
\label{KMSCCR}
s(f,g):=\int{p\over 1-e^{-p}}\widehat{f}(p)\,\ol\widehat{g}(p).\,dp
=\langle f|g\rangle_\omega\,.
\end{equation}
\item[(ii)] This quasi--free state $\omega$ on $\CCR$ extends to a
KMS--functional on $\pi_\omega(\CCR)'',$ hence on $\al R.(\al S.(\R),\sigma),$
where it is defined by Equation~(\ref{QFresolvent}).
The time evolution
used for the KMS--condition is translation of test functions $f.$
\end{itemize}
\end{teo}
Next, we would like to define a graded KMS--functional $\psi$
on $\ClifS$ with $\dom\psi=\hbox{*-alg}\set c(f),{f\in\al S.(\R)}..$
By the last part of Proposition~\ref{KMSbasics}, it suffices to define
$\psi$ and check its KMS-properties on the monomials
$c(f_1)\cdots c(f_n)\,.$ Recall that a quasi--free functional on
the Clifford algebra
is uniquely defined by its two point functional and the relations:
\begin{eqnarray}
\label{QFodd}
\psi(c(f_1)\cdots c(f_{2k+1}))&=& 0  \\[1mm]
\label{QFeven}
\psi(c(f_1)\cdots c(f_{2k}))&=&
(-1)^{\left({k\atop 2}\right)}\sum_P(-1)^P\prod_{j=1}^k
\psi\Big(c(f\s P(j).)\,c(f\s P(k+j).)\Big)
\end{eqnarray}
where $k\in\N$ and $P$ is any permutation of $\{1,\,2,\ldots,\,2k\}$
such that $P(1)<\cdots<P(k)$ and $P(j)<P(k+j)$ for $j=1,\ldots,\,k\,$
(cf. p89 in~\cite{PR}). Using this formula, we define
a quasi--free functional $\psi$ with two point function
\begin{eqnarray}
\theta(f,g):=
\psi(c(f)c(g))&=&\lim_{\varepsilon\to 0^+}\int{p\over 1-e^{-p}}\,
{p\over p^2+\varepsilon^2}\,{\widehat{f}(p)}\,\ol\hat{g}(p).\,dp
\nonumber \\[1mm]
\label{QF2ptfn}
&=&\lim_{\varepsilon\to 0^+}\Big( \int_{-\infty}^{-\varepsilon}
+\int_{\varepsilon}^\infty\Big){1\over 1-e^{-p}}\,
\,{\widehat{f}(p)}\,\ol\hat{g}(p).\,dp \\[1mm]
&=&\lim_{\varepsilon\to 0^+}\Big( \int_{-\infty}^{-\varepsilon}
+\int_{\varepsilon}^\infty\Big)\left({1\over p}\right){p\over 1-e^{-p}}\,
\,{\widehat{f}(p)}\,\ol\hat{g}(p).\,dp \nonumber\\[1mm]
&=&\al P.\left({1\over p}\right)(G) \nonumber
\end{eqnarray}
where $\al P.\left({1\over p}\right)$ denotes the distribution
consisting of 
the Cauchy Principal Part integral of $1/p\,,$ and $G(p):={p\over 1-e^{-p}}\,
{\widehat{f}(p)}\,\ol\hat{g}(p).$ which is differentiable
everywhere and of fast decay since 
${p\over 1-e^{-p}}$ is differentiable and of linear growth,
 and $f,\,g$ are real-valued Schwartz functions.
Thus $\theta(f,g)$ is well-defined for all $f,\,g\in\al S.(\R)\,.$

We mention as an aside that the quasifreeness of a graded KMS--functional
$\psi$ on the Clifford algebra and the formula for its two--point
function are a consequence of the graded KMS--condition, as can
be shown by similar arguments as in \cite{RoSiTe}.

This quadratic form $\theta$ is unbounded and not positive definite,
because $(1-e^{-p})^{-1}$ is unbounded and not positive. % A natural domain
% for $\theta$ is 
% \[
% \dom\theta:=\set f\in\al S.(\R),{\lim_{\varepsilon\to 0^+}
% \Big( \int_{-\infty}^{-\varepsilon}
% +\int_{\varepsilon}^\infty\Big){e^{-sp}\over |1-e^{-p}|}\,
% \left|\widehat{f}(p)\right|^2dp<\infty\,,\quad s\in[0,1]}..
% \]
% (where the factor $e^{-sp}$ is necessary to define KMS-functions).
$\theta$ has the following useful properties.
\begin{teo}
\label{ThetaProp}
Let $f,\,g\in\al S.(\R)\,,$ then
\begin{itemize}
\item[(i)] $\theta(f,g)=(g,\,(P+T)f)$ where $P=2\pi\times\hbox{projection}$
onto positive spectrum of $D:=id/dx,$ and $T$ is an unbounded operator
given explicitly by 
\[
(g,Tf)=2i\int dx\,\int dy\,f(x)\,g(y)\,(x-y)\int_0^\infty dp\,
\ln(1-e^{-p})\,\cos(p(x-y))\;.
\]
Moreover $P_JTP_J$ is trace--class and selfadjoint
 for all compact intervals $J\subset\R$
where $P_J$ is the projection onto $L^2(J)\subset L^2(\R)\,.$
\item[(ii)]
For $z\in S=\R+i[0,1]$ define 
\[
G(z):=\theta(f,g_z):=
\lim_{\varepsilon\to 0^+}\Big( \int_{-\infty}^{-\varepsilon}
+\int_{\varepsilon}^\infty\Big){e^{ipz}\over 1-e^{-p}}\,
\,{\widehat{f}(p)}\,\ol\hat{g}(p).\,dp \,.
\]
Then $G$ is continuous on $S,$ analytic on its interior, and satisfies
\[
\big|G(t+is)\big|\leq A+B|t|\quad\hbox{for}\;t\in\R,\;
s\in[0,1]
\]
and constants $A,\,B\,.$
\end{itemize}
\end{teo}
Using these properties, one can now establish that:
\begin{teo}
\label{QFCAR}
The quasi--free functional $\psi$ with two point functional $\theta$
and domain $\dom\psi=\hbox{*-alg}\set c(f), {f\in\al S.(\R)}.\,,$
is a graded KMS--functional on 
$\ClifS$ where time evolution is given by translation.
\end{teo}
Thus by Proposition~\ref{KMStensor} we have a KMS--functional
$\varphi=\psi\otimes\omega$ on $\al A.$ with domain
\[
\dom\varphi=\span\set C\otimes R,C\in\dom\psi,\;{R\in 
\al R.(\al S.(\R),\sigma)}.\,.
\]
What makes this KMS--functional interesting, is that it satisfies 
supersymmetry, i.e. 
\begin{teo}
\label{KMSfSUSY}
For the quasifree functional $\varphi$ above, we have that
\begin{itemize}
\item[(i)] $\al A._0\subset\dom\varphi\,,$
\item[(ii)] $\varphi(\delta(A))=0$ for all $A\in\al D._S$
\item[(iii)] $\varphi\big(B
M_A\overline{\delta}_0(A)C\big)=-i{d\over dt}
\varphi\big(BM_A\,\alpha_t(A)C\big)\Big|_0=\varphi\big(BM_A\delta^2(A)C\big)$
for all $A\in\al D._S$ and $B,C\in\al A._0\,$ where $M_A$ is a monomial of resolvents
$\rlf$ such that ${M_A\delta(A)}\in\al D._S$ (as in Definition~\ref{deltasquared}). 
\end{itemize}
\end{teo}
Whilst the functional $\varphi$ is unbounded, it is locally bounded in the 
sense of the theorem below. For the local algebras, let $J\subset\R$
be a bounded interval and define
\begin{eqnarray*}
\al A.(J)&:=&{\rm C}^*\set{c(f),\;\rlf},{\supp f\subseteq J,\;f\in\al S.(\R),\;
\lambda\in\R\backslash 0}.  \\[1mm]
\al A._0(J)&:=&\hbox{*-alg}\set{c(f),\;\rlf},{\supp f\subseteq J,\;f\in\al S.(\R),\;
\lambda\in\R\backslash 0}.
\end{eqnarray*}
hence $\al A._0(J)$ is a dense *-algebra of $\al A.(J),$ and $\al A._0(J)\subset
\dom\varphi\,.$ Then:
\begin{teo}
\label{LocBd}
For the quasifree functional $\varphi$ above, and a bounded interval $J\subset\R$ 
 we have that
\[
\left\|\varphi\restriction\al A._0(J)\right\|\leq\exp(K|J|^2)
\]
where $K$ is a constant (independent of $J$), and $|J|$ is the length of $J\,.$
\end{teo}
Thus $\varphi$ is bounded on all the local algebras $\al A._0(J)\,.$

\section{The JLO--cocycle.}
\label{JLOsection}

From an assumed supersymmetry structure on a C*-algebra and a
KMS-functional, Jaffe, Lesniewski and Osterwalder~\cite{JLO}
and Kastler~\cite{Ka} constructed 
with a Chern character formula an entire cyclic cocycle in the sense of 
Connes~\cite{Co}.
Their assumed supersymmetry assumptions are too restrictive to
include quantum field theories on noncompact spacetimes.
Here we want to show that we can adapt the JLO cocycle 
formula to produce a well-defined entire cyclic cocycle
for our model, using the KMS-functional in
the preceding section.
We first need to make sense of the Chern character formula:
\begin{eqnarray}
\tau_n(a_0,\ldots,a_n)&:=& i^{\epsilon_n}\int_{\sigma_n}
\varphi\Big(a_0\,\alpha_{is_1}\big(\delta\gamma(a_1)\big)\,
\alpha_{is_2}\big(\delta(a_2)\big)\,\alpha_{is_3}\big(\delta\gamma(a_3)\big)
\cdots \nonumber  \\[1mm]
\label{ChCa}
& &\qquad\qquad\cdots\alpha_{is_n}\big(\delta\gamma^n(a_n)\big)\Big)\,
ds_1\cdots ds_n\,,\qquad a_i\in\al D._S,
\end{eqnarray}
where $\epsilon_n:=n\,{\rm mod}\, 2,$  $\varphi$ is the
graded KMS--functional above w.r.t. the data $\gamma,\,\alpha,\,\delta$
above, and  
\[
\sigma_n:=\set{\bf s}\in\R^n,0\leq s_1\leq s_2\leq\cdots\leq s_n\leq 1.\,.
\]
Since only $\alpha:\R\to\aut\al A.$ is given, the expressions 
$\alpha_{is},\;s\in\R$ in the formula are undefined, and need to
be interpreted. Let $b_0,\ldots,b_n\in\al A._0\subset\dom\varphi,$ then 
using the KMS--property of $\varphi,$ the function
\[
Q(t_1,\ldots,t_n):=\varphi\big(b_0\alpha_{t_1}\big(b_1\alpha_{t_2}(b_2\cdots
\alpha_{t_n}(b_n)\cdots\big)\big)
=\varphi\big(b_0\alpha_{t_1}(b_1)\cdots\alpha_{t_1+\cdots+ t_n}(b_n)\big)
\]
can be analytically extended in each variable $t_j$ into the strip
$\set z_j,0\leq{\rm Im}\,z_j\leq 1.$ keeping the other varables real.
This produces $n$ functions $Q^j:T_j\to\C$ where
$T_j:= \set{\bf z}\in \C^n,z_j\in\R+i[0,1],\;z_k\in\R\;\hbox{for}\;k\not=j.$
such that $Q^i\restriction\R^n=Q^j\restriction\R^n$ for all $i,j\,.$
The sets $T_j$ are {\it flat tubes,} i.e. of the form
$T_B=\R^n+iB$ where the basis $B\subset\R^n$ is of dimension less than $n.$
To continue, we now need the {\it Flat Tube Theorem}~\cite{BrBu}:
\begin{teo}
\label{FTT}
Let $T_{B_1},\,T_{B_2}\subset\C^n$ be two flat tubes whose bases $B_1,\,B_2$
are convex, with closures which contain $0$ and are star-shaped 
w.r.t. $0.$ Let $F_1,\,F_2$
be any two functions analytic in $T_{B_1},\,T_{B_2}$ respectively,
with continuous boundary values on $\R^n$ and
such that $F_1\restriction\R^n=F_2\restriction\R^n\,.$
Then there is a unique function $F$ extending $F_1$ and $F_2$ analytically 
into the tube
$T\s\wh{B_1\cup B_2}.$ (where $\wh{B_1\cup B_2}$ is the convex hull
of ${B_1\cup B_2})$  and with continuous boundary values on $\R^n\,.$
We have that $T\s\wh{B_1\cup B_2}.=\bigcup\limits_{0\leq\lambda\leq 1}T_{B_\lambda}$
where $B_\lambda:=(1-\lambda)B_1+\lambda B_2\,.$
\end{teo}
Using this inductively, we can extend $Q$ by analytic continuation
into the tube $\al T._n:=\R^n+i\Sigma_n$ where $\Sigma_n$ is the
convex hull of the unit intervals on the axes, i.e. the simplex
$\Sigma_n:=\set{\bf s}\in\R^n,0\leq s_i\;\forall\,i,\;\;s_1+\cdots+s_n\leq1.\,.$
So we will interpret 
\[
\varphi\big(b_0\alpha_{z_1}(b_1)\cdots\alpha_{z_1+\cdots+ z_n}(b_n)\big)
:=Q(z_1,\ldots,z_n)
\]
for ${\bf z}\in\al T._n$ to be this unique analytic continuation.
The change of variables $w_1:=z_1,\ab\;w_2:=z_1+z_2,\ldots,\,
w_n:=z_1+\cdots+z_n$ defines an invertible complex linear
map $W:\R^n+i\Sigma_n\to\R^n+i\sigma_n,$ so both $W$ and $W^{-1}$
are analytic, and hence
\[
\big(Q\circ W^{-1})(w_1,\ldots,\,w_n)=:
\varphi\big(b_0\alpha_{w_1}(b_1)\cdots\alpha_{w_n}(b_n)\big)
\] 
is analytic on $\R^n+i\sigma_n\,.$
In particular
% \begin{eqnarray*}
% & &
\[
\int_{\Sigma_n}\varphi\big(b_0\alpha_{ir_1}(b_1)\cdots\alpha_{ir_1+\cdots+ ir_n}(b_n)\big)
\,dr_1\cdots dr_n % \\[1mm]
% & &\qquad
=\int_{\sigma_n}
\varphi\big(b_0\alpha_{is_1}(b_1)\cdots\alpha_{is_n}(b_n)\big)
\,ds_1\cdots ds_n\;
\]
% \end{eqnarray*}
by the change of variables $s_1=r_1,\;s_2=r_1+r_2,\ldots,s_n=r_1+\cdots+r_n\,.$
By the substitutions $a_0=b_0,$ $b_1=\delta\gamma(a_1),\ldots, b_n=\delta\gamma^n(a_n)$
into this formula, we arrive at a consistent interpretation of the 
Chern character formula~(\ref{ChCa}). 

Let us recall from \cite{JLO,JLOK,Co} the definition of an entire cyclic cocycle.
\begin{defi}
\label{EntCyCoc}
Equip $\al D._S$ with the Sobolev norm $\|a\|_*=\|a\|+\|\delta a\|,$ and
for any *--algebra $\al D.\subseteq\al D._S$ let
$\al C.^n(\al D.)$ denote the space of $(n+1)\hbox{--linear}$ functionals
on $\al D.$ which are continuous w.r.t. the norm $\|\cdot\|_*\,,$
and let $\|\cdot\|_*$ denote also the norm on $\al C.^n(\al D.)$
w.r.t. the norm $\|\cdot\|_*$ on $\al D.\,.$
Define the space of cochains $\al C.(\al D.)$ to be the space
of sequences $\rho=(\rho_0,\rho_1,\ldots)$ where $\rho_n\in\al C.^n(\al D.)$
which satisfy the entire analyticity condition:
\begin{equation}
\label{EntCon}
\lim_{n\to\infty}n^{1/2}\|\rho_n\|_*^{1/n}=0\;.
\end{equation}
The entire cyclic cohomology is defined by a coboundary operator 
$\partial=b+B$ on $\al C.(\al D.)$ for operators
\begin{eqnarray}
b:\al C.^n(\al D.)\to\al C.^{n+1}(\al D.)\;,\qquad\quad
B:\al C.^{n+1}(\al D.)\to\al C.^n(\al D.) \nonumber\\[1mm]
\label{dbB}
\hbox{i.e.}\qquad
(\partial\rho)_n(a_0,\ldots,a_n)=(b\rho_{n-1})(a_0,\ldots,a_n)+
(B\rho_{n+1})(a_0,\ldots,a_n)
\end{eqnarray}
where $b$ and $B$ are given by:
\begin{eqnarray}
(b\rho_n)(a_0,\ldots,a_{n+1})&=&
\sum_{j=0}^n(-1)^j\rho_n(a_0,\ldots,a_ja_{j+1},\ldots,a_{n+1}) \nonumber\\[1mm]
\label{Defb}
& &\qquad\qquad+(-1)^{n+1}\rho_n(\wt{a}_{n+1}\cdot a_0,a_1,\ldots,a_n)\\[1mm]
(B\rho_n)(a_0,\ldots,a_{n-1})&=&
\rho_n(\un,a_0,\ldots,a_{n-1})+(-1)^{n-1}\rho_n(a_0,\ldots,a_{n-1},\un)
\nonumber \\[1mm]
& &+\sum_{j=1}^{n-1}(-1)^{(n-1)j}\Big[\rho_n\left(\un,\gamma(a_{n-j}),\ldots,
\gamma(a_{n-1}),a_0,\ldots,a_{n-j-1}\right)  \nonumber\\[1mm]
\label{DefB}
& &\quad\qquad+(-1)^{n-1}\rho_n\left(\gamma(a_{n-j}),\ldots,\gamma(a_{n-1}),
a_0,\ldots,a_{n-j-1},\un\right)\Big] \\[1mm]
\hbox{where}\qquad& & \quad
\wt{a}:=\cases{\gamma(a) & if $\rho_n\in\al C._+^n(\al D.)$ \cr
a & if $\rho_n\in\al C._-^n(\al D.)$ \cr}
\nonumber
\end{eqnarray}
where $\al C._+^n(\al D.)$ denotes the even part under $\gamma,$ and 
$\al C._-^n(\al D.)$ the odd part. The {\bf entire cyclic cocycles}
are those $\rho\in\al C.(\al D.)$ for which $\partial\rho=0,$ i.e.
\begin{equation}
\label{CocEq}
(b\rho_{n-1})(a_0,\ldots,a_n)=-(B\rho_{n+1})(a_0,\ldots,a_n)
\,,\qquad n=1,2,\ldots
\end{equation}
\end{defi}
Below we will use the even part of $\tau$ to define an entire
cyclic cocycle.

\begin{teo}
\label{CoBd}
For $\tau_n$ defined in Equation~(\ref{ChCa}) we have that 
\[
\left|\tau_n(a_0,\ldots,a_n)\right|\leq
{A\over n!}\,e^{Bn}\,\|a_0\|_*\cdots\|a_n\|_*
\]
for all $a_i\in\al D._k:=\hbox{*--alg}\set{\un,\,R(1,f),\,\zeta(f)},
{\rm supp}(f)\subseteq [-k,\,k].\subset\al D._S$ 
and where $\|a\|_*:=\|a\|+\|\delta a\|$
as above, and for some
 constants  $A$ and $B$ which depend on $k>0$ but are independent of $n.$
Thus condition ~(\ref{EntCon}) holds for $\tau,$ i.e.
\[
\lim_{n\to\infty}n^{1/2}\|\tau_n\|_*^{1/n}=0\;.
\]
\end{teo}
Using this, we can now prove that:
\begin{teo}
\label{TCycCoc}
The sequence $\wt\tau:=(\tau_0,0,-\tau_2,0,\tau_4,0,-\tau_6,\ldots)\in \al C.(\al D._k)$
defines an entire cyclic cocycle for each $k>0\,,$ i.e.
\[
(b\tau_{n-1})(a_0,\ldots,a_n)=(B\tau_{n+1})(a_0,\ldots,a_n)
\,,\qquad n=1,3,5,\ldots
\]
and the entire analyticity condition holds.
\end{teo}
It is possible to have taken the choice $\partial=b-B$ for the cyclic coboundary 
operator above;- this is in fact done in \cite{JO1}, and would have led
to the cyclic cocycle ${(\tau_0,0,\tau_2,0,\tau_4,0,\ldots)}$ instead
of $\wt\tau$ above.

Note that whilst we have obtained entire cyclic cocycles on each compact set $[-k,\,k]$
these do not define an entire cyclic cocycle on
$\al D._{\rm comp}:=\hbox{*--alg}\set{\un,\,R(1,f),\,\zeta(f)},
{\rm supp}(f)\;\hbox{is compact}.\subset\al D._S$
because one can choose a sequence $\{a_0,\,a_1,\ldots\}$ with 
$a_j\in\al D._{k_j}$ where $k_j$ grows sufficiently fast so that through
the dependencies of the constants $A$ and $B$ 
in Theorem~\ref{CoBd} on $k_j$ the entire analytic condition fails.
One expects to use an inductive limit argument, to define an index
on $\al D._{\rm comp}$ from the indices on the $\al D._k\,.$

From this cyclic cocycle we can calculate an index for this quantum field theory,
but its physical significance is presently unclear, though one would expect it to remain 
stable under deformations. This type of index is discussed in more detail in
Longo~\cite{Lo}.

\section{Conclusions}

In this paper we have explored how supersymmetric quantum fields 
can be treated in a C*-algebra setting, avoiding the obstructions
found by Kishimoto and Nakamura~\cite{KiNa}
and by Buchholz and Longo \cite{BuLo}. We did this in detail 
for a simple one--dimensional model. 

In order to establish a reasonable
domain of definition for the super--derivation, we found it 
necessary to analyze a notion of ``mollifiers'' for the 
quantum fields and to introduce a corresponding C*-algebra,
the resolvent algebra. The full algebra $\al A.$ defining the model
can then be taken as the tensor product of this resolvent algebra and 
the familiar CAR--algebra.

The super--derivation is 
defined on a subalgebra which is weakly dense
in  $\al A.$ in all representations of physical interest;
alternatively, one can define it on a 
norm dense subalgebra of  $\al A.$ with range in a *--algebra ${\cal E}$ of
bounded and unbounded 
operators which are affiliated with  $\al A.$ in the sense of
\cite{Georg}. Similarly, the basic supersymmetry relation 
can either be formulated in a mollified form on some weakly dense 
domain or, alternatively, as a relation between maps 
which have been extended to the *--algebra  ${\cal E}$.
These findings reveal some basic features 
of supersymmetric quantum field theories which have to be
taken into account in a general C*-framework covering such theories.
The tools developed here should also be useful in other areas
of quantum field theory where one needs to use graded derivations, e.g.
in BRS--constraint theory.

We also exhibited in the present model 
graded KMS--functionals for arbitrary positive
temperatures which are supersymmetric. In accordance
with the general results in \cite{BuLo}, these functionals
are unbounded. Yet their restrictions to any local subalgebra 
of the underlying C*-algebra $\al A.$ are bounded. It is an
interesting question whether these functionals are also 
locally normal with respect to the vacuum representation
of the theory, as one would heuristically expect.

The KMS--functionals were then employed to 
define cyclic cocycles. In view of the fact that
the domain of definition of these functionals does not contain
analytic elements with regard to the time evolution, the 
strategy outlined in \cite{Ka,JLO} could not be applied here.
That these cocycles can be constructed, nevertheless,
is due to the fact that the functionals inherit 
sufficiently strong analyticity properties from the KMS--condition
which allow one to perform the necessary complex integrations.
Moreover, the resulting cocycles are entire on all local
algebras.

These functionals may thus be taken as an input for a quantum
index theory as suggested by  Longo~\cite{Lo}. 
Such an index should be stable under deformations, and one can easily
think of possible deformations of our model, e.g. deform the 
supersymmetry generator $Q$ by an 
appropriate function $M\,:$
\begin{eqnarray*}
Q_M&:=&\int M(x)\,j(x)\,c(x)\,dx \\[1mm]
\hbox{so}\qquad
\delta_M(c(f))&=&j(Mf),\qquad \delta_M(j(f))=c(Mf') \\[1mm]
\hbox{thus:}\qquad
\delta_M^2&=:&\delta_{0M}
\end{eqnarray*}
defines the new generator for time evolution.

\section{Appendix}

In this appendix we give proofs of the statements in the main 
body of the text.
\subsection*{Proof of Theorem~\ref{Relemen}}
(i) By (\ref{Resolv}) we have that $i(\mu-\lambda)\rlf R(\mu,f)=
\rlf-R(\mu,f)=-\big(R(\mu,f)-\rlf\big)=i(\mu-\lambda)R(\mu,f)\rlf\,,$
i.e. ${[\rlf,\,R(\mu,f)]}=0\,.$\chop
(ii) The Fock representation is a subrepresentation of $\pi_S$, 
and the resolvents of the fields $\varphi(f)$
give the Fock representation induced on
$\al R.(X,\sigma),$ i.e. $\pi(\rlf)=(i\lambda-\varphi(f))^{-1}\,.$
Since this is nonzero, $\rlf\not=0$ for all nonzero $f$ and $\lambda\,.$
Now by $\rlf^*=R(-\lambda,f)$ we get
\[
2|\lambda|\|\rlf\|^2=\|2\lambda\rlf\rlf^*\|=\|\rlf-\rlf^*\|
\leq 2\|\rlf\|\,.
\]
Thus by $\rlf\not=0$ we find $\|\rlf\|\leq 1/|\lambda|\,.$ Now
\[
\|\rlf\|\geq\|\pi(\rlf)\|=\|(i\lambda-\varphi(f))^{-1}\|
=\sup_{t\in\sigma(\varphi(f))}\Big|{1\over i\lambda-t}\Big|=
{1\over|\lambda|}
\]
using the fact that the spectrum $\sigma(\varphi(f))=\R\,.$
Thus $\|\rlf\|=1/|\lambda|\,.$\chop
(iii) Rearrange equation~(\ref{Resolv}) to get: 
\[
\rlf\big(\un-i(\lambda_0-\lambda)R(\lambda_0,f)\big)=R(\lambda_0,f)\,.
% \|\rlf-R(\mu,f)\|=|\lambda-\mu|\cdot\|\rlf R(\mu,f)\|
% \leq|\lambda-\mu|\big/|\lambda\mu|
\]
Now by (ii) above, if $\big|\lambda_0-\lambda\big|<\big|\lambda_0\big|$
then ${\big\|i(\lambda_0-\lambda)R(\lambda_0,f)\big\|}<1,$ and hence
${\big(\un-i(\lambda_0-\lambda)R(\lambda_0,f)\big)^{-1}}$ exists, and is
given by a norm convergent power series in ${i(\lambda_0-\lambda)R(\lambda_0,f)}\,.$
That is, we have that
\[
\rlf=R(\lambda_0,f)\big(\un-i(\lambda_0-\lambda)R(\lambda_0,f)\big)^{-1}
=\sum_{n=0}^\infty(\lambda_0-\lambda)^n\,R(\lambda_0,f)^{n+1}i^n
\]
when $\big|\lambda_0-\lambda\big|<\big|\lambda_0\big|,$
as claimed. \chop
% from which it is clear that $\rlf$ is norm continuous in $\lambda$
% for nonzero $\lambda\,.$ We also get from equation~(\ref{Resolv})
% that
% \begin{eqnarray}
% & &
% \left\|\big(R(\lambda+h,f)-\rlf)\big)\big/h+iR(\lambda+h,f)\rlf\right\|=0
% \nonumber\\[1mm]
% \label{R-ODE}
% \hbox{so} & & {d\over d\lambda}\rlf=-i\rlf^2
% \end{eqnarray}
% and hence $\rlf$ is smooth in $\lambda,$ determined by the ODE~(\ref{R-ODE}),
% and one value of $\rlf,$  $\lambda\not=0$ (making use of equation~(\ref{Rinvol})
% to pass through $\lambda=0\,).$
% Consider the series $S(\lambda):={\sum\limits_{n=0}^\infty i^n(\lambda_0-\lambda)^n
% R(\lambda_0,f)^{n+1}}$ for $\lambda\not=0\not=\lambda_0.$ It converges in norm
% for $|\lambda_0-\lambda|<|\lambda_0|$ and
% \[
% {d\over d\lambda}S({\lambda})=\sum_{n=1}^\infty i^nn(\lambda_0-\lambda)^{n-1}
% R(\lambda_0,f)^{n+2}=iS(\lambda)^2\,.
% \]
% For the last inequality we used  spectral theory, and the fact that 
% $S(\lambda)=h(R(\lambda_0,f))$ where $h(t):=t\big/[1-(\lambda-\lambda_0)it]\,.$
% Then we have ${d\over d\lambda}h(t)=ih(t)^2$ so that from the series
% expansions we get ${d\over d\lambda}S({\lambda})=iS(\lambda)^2\,.$
% Since $S(\lambda_0)=R(\lambda_0,f),$ we conclude that for 
% $|\lambda-\lambda_0|<|\lambda_0|$ we have 
% \[
% \rlf=S(\lambda)=\sum_{n=0}^\infty(\lambda_0-\lambda)^ni^nT(\lambda_0,f)^{n+1}\,.
% \]
(iv) From equation~(\ref{Rhomog}) we get $R(\lambda,tf)={1\over\lambda}
R(1,{t\over\lambda}f)={1\over t}R({\lambda\over t},f)$ so
\begin{eqnarray*}
R(\lambda,sf) &-& R(\lambda,tf)=\f 1,s.R(\f\lambda,s.,f)-
\f 1,t.R(\f\lambda,t.,f)  \\[1mm]
&=& \f 1,s.\left(R(\f\lambda,s.,f)-R(\f\lambda,t.,f)\right)
+\left(\f 1,s.-\f 1,t.\right)R(\f\lambda,t.,f) \\[1mm]
&=&\f i\lambda,s.\left(\f 1,t.-\f 1,s.\right)R(\f\lambda,s.,f)
R(\f\lambda,t.,f)+\left(\f 1,s.-\f 1,t.\right)R(\f\lambda,t.,f)\,.
\end{eqnarray*}
Thus $\left\|R(\lambda,sf) - R(\lambda,tf)\right\|\leq\left|\f 1,s.
-\f 1,t.\right|\,2\left|\f t,\lambda.\right|$
from which continuity away from zero is clear.\chop
(v) This follows directly from equation~(\ref{Rccr})
by interchanging $\lambda$ and $f$ with $\mu$ and $g$ resp.\chop
(vi)
Recall that we have a faithful  (strongly regular) representation $\pi_S$
of $\al R.(X,\sigma)$ which is an extension of a  regular representation 
of the Weyl algebra  $\CCRX$ such that $\pi\s S.\left(\CCRX\right)''
\supset\pi\s S.\left(\al R.(X,\sigma)\right)$ and such that each 
$\pi_S(\rlf)$ is the resolvent of the generator $j\s\pi_S.(f)$ of 
the one-parameter group $t\to\pi_S(\delta\s tf.).$
 Let $T\in{\rm Sp}(X,\sigma)$ then 
 it defines an automorphism of $\CCRX$ by $\alpha_T(\delta_f)=
 \delta\s Tf.$ which preserves the set of strongly regular  states,
 and in fact defines a bijection on the set of strongly regular states
 by $\omega\to\omega\circ\alpha_T.$
 Now $\pi_S$ is the direct sum of the GNS--representations of all 
 the strongly regular states, and hence $\pi_S\circ\alpha_T$
 is just $\pi_S$ where its direct summands have been permuted.
 Such a permutation of direct summands can be done by 
 conjugation of a unitary, thus
 $\pi_S$ is unitarily equivalent to $\pi_S\circ\alpha_T,$
 and so we can extend $\alpha_T$ by unitary conjugation to
 $\pi\s S.\left(\CCRX\right)''.$
 By equation~(\ref{Laplace1}) we get that
 $\alpha_T\left(\pi_S(\rlf)\right)=\pi_S(R(\lambda,Tf)),$
 and hence $\alpha_T$ preserves $\al R.(X,\sigma),$
 so defines an automorphism on it.
% the map $\rlf\to R(\lambda,Tf)$
% is a permutation of the generating elements 
% $\set{\rlf},\lambda\in\R\backslash 0,\;f\in X\backslash 0.$
% of $\al R._0$
% which obviously preserves the relations
% (\ref{Rinvol}) to (\ref{Rsum}), hence defines a *--automorphism of 
% $\al R._0\,.$ Since $T$ is invertible and linear it maps strongly regular
% representations to strongly regular ones. So composition with $\alpha$
% preserves the set of strongly regular representations, and hence 
% $\pi\s S.\circ\alpha$ is unitarily equivalent to $\pi\s S..$
% Thus $\alpha$ preserves $\ker\pi\s S.$ and hence defines an automorphism
% of $\al R.(X,\sigma)\,.$
\subsection*{Proof of Theorem~\ref{RegThm}}

(i) Observe that by Theorem 1 p216 of Yosida~\cite{Yos}, we deduce from 
${\ker\pi(R(1,f))}=\{0\}$ that $\pi(\rlf)$ is the resolvent of $j_\pi(f),$
i.e. we have now for all $\lambda\not=0$ that
$j_\pi(f)=i\lambda\un-\pi(\rlf))^{-1}\,.$ Then
\begin{eqnarray*}
j_\pi(tf) &=& i\un-\pi(R(1,tf))^{-1}=i\un-t\pi(R(\f 1,t.,f))^{-1} \\[1mm]
&=& t\left(i\f 1,t.{\un}-\pi\big(R(\f 1,t.,f)\big)^{-1}\right)
=t\,j_\pi(f)\,.
\end{eqnarray*}
Thus
\begin{eqnarray*}
j_\pi(f)^* &=& \left(i\un-\pi(R(1,f))^{-1}\right)^*
\supseteq -i\un-\left(\pi(R(1,f))^{-1}\right)^*\\[1mm]
&=&-i\un-\pi(R(1,f)^*)^{-1}
=-i\un-\pi(R(-1,f))^{-1} \\[1mm]
&=&-i\un+\pi(R(1,-f))^{-1} = -j_\pi(-f)=j_\pi(f)
\end{eqnarray*}
and hence $j_\pi(f)$ is symmetric. To see that it is selfadjoint
note that:
\[
\ran\left(j_\pi(f)\pm i\un\right)=\ran\left(-\pi\big(R(\pm 1,f)\big)^{-1}\right)
=\dom\left(\pi\big(R(\pm 1,f)\big)\right)=\al H._\pi
\]
hence the deficiency spaces $\left(\ran\left(j_\pi(f)\pm i\un\right)\right)^\perp
=\{0\}$ and so $j_\pi(f)$ is selfadjoint.\chop
For the domain claim, recall that 
$\dom j_\pi(f)=\ran\pi\big(R(1,f)\big)\,.$
So 
\begin{eqnarray*}
\pi(\rlf)\dom j_\pi(h)&=&\pi(\rlf)\pi\big(R(1,h)\big)\al H._\pi  \\[1mm]
&=&\pi\left(R(1,h)\rlf+i\sigma(f,h)R(1,h)\rlf^2R(1,h)\right)\al H._\pi \\[1mm]
&\subseteq&\pi\left(R(1,h)\right)\al H._\pi=\dom j_\pi(h).
\end{eqnarray*}
(ii) Let $j_\pi(f)=\int\lambda dP(\lambda)$ be the spectral resolution of $j_\pi(f).$
Then $\pi(R(\mu,f))=\int{1\over i\mu-\lambda}dP(\lambda)$ hence
\[
i\mu\pi(R(\mu,f))\psi=\int{i\mu\over i\mu-\lambda}\,dP(\lambda)\psi\quad\forall\;\psi
\in\al H._\pi\,.
\]
Since $\left|{i\mu\over i\mu-\lambda}\right|<1$ (for $\mu\in\R\backslash 0)$
 the integrand is dominated by $1$ which is an $L^1\hbox{--function}$
with respect to $dP(\lambda),$ and as we have pointwise that
$\lim\limits_{\mu\to\infty}{i\mu\over i\mu-\lambda}=1,$ we can apply
the dominated convergence theorem to get that
\[
\lim_{\mu\to\infty}i\mu\pi(R(\mu,f))\psi=\int dP(\lambda)\psi=\psi\,.
\]
(iii) $i\pi(R(1,sf))\psi=\int{i\over i-s\lambda}\,dP(\lambda)\psi\to\psi$ as $s\to0$ by the same
argument as in (ii)\,.\chop
(iv) Let $\al D.:={\pi\big(R(1,f)R(1,h)\big)\al H._\pi},$ then by definition
$\al D.\subseteq\ran\pi(R(1,f))=\dom j_\pi(f)\,.$
Moreover ${\pi\big(R(1,f)R(1,h)\big)\al H._\pi} =
{\pi\big(R(1,h)[R(1,f)+i\sigma(f,h)R(1,f)^2R(1,h)]\big)\al H._\pi}\subseteq\ran\pi(R(1,h))=
\dom j_\pi(h),$
i.e. $\al D.\subseteq\dom j_\pi(f)\cap\dom j_\pi(h)\,.$
That $\al D.$ is dense, follows from (iii) of this theorem, using
\[
\lim_{s\to 0}\lim_{t\to 0}\pi\big(R(1,sf)R(1,th)\psi=-\psi
\]
for all $\psi\in\al H._\pi\,,$ as well as $sR(1,sf)=R(1/s,\,f)$
and the fact mentioned before (cf. Theorem~1 p216
in~\cite{Yos}) that
all $\pi(\rlf)$ have the same range for $f$ fixed.\chop
Let $\psi\in\al D.,$ i.e. $\psi=\pi\big(R(1,f)R(1,h)\big)\varphi$ for some
$\varphi\in\al H._\pi\,.$ Then
\begin{eqnarray*}
& &\pi\big(R(1,h)R(1,f)\big)\big[j_\pi(f),j_\pi(h)\big]\psi  \\[1mm]
& &\qquad\quad=\pi\big(R(1,h)R(1,f)\big)\big[\pi(R(1,f))^{-1},\pi(R(1,h))^{-1}\big]
\pi\big(R(1,f)R(1,h)\big)\varphi  \\[1mm]
& & = \pi\big(R(1,f)R(1,h)-R(1,h)R(1,f)\big)\varphi
= i\sigma(f,h)\pi\big(R(1,h)R(1,f)^2R(1,h)\big)\varphi \\[1mm]
& & =  i\sigma(f,h)\pi\big(R(1,h)R(1,f)\big)\psi\;.
\end{eqnarray*}
Since $\ker\pi\big(R(1,h)R(1,f)\big)=\{0\}$ it follows that 
$\big[j_\pi(f),j_\pi(h)\big]=i\sigma(f,h)$ on $\al D.\,.$\chop
(v) From Equation~(\ref{Rhomog}) we have that
\[
\pi(\rlf)=(\un i\lambda-j_\pi(f))^{-1}=\f 1,\lambda.\pi\big(R(1,\f 1,\lambda.f)\big)
=\f 1,\lambda.\cdot\left(i{\un}-j_\pi(\f 1,\lambda.f)\right)^{-1}
\]
and hence that $j_\pi(f)=\lambda \,j_\pi(\f 1,\lambda.f),$ i.e.
$j_\pi(\lambda f)=\lambda j_\pi(f)$ for all $\lambda\in\R\backslash 0\,.$
In equation~(\ref{Rsum}):
\[
\pi\left(\rlf R(\mu,g)\right)= \pi\left(R(\lambda+\mu,\,f+g)[\rlf+R(\mu,g)
+i\sigma(f,g)\rlf^2R(\mu,g)]\right)
\]
multiply on the left by $i(\lambda+\mu)\un-j_\pi(f+g)$ and apply to
${(i\mu\un-j_\pi(g))}{(i\lambda\un-j_\pi(f))}\psi,$ $\psi\in\al D.$ to get
\[
\big(i(\lambda+\mu)\un-j_\pi(f+g)\big)\psi
=\left((i\mu\un-j_\pi(g))+(i\lambda\un-j_\pi(f))\right)\psi
\]
making use of $\left[(i\mu\un-j_\pi(g)),
(i\lambda\un-j_\pi(f))\right]\psi=i\sigma(g,f)\psi\,.$
Thus $j_\pi(f+g)=j_\pi(f)+j_\pi(g)$ on $\al D..$\chop
(vi) From the spectral resolution for $j_\pi(f)$ we have trivially that on $\dom j_\pi(f)$
\[
j_\pi(f)\pi(R(\mu,f))=\pi(R(\mu,f))j_\pi(f)=\int{\lambda\over i\mu\un-\lambda}dP(\lambda)
=i\mu\pi(R(\mu,f))-\un\,.
\]
(vii) Let $\psi\in\dom j_\pi(f)=\ran\pi(\rlf),$ i.e. $\psi=\pi(\rlf)\varphi$ for some
$\varphi\in\al H._\pi.$ Then
\begin{eqnarray*}
\pi(\rlf)\big[j_\pi(f),\,\pi(R(\lambda,g))\big]\psi &=&
\pi(\rlf)\big[j_\pi(f),\,\pi(R(\lambda,g))\big]\pi(\rlf)\varphi  \\[1mm]
&=& \pi\left(\big[R(\lambda,g),\rlf\big]\right)\varphi
=i\sigma(g,f)\pi\left(\rlf R(\lambda,g)^2\rlf\right)\varphi  \\[1mm]
&=& i\sigma(g,f)\pi\left(\rlf R(\lambda,g)^2\right)\psi\;.
\end{eqnarray*}
Since $\ker\pi(\rlf)=\{0\}\;,$ it follows that
\[
\big[j_\pi(f),\,\pi(R(\lambda,g))\big]= i\sigma(g,f)\pi\left( R(\lambda,g)^2\right)
\]
on $\dom j_\pi(f)\,.$\chop
(viii) We first prove the second equality.
Let $\psi,\,\varphi\in\wt\al D.:={\rm Span}\set{\chi\s[-a,a].\big(j_\pi(f)\big)\al H._\pi},a>0.$
where $\chi\s[-a,a].$ indicates the characteristic function of ${[-a,a]},$ and note that
$\wt\al D.$
is a dense subspace.  Since ${\left\|j_\pi(f)^n\restriction\chi\s[-a,a].\big(j_\pi(f)\big)\al H._\pi\right\|}
\leq a^n\,,$ $n\in\N,$ we can use the exponential series, i.e.
\[
W(f)\psi:=\exp\big(ij_\pi(f)\big)\psi
=\sum_{n=0}^\infty{\big(ij_\pi(f)\big)^n\over n!}\psi \qquad\forall\;\psi\in\wt\al D.\;.
\]
By the usual rearrangement of series we then have
\[
\left(\varphi,\,W(f)\pi\big(R(\lambda,h)\big)W(f)^*\psi\right)=
\sum_{n=0}^\infty{1\over n!}\left(\varphi,\,\big({\rm ad}\,ij_\pi(f)\big)^n\big(\pi(R(\lambda,h))
\big)\psi\right)
\]
for all $\varphi,\,\psi\in\wt\al D.\,.$ Using part (vii) we have 
\begin{eqnarray*}
\big({\rm ad}\,ij_\pi(f)\big)\big(\pi(R(\lambda,h)^k)\big)
&=&k\,\sigma(f,h)\pi\big(R(\lambda,h)^{k+1}\big) \\[1mm]
\hbox{thus}\qquad\qquad\big({\rm ad}\,ij_\pi(f)\big)^n\big(\pi(R(\lambda,h))
&=& n!\,\sigma(f,h)^n\pi\big(R(\lambda,h)^{n+1}\big) \\[1mm]
\hbox{so}\qquad\qquad 
\left(\varphi,\,W(tf)\pi\big(R(\lambda,h)\big)W(tf)^*\psi\right)
&=& \sum_{n=0}^\infty t^n\sigma(f,h)^n\left(\varphi,\,\pi(R(\lambda,h)^{n+1})
\psi\right) \\[1mm]
&=& \left(\varphi,\,\pi(R(\lambda+it\sigma(f,h),\,h))
\psi\right)
\end{eqnarray*}
whenever $\big|t\sigma(f,h)\big|<|\lambda|$ and
where we made use of the Von Neumann series (Theorem~\ref{Relemen}(iii)) in the last step.
Since the operators involved are bounded and $\wt\al D.$ is dense, it follows that
$W(tf)\pi\big(R(\lambda,h)\big)W(tf)^*=\pi(R(\lambda+it\sigma(f,h),\,h))$
for $\big|t\sigma(f,h)\big|<|\lambda|.$ By analyticity in $\lambda$ this can be
extended to complex $\lambda$ such that $\lambda\not\in i\R.$ Using the group property of
$t\mapsto W(tf)$ we then obtain for $\lambda\in\R\backslash 0$ that
\begin{equation}
\label{AdWR}
W(f)\pi\big(R(\lambda,h)\big)W(f)^*=\pi(R(\lambda+i\sigma(f,h),\,h))\;.
\end{equation}
To prove the first equation, let us write $W(h)$ in terms of resolvents.
Note that $\lim\limits_{n\to\infty}(1-it/n)^{-n}=e^{it},$ $t\in\R$ and so by the 
bound: $\sup\limits_{t\in\R}\left|(1-it/n)^{-n}\right|=\sup\limits_{t\in\R}\left(1+t^2/n^2\right)^{-n}
=1,$ it follows from spectral theory (cf. Theorem~VIII.5(d), p262 in \cite{RS1}) that
\[
W(h)=e^{ij_\pi(h)}=\lim_{n\to\infty}\left(1-ij_\pi(h)/n\right)^{-n}
=\lim_{n\to\infty}\pi\left(iR(1,-h/n)\right)^n
\]
in strong operator topology. Apply equation~(\ref{AdWR}) to this to get
\begin{eqnarray*}
W(f)W(h)W(f)^* &=& \slim_{n\to\infty}\pi\left(iR(1+i\sigma(f,-\f h,n.),-\f h,n.)
\right)^n \\[1mm]
&=& \slim_{n\to\infty}\left(\un-i\big(\sigma(f,h)+j_\pi(h)\big)\big/n\right)^{-n}  \\[1mm]
&=&\exp[i\sigma(f,h)+ij_\pi(h)]=e^{i\sigma(f,h)}W(h)
\end{eqnarray*}
as required. Now
\[
W(f)\al D.=W(f)\pi\big(\rlf R(\mu,h)\big)\al H._\pi
=\pi\big(\rlf R(\mu+i\sigma(f,h),h)\big)\al H._\pi=\al D.
\]
hence we conclude that $\al D.$ is a core for $j_\pi(f)$ (cf.  Theorem~VIII.11, p269
in \cite{RS1}).

\subsection*{Proof of Proposition~\ref{MasterLemma}}

% (I)  \XP Simple minded proof (needing fewer assumptions).!
Recall Equation~(\ref{QFresolvent})
\begin{eqnarray*}
& & \omega(R(\lambda_1,f_1) \cdots R(\lambda_n,f_n))  \nonumber\\[1mm]
& & =  (-i)^n  
\int_0^\infty \! dt_1 \dots  \int_0^\infty \! dt_n \,
\exp\Big(-\sum_k t_k \lambda_k
- \sum_{k<l} \, t_k t_l \langle f_k | f_l \rangle_\omega 
-\hlf \sum_l \,  t_l^2 \langle f_l | f_l \rangle_\omega\Big)\,
\end{eqnarray*}
then the integrand $F(\bx)$
is differentiable as a function of $\bx\in\R^n,$ because by 
assumption all $x_k,\,x_l\mapsto\langle f_k(x_k) | f_l(x_l) \rangle_\omega$ 
are continuously differentiable. Thus by the
chain rule we obtain for its partial derivatives that 
\begin{eqnarray}
{\partial\over\partial x_r}\,F(\bx)&=&
{\partial\over\partial x_r}\Big[-\sum_k t_k \lambda_k
- \sum_{k<l} \, t_k t_l \langle f_k | f_l \rangle_\omega 
-\hlf \sum_l \,  t_l^2 \langle f_l | f_l \rangle_\omega\Big]\,F(\bx)
\nonumber \\[1mm]
&=&\Big[-\sum_{k=1}^{r-1}t_rt_k\,{\partial\over\partial x_r}\,
\langle f_k(x_k) | f_r(x_r) \rangle_\omega
-\hlf\,t_r^2\,{\partial\over\partial x_r}\,
\langle f_r(x_r) | f_r(x_r) \rangle_\omega \nonumber\\[1mm]
\label{partialxr}
&&\qquad\quad-\sum_{k=r+1}^nt_rt_k\,{\partial\over\partial x_r}\,
\langle f_r(x_r) | f_k(x_k) \rangle_\omega\Big]\,F(\bx)
\\[1mm]
&=&-\sum_{k=1}^{r-1}\,{\partial\over\partial x_r}\,
\langle f_k(x_k) | f_r(x_r) \rangle_\omega\,
{\partial^2\over\partial \lambda_r\partial\lambda_k}\,F(\bx)
-\hlf\,{\partial\over\partial x_r}\,
\langle f_r(x_r) | f_r(x_r) \rangle_\omega
{\partial^2\over\partial \lambda_r^2}\,F(\bx)\nonumber \\[1mm]
\label{partiallambda}
&&\qquad\quad -\sum_{k=r+1}^n{\partial\over\partial x_r}\,
\langle f_r(x_r) | f_k(x_k) \rangle_\omega
{\partial^2\over\partial \lambda_r\partial\lambda_k}\,F(\bx)
\end{eqnarray}
making use of $t\exp(-t\lambda)=
-{\partial\over\partial \lambda}\exp(-t\lambda)\,.$
In the middle step (\ref{partialxr}), all the terms are integrable functions
in the $t_i\hbox{--variables,}$ and  bounded
by an integrable function (uniformly in $x_r$). To see this, 
recall that $\langle \cdot |\cdot \rangle_\omega$ is positive semidefinite
and hence ${\left|t_it_jF(\bx)\right|}\leq{t_it_j
\exp\Big(-\sum_k t_k \lambda_k\Big)}$ and this is integrable in the 
(positive) variables $t_1,\ldots,\,t_n$ because $\lambda_k>0$ for all $k.$
Thus we can use the dominated convergence theorem for derivatives
(cf. Theorem~2.7 in~\cite{Foll}) to conclude that
 ${\omega(R(\lambda_1,f_1) \cdots R(\lambda_n,f_n))}$
is differentiable in all $x_i\hbox{--variables,}$ and the partial derivatives
can be taken into the integral to give via (\ref{partiallambda}):
\begin{eqnarray*}
&&\!\!\!\!\! \frac{\partial}{\partial x_r} \,
\omega\Big(R(\lambda_1, f_1(x_1)) \cdots R(\lambda_n, f_1(x_n))\Big) 
\\[1mm]
&=& -  \sum_{k=1}^{r-1}
  \frac{\partial }{\partial x_r}\Big(\big\langle f_k(x_k) | 
f_r(x_r) \big\rangle_\omega  \Big)
\int_0^\infty \! dt_1 \dots  \int_0^\infty \! dt_n \,
{\partial^2\over\partial \lambda_r\partial\lambda_k}\,F(\bx) \\[1mm]
&&\qquad-\hlf\frac{\partial }{\partial x_r}\Big(\big\langle f_r(x_r) | 
f_r(x_r) \big\rangle_\omega  \Big)
\int_0^\infty \! dt_1 \dots  \int_0^\infty \! dt_n \,
{\partial^2\over\partial \lambda_r^2}\,F(\bx)  \\[1mm]
% \, \omega\Big(R(\lambda_1, f_1(x_1)) \cdots R(\lambda_n, f_1(x_n))\Big)\\[1mm] 
&&\qquad -  \sum_{k=r+1}^n  
  \frac{\partial }{\partial x_r}\Big(
\big\langle f_r(x_r) | 
  f_k(x_k)
\big\rangle_\omega  \Big)
\int_0^\infty \! dt_1 \dots  \int_0^\infty \! dt_n \,
{\partial^2\over\partial \lambda_r\partial\lambda_k}\,F(\bx)\,.
% \frac{\partial}{\partial \lambda_k}
\end{eqnarray*}
We need to argue that we can take the partial derivatives w.r.t.
the $\lambda_i\hbox{'s}$ through the integrals above.
Now if we have that $\langle f_r(x_r) | f_r(x_r) \rangle_\omega=0,$
then each integrand 
factorises into a  $t_r\hbox{--dependent}$ and a $t_r\hbox{--independent}$ part.
Then the integrand w.r.t. the 
$t_r\hbox{--variable}$ is of the form $t_r^k\exp\big(-t_r\lambda_r\big)$
for $k=0,\,1,\,2\,,$ and so we get explicitly from the Laplace transforms that
we can take $\partial\big/\partial\lambda_r$ through the $t_r\hbox{--integral.}$
This takes care of the part of the integral corresponding to those
variables $t_r$ for which $\langle f_r(x_r) | f_r(x_r) \rangle_\omega=0\,.$
The remaining factor $\wt{F}(\bx)$ of $F(\bx)$ depends only on variables $t_r$ for which
we have that $\langle f_r(x_r) | f_r(x_r) \rangle_\omega>0\,.$
Then ${\left|{\partial^2\over\partial \lambda_r\partial\lambda_k}\,\wt{F}(\bx)\right|}
\leq {t_rt_k\exp\Big(-\hlf \sum_l \,  t_l^2 \langle f_l | f_l \rangle_\omega\Big)}$   
which is integrable w.r.t. the remaining variables, and
likewise we also get a dominating function for the first derivatives.
Thus by dominated convergence (uniformly in the $\lambda\hbox{--variables}$) we can
take the partial derivatives in $\lambda_i$ through the integral in
in the remaining variables. 
Thus we get
\begin{eqnarray*}
&&\!\!\!\!\! \frac{\partial}{\partial x_r} \,
\omega\Big(R(\lambda_1, f_1(x_1)) \cdots R(\lambda_n, f_1(x_n))\Big) 
\\[1mm]
&=& -  \sum_{k=1}^{r-1}
  \frac{\partial }{\partial x_r}\Big(\big\langle f_k(x_k) | 
f_r(x_r) \big\rangle_\omega  \Big)
\, \frac{\partial^2}{\partial \lambda_r\partial\lambda_k}
% \frac{\partial}{\partial \lambda_k}
\, \omega\Big(R(\lambda_1, f_1(x_1)) \cdots R(\lambda_n, f_1(x_n))\Big)\\[1mm]
&&\qquad-\hlf\frac{\partial }{\partial x_r}\Big(\big\langle f_r(x_r) | 
f_r(x_r) \big\rangle_\omega  \Big)
\, \frac{\partial^2}{\partial \lambda_r^2} % \frac{\partial}{\partial \lambda_r}
\, \omega\Big(R(\lambda_1, f_1(x_1)) \cdots R(\lambda_n, f_1(x_n))\Big)\\[1mm] 
&&\qquad -  \sum_{k=r+1}^n  
  \frac{\partial }{\partial x_r}\Big(
\big\langle f_r(x_r) | 
  f_k(x_k)
\big\rangle_\omega  \Big)
\, \frac{\partial^2}{\partial \lambda_r\partial \lambda_k} 
% \frac{\partial}{\partial \lambda_k}
\, \omega\Big(R(\lambda_1, f_1(x_1)) \cdots R(\lambda_n, f_1(x_n))\Big)
\end{eqnarray*}

\subsection*{Proof of Proposition~\ref{KMSbasics}}

To prove this theorem, we first need to establish the following lemma.
\begin{lem}
\label{AnalFn}
For the strip $S:=\R+i[0,1[\subset\C,$ let $F: S\to\C$
be a continuous function, analytic on the interior of $S,$ which satisfies
for some $C>0$ and $\lambda\in\C,$ $|\lambda|=1$ the conditions:
\begin{eqnarray*}
\big|F(t+is)\big| &\leq& C(1+|t|)^N\quad\forall\,t\in\R,\;s\in(0,1) \\[1mm]
\hbox{and}\qquad\qquad F(t+i)&=&\lambda\,F(t)\quad\forall\,t\in\R\;.
\end{eqnarray*}
Then $F=0$ if $\lambda\not=1$ and $F=\hbox{constant}$ if $\lambda=1\,.$
\end{lem}
\begin{beweis}
Note that $\C$ is covered by the strips $S_n:= S+in.$ We define
$G:\C\to\C$ by $G(z):=\lambda^nF(z-in)$ whenever $z\in S_n$
This is consistent, because on the joining lines 
$\R+in=S_n\cap S_{n-1}$ we have that
$G(t+in)=\lambda^nF(t)=\lambda^{n-1}F(t+i)\,.$ By the continuity
of $F$ on $S_0= S,$ $G$ is continuous. Now $G$ is analytic on the interior
of each $S_n$ and continuous on the boundary, 
i.e. continuous on the lines $\R+in$ and analytic on either side of them. 
So it follows from a
well-known theorem of analytic continuation that  $G$ is analytic on the lines
$\R+in$ (cf.~\cite{Ne} p183) hence entire. 
Moreover $G(z+i)=\lambda G(z)$ for all $z\,.$
Let $\Gamma$ be a closed anticlockwise circle of radius $R$ centered at the fixed point
$z_0\,.$ If $z\in \Gamma\cap S_n$ then
\begin{eqnarray*}
\big|G(z)\big|&=& \big|\lambda^nF(z-in)\big|=\big|F(z-in)\big| \\[1mm]
&\leq&C\big(1+|{\rm Re}(z-in)|\big)^N=C\big(1+|{\rm Re}(z)|\big)^N \\[1mm]
&\leq&C\big(1+\big(R+|{\rm Re}(z_0)\big)|\big)^N 
\leq C\big(1+R+|z_0|\big)^N
\end{eqnarray*}
which is independent of $n,$ i.e. $|G(z)|\leq C\big(1+R+|z_0|\big)^N$ for all
$z\in\Gamma\,.$ Applying this to the Cauchy integral formula:
\begin{eqnarray*}
G^{(k)}(z_0)&=&{k!\over 2\pi i}\int_\gamma{G(z)\over(z-z_0)^{k+1}}\,dz\qquad\quad
\hbox{we find:} \\[1mm]
\left|G^{(k)}(z_0)\right|&\leq&k!R\sup_{z\in\Gamma}{\big|G(z)\big|\over R^{k+1}}
\leq k!\,C\,{\big(1+R+|z_0|\big)^N\over R^k}\;.
\end{eqnarray*}
When $k>N+1$ this goes to zero as $R\to\infty,$ hence $G^{(k)}(z_0)=0$ for all
 $k>N+1\,.$ This is true for all $z_0\in\C,$ so $G$ is a polynomial.
However only a constant polynomial can satisfy  $G(z+i)=\lambda G(z)$
(or else it has infinitely many zeroes), hence $G$ is a constant, and if 
$\lambda\not=1,$ the only possible constant is zero. 
\end{beweis}
(i) The KMS-condition for $A=\un$ reads $F\s\un,B.(t)=\varphi\big(\alpha_t(B)\big)
=F\s\un,B.(t+i)\,.$ Thus by lemma~\ref{AnalFn} it follows that
$F\s\un,B.$ is constant, hence that $\varphi\big(\alpha_t(B)\big)$
is independent of $t\,.$\chop
(ii) Let $\gamma(A)=-A\in\dom\varphi\,,$ then 
$F\s A,\un.(t+i)=\varphi(\gamma(A))=-\varphi(A)=-F\s A,\un.(t)$
so by lemma~\ref{AnalFn} we have $0=F\s A,\un.=\varphi(A)\,.$
For any $B\in\dom\varphi$ decompose $B=B_++B_-$ into $\gamma\hbox{-even}$ and odd parts
then we get $\varphi(B)=\varphi(B_+)=\varphi(\gamma(B))\,.$\chop
Finally, let a functional $\varphi$ satisfy the graded KMS-condition on a
set $Y\subset\dom\varphi\,.$ Let $A,\,B\in\span Y,$ i.e.
$A=\sum\limits_i\lambda_iA_i\,,$ $B=\sum\limits_j\mu_jB_j$ for $A_i,\,B_j\in Y\,,$
and $\lambda_i,\,\mu_j\in\C\,.$
Then
\[
F\s A,B.(t):=\varphi\big(A\alpha_t(B)\big)=\sum_{i,j}\lambda_i\mu_j\,
\varphi\big(A_i\alpha_t(B_j)\big)= \sum_{i,j}\lambda_i\mu_j\,F\s A_i,B_j.(t)\,.
\]
Since $\varphi$ is $\gamma\hbox{--KMS}$ on $Y$ the $F\s A_i,B_j.$ are $\gamma\hbox{--KMS}$
functions. Thus $F\s A,B.$ is continous on $S,$ analytic on its interior, and
\begin{eqnarray*}
F\s A,B.(t+i)&=&\sum_{i,j}\lambda_i\mu_j\,F\s A_i,B_j.(t+i)
=\sum_{i,j}\lambda_i\mu_j\, \varphi\big(\alpha_t(B_j)\gamma(A_i)\big) \\[1mm]
&=& \varphi\big(\alpha_t(B)\gamma(A)\big) \;,\\[1mm]
\left|F\s A,B.(t+is)\right|&\leq&\sum_{i,j}|\lambda_i\mu_j|\,
\left|F\s A_i,B_j.(t+is)\right| \\[1mm]
&\leq&\sum_{i,j}|\lambda_i\mu_j|\,C_{ij}(1+|t|)^{N_{ij}}
\leq C(1+|t|)^N
\end{eqnarray*}
where $C:=\sum\limits_{i,j}|\lambda_i\mu_j|\,C_{ij}$ and 
$N:=\mathop{\rm max}\limits_{ij}(N_{ij})\,.$
So $\varphi$ is $\gamma\hbox{--KMS}$ on $\span Y\,.$

\subsection*{Proof of Proposition~\ref{KMStensor}}

Since $\dom\psi$ and $\al B.$ are dense *--algebras, it follows that 
 $\dom\varphi:={\span
\set C\otimes B,C\in\dom\psi,\;B\in{\al B.}.}$
is a dense *-algebra of $\al A.=\al C.\otimes\al B.,$ and that it is 
invariant w.r.t. $\gamma\otimes\iota$ and $\sigma\otimes\beta\,.$ 
Thus by the last part of Proposition~\ref{KMSbasics}
it suffices to verify the KMS--property on
$\set C\otimes B,C\in\dom\psi,\;B\in{\al B.}.\,.$
Let $A_i:=C_i\otimes B_i\,,$ $i=1,2$ for $C_i\in\dom\psi\subset\al C.,$ and
$B_i\in\al B.\,.$ Consider for $t\in\R$ the function
\begin{eqnarray*}
F\s A_1,A_2.(t)&:=& \varphi\left(A_1(\sigma\otimes\beta)_t(A_2)\right) \\[1mm]
&=&\varphi\big((C_1\otimes B_1)(\sigma_t(C_2)\otimes\beta_t(B_2))\big)  \\[1mm]
&=&\psi\big(C_1\,\sigma_t(C_2)\big)\,\omega\big(B_1\,\beta_t(B_2)\big)
=F^\psi\s C_1,C_2.(t)\,F^\omega\s B_1,B_2.(t)
\end{eqnarray*} 
where $F^\psi\s C_1,C_2.$ and $F^\omega\s B_1,B_2.$ are the KMS--functions of 
$\psi$ and $\omega$ resp.. Thus, using their analytic properties, it follows that
$F\s A_1,A_2.$ extends to an function on the strip
$S=\R+i[0,1]\,,$ analytic on its interior and continuous on
the boundary, given by $F\s A_1,A_2.(z)=F^\psi\s C_1,C_2.(z)\,F^\omega\s B_1,B_2.(z)\,.$
Moreover
\begin{eqnarray*}
F\s A_1,A_2.(t+i)&=&F^\psi\s C_1,C_2.(t+i)\,F^\omega\s B_1,B_2.(t+i)
=\psi\big(\sigma_t(C_2)\,\gamma(C_1)\big)\,\omega\big(\beta_t(B_2)\,B_1\big) \\[1mm]
&=& \varphi\left(\sigma_t(C_2)\gamma(C_1)\otimes\beta_t(B_2)\,B_1\right)
=\varphi\left((\sigma\otimes\beta)_t(A_2)(\gamma\otimes\iota)(A_1)\right)
\end{eqnarray*}
and thus $F\s A_1,A_2.$ will be a $(\gamma\otimes\iota)\hbox{--KMS}$ function 
if the tempered growth property also holds. We have 
${\big|F^\psi\s C_1,C_2.(t+is)\big|}\leq K(1+|t|)^N$ for a constant $K$ and $N\in\N$
depending on $C_i\,.$ Now as $\omega$ is a state we have from the KMS--property that
${\big|F^\omega\s B_1,B_2.(t+is)\big|}\leq{\|B_1\|\,\|B_2\|}$ for $s=0,\,1\,.$
So by the maximum modulus principle (apply it after first mapping $S$ to the unit disk
by the Schwartz mapping principle) it follows that
${\big|F^\omega\s B_1,B_2.(z)\big|}\leq{\|B_1\|\,\|B_2\|}$ for all $z\in S\,.$
Thus for $t\in\R,$ $s\in[0,1]$ we have
\[
\big|F\s A_1,A_2.(t+is)\big|\leq \|B_1\|\,\|B_2\|\, K(1+|t|)^N
\]
and so the tempered growth property holds for $F\s A_1,A_2.\,.$
Thus $\varphi$ is a $(\gamma\otimes\iota)\hbox{--KMS}$ functional.

\subsection*{Proof of Theorem~\ref{KMSresolv}}

(i) To prove there is a quasi--free state on $\CCR$ defined by
$\omega(\delta_f):=\exp[-s(f,f)/2]\,,$ $f\in\al S.(\R,\R)$ where
${s(f,f)}:={\int{p\over 1-e^{-p}}\left|\widehat{f}(p)\right|^2dp}\,,$
it suffices to show that 
${|\sigma(f,h)|^2}\leq{4\,s(f,f)\,s(h,h)}$ by~\cite{MaVe}.
\begin{eqnarray*}
|\sigma(f,h)|^2&=&\Big|\int ip\,\widehat{f}(p)\,\ol\widehat{h}(p).\,dp\Big|^2
\leq \Big(\int\left|p\,\widehat{f}(p)\,\widehat{h}(p)\right|dp\Big)^2 \\[1mm]
&=&\Big(2\int_0^\infty p \left|\widehat{f}(p)\,\widehat{h}(p)\right|dp\Big)^2
\qquad\quad\hbox{as $\big|p\widehat{f}(p)\,\widehat{h}(p)\big|$ is even by 
$\widehat{f}(-p)=\ol\widehat{f}(p).$}\\[1mm]
&\leq&4\int_0^\infty p\big|\widehat{f}(p)\big|^2dp\,
\int_0^\infty k\big|\widehat{h}(k)\big|^2dk\qquad\qquad\hbox{by Cauchy--Schwartz}\\[1mm]
&<&4\int_0^\infty{ p\over 1-e^{-p}}\big|\widehat{f}(p)\big|^2dp\,
\int_0^\infty{ k\over 1-e^{-k}}\big|\widehat{h}(k)\big|^2dk\quad\qquad
\hbox{as $p<{ p\over 1-e^{-p}}$ for $p>0$} \\[1mm]
&<&4\int_{-\infty}^\infty{ p\over 1-e^{-p}}\big|\widehat{f}(p)\big|^2dp\,
\int_{-\infty}^\infty{ k\over 1-e^{-k}}\big|\widehat{h}(k)\big|^2dk\quad\qquad
\hbox{as $0<{ p\over 1-e^{-p}}$ for all $p$} \\[1mm]
&=&4\,s(f,f)\,s(h,h)
\end{eqnarray*}
(ii)
Next we need to show that this quasi--free state $\omega$ on $\CCR$ extends to a
KMS--functional on $\pi_\omega(\CCR)'',$ and hence on $\al R.(\al S.(\R),\sigma).$
We first prove that $\omega$ is KMS on the *--algebra
$\Delta_c
% ${\Delta(\al S.(\R),\sigma)}
:={\span\set\delta_f,{f\in\al S.(\R),\;\supp \widehat{f}\;\hbox{compact}}.}\,.$
Let $\delta_f,\,\delta_h\in\Delta_c$ and
consider
\begin{eqnarray*}
F_1(t)&:=&\omega\big(\delta\s f.\alpha_t(\delta\s h.)\big)=
e^{-i\sigma(f,h_t)/2}\omega(\delta\s f+h_t.) \\[1mm]
&=&\exp\big[-\f i,2.\,\sigma(f,h_t)-\f 1,2.\,s(f+h_t,f+h_t)\big]\\[1mm]
&=&\exp\Big[\f 1,2.\int\Big(-p\,\widehat{f}\,\ol\widehat{h_t}.
-{p\over 1-e^{-p}}
\big|\widehat{f}+\widehat{h_t}\big|^2\Big)\,dp\Big] \\[1mm]
&=&\exp\Big[\f 1,2.\int p \Big({|\widehat{f}|^2+|\widehat{h}|^2\over e^{-p}-1}
-{2-e^{-p}\over 1-e^{-p}}\,e^{ipt}\widehat{f}\,\ol\widehat{h}.
-{e^{-ipt}\over 1-e^{-p}}\,\ol\widehat{f}.\,\widehat{h}\Big)\,dp\Big]
\end{eqnarray*}  
 where we used $\widehat{h_t}(p)=e^{-ipt}\widehat{h}(p)\,.$
 Now put $K:={\exp\Big[\hlf\int p{|\widehat{f}|^2+|\widehat{h}|^2\over e^{-p}-1}\,dp\Big]\,,}$
 and substitute $p\to-p$ in the last integral, using $\widehat{f}\,\ol\widehat{h}.(-p)
 =\ol\widehat{f}.\,\widehat{h}(p)$ to get:
 \[
 F_1(t)=K\,\exp\Big[-\int{pe^{ipt}\over 1-e^{-p}}\,\widehat{f}\,\ol\widehat{h}.(p)\,dp\Big]\;.
 \]
 By a similar calculation we find $\omega\big(\alpha_t(\delta\s h.)\delta\s f.\big)=F_1(t+i)\,$
 and this suggests that we define the KMS--function for $z\in S=\R+i[0,1]$ by:
 \begin{equation}
 \label{KMSdeltas}
 F(z):=K\,\exp\Big[-\int{pe^{ipz}\over 1-e^{-p}}\,\widehat{f}\,\ol\widehat{h}.(p)\,dp\Big]\;.
 \end{equation}
Let $z=x+iy\in S=\R+i[0,1],$ then we know that the integral exists for $y=0,\,1\,.$
For $y\in(0,1)$ the function $\left|{p\,e^{ipz}\over 1-e^{-p}}\right|={p\,e^{-py}\over 1-e^{-p}}$
is bounded, so the integral exists for all $z\in S,$ hence the definition~(\ref{KMSdeltas}) makes sense
for $F\,.$ It is however not clear that $F$ is analytic.
However, recall that by assumption $\supp \wh{f}$ and $\supp \wh{h}$ are compact, 
then it follows from the dominated convergence theorem that $F$ is continuous on $S$
and as the differential w.r.t. $z$ of the integrand is continuous in $p,$
it also implies that $F$ is analytic on the interior of $S.$
Thus $F$ is the KMS--function for $\omega,$ hence by the last part of  
 Proposition~\ref{KMStensor}, $\omega$ is a KMS--state on $\Delta_c\,.$

Next, we prove that $\pi_\omega(\Delta_c)$ is strong operator dense in
$\pi_\omega(\CCR)''\,.$
It suffices to show that for each $f\in\al S.(\R)$ there is a sequence
$\{f_n\}\subset\al S.(\R)$ with Fourier transforms of compact support, such that
${\omega\big(\delta_g\,\delta\s f_n.\,\delta_h\big)}\to
{\omega\big(\delta_g\,\delta\s f.\,\delta_h\big)}$
for all $g,\,h\in\al S.(\R)$ as $n\to\infty\,.$
Fix an $f\in\al S.(\R)$ and define $f_n\in\al S.(\R)$ by its 
Fourier transform $\widehat{f_n}=K_n\cdot\widehat{f}$ where 
each $K_n:\R\to[0,1]$ is a  smooth bump function of compact support which is $1$
on ${[-n,n]}$ and zero on ${\R\backslash[-n-1,\,n+1]}\,.$ Then
\begin{eqnarray*}
\omega\big(\delta_g\,\delta\s f_n.\,\delta_h\big) &=& e^{-i\sigma(g+f_n,\,f_n+h)/2}
\omega\big(\delta\s g+f_n+h.\big) \\[1mm]
&=& \exp\left[-\f i, 2.\,\sigma(g+f_n,\,f_n+h)-\hlf\,s\big( g+f_n+h,\, g+f_n+h\big)\right] \\[1mm]
&=& \exp\Big[\hlf\int\Big(-p(\wh{g}+\wh{f_n})(\ol\wh{f_n}.+\ol\wh{h}.)
-{p\over 1-e^{-p}}\big|\wh{g}+\wh{f_n}+\wh{h}\big|^2\Big)\,dp\Big] \\[1mm]
&\maprightt\infty,n.&
\exp\Big[\hlf\int\Big(-p(\wh{g}+\wh{f})(\ol\wh{f}.+\ol\wh{h}.)
-{p\over 1-e^{-p}}\big|\wh{g}+\wh{f}+\wh{h}\big|^2\Big)\,dp\Big] \\[1mm]
&=& \omega\big(\delta_g\,\delta\s f.\,\delta_h\big) 
\end{eqnarray*}
using dominated convergence as $f$ is $L^1\,.$

Next, we want to use the strong operator denseness of $\pi_\omega(\Delta_c)$
in $\pi_\omega(\CCR)''$ to show that $\omega$ is KMS on all of
$\pi_\omega(\CCR)''\,.$
Let $A,\,B\in\pi_\omega(\CCR)''$ be selfadjoint, then by the Kaplansky density theorem
(cf. Theorem~5,3.5 p329 in~\cite{KaRiI}) it follows that there
are sequences $\{A_n\},\;\{B_n\}\subset\pi_\omega(\Delta_c)$ of
selfadjoint elements such that 
$\|A_n\|\leq\|A\|,$ $\|B_n\|\leq\|B\|$ and
$A_n\to A,$  $B_n\to B,$  in
strong operator topology. Now $\omega\circ\alpha_t=\omega$ because
$s(f_t,\,f_t)=s(f,f),$ and thus there is an implementing unitary
$U_t\in\al B.(\al H._\omega)$ such that
${U_t\pi_\omega(D)U_t^*}=\pi_\omega(\alpha_t(D))$ for $D\in\CCR$
and $U_t\Omega_\omega=\Omega_\omega\,.$
Abbreviate the KMS--functions $F_n(z):=F\s A_n,B_n.(z)$ of $\omega,$
then for all $t\in\R\,:$
\begin{eqnarray*}
F_n(t)-F_k(t)&=&\omega\big(A_n\,\alpha_t(B_n)\big)-
\omega\big(A_k\,\alpha_t(B_k)\big)\\[1mm]
&=&\big(A_n\Omega_\omega,\,U_tB_n\Omega_\omega\big)
-\big(A_k\Omega_\omega,\,U_tB_k\Omega_\omega\big) \\[1mm]
&=&\big(U_t^*A_n\Omega_\omega,\,(B_n-B_k)\Omega_\omega\big)
+\big(U_t^*(A_n-A_k)\Omega_\omega,\,B_k\Omega_\omega\big) \\[1mm]
\hbox{Thus}\qquad
\left|F_n(t)-F_k(t)\right| &\leq&
\big\|(B_n-B_k)\Omega_\omega\|\cdot\|A_n\|+
\big\|(A_n-A_k)\Omega_\omega\|\cdot\|B_k\|\\[1mm]
&\leq& \big\|(B_n-B_k)\Omega_\omega\|\cdot\|A\|+
\big\|(A_n-A_k)\Omega_\omega\|\cdot\|B\|
\end{eqnarray*}

and this is independent of $t.$ Similarly
\[
\left|F_n(t+i)-F_k(t+i)\right|
\leq\big\|(A_n-A_k)\Omega_\omega\|\cdot\|B\|+
\big\|(B_n-B_k)\Omega_\omega\|\cdot\|A\|\;.
\]
Now $F_n-F_k$ is a analytic function on the strip $S,$
so by combining The Riemann mapping theorem with the maximum modulus
principle we have that $|F_n-F_k|$ takes its maximum on the boundary of $S\,.$
Thus by the inequalities above, $F_n-F_k$ converges uniformly to zero,
hence the uniform limit $F:=\lim\limits_nF_n$ exists and is 
analytic and bounded on $S.$ Since
${\omega\big(A_n\alpha_t(B_n)\big)}={\big(A_n\Omega_\omega,\,U_tB_n\Omega_\omega\big)}
\to{\big(A\Omega_\omega,\,U_tB\Omega_\omega\big)}
={\omega\big(A\alpha_t(B)\big)}$
it follows that $F$ is the KMS--function $F\s A,B.$ of $\omega.$
Thus $\omega$ is KMS on the selfadjoint part, hence on all of $\pi_\omega(\CCR)''\,.$

Since $\omega$ is quasifree, it is strongly regular, hence the 
resolvents of the generators of the one-parameter groups 
$t\to\pi_\omega(\delta\s tf.)$ will provide a representation
of 
 $\al R.(\al S.(\R),\sigma),$
where it is defined by Equation~(\ref{QFresolvent})
via spectral theory.

\subsection*{Proof of Theorem~\ref{ThetaProp}}

 For $z\in S=\R+i[0,1]$ consider 
\begin{eqnarray}
G(z)&:=&
\lim_{\varepsilon\to 0^+}\Big( \int_{-\infty}^{-\varepsilon}
+\int_{\varepsilon}^\infty\Big){e^{ipz}\over 1-e^{-p}}\,
\,{\widehat{f}(p)}\,\ol\hat{g}(p).\,dp \nonumber\\[1mm]
&=&\lim_{\varepsilon\to 0^+}\int_\varepsilon^\infty
\Big({e^{ipz}\over 1-e^{-p}}\,{\widehat{f}(p)}\,\ol\hat{g}(p).
+{e^{-ipz}\over 1-e^{p}}\,\ol{\widehat{f}(p)}.\,{\hat{g}(p)}
\Big)\,dp\qquad\hbox{by $\widehat{f}(-p)=\ol\widehat{f}(p).$}\nonumber\\[1mm]
&=&\int_0^\infty
\Big[\Big(e^{ipz}+{e^{-p}e^{ipz}\over 1-e^{-p}}\Big)
\,{\widehat{f}(p)}\,\ol\hat{g}(p).
- {e^{-p}e^{-ipz}\over 1-e^{-p}}\,\ol{\widehat{f}(p)}.\,{\hat{g}(p)}
\Big]\,dp \nonumber \\[1mm]
&=&(g,F(D)f) +\int_0^\infty dp\int dx\int dy\,f(x)\,g(y){e^{-p}\over 1-e^{-p}}
\Big(e^{ipz}e^{ip(y-x)}-e^{-ipz}e^{ip(x-y)}\Big)\nonumber\\[1mm]
\label{GIntegr}
&=&(g,F(D)f) +2i\int_0^\infty dp\int dx\int dy\,f(x)\,g(y)\,{e^{-p}\over 1-e^{-p}}
\,\sin p(z+y-x)
\end{eqnarray}
where $D=id/dx$ and $F(p):=e^{ipz}\chi\s[0,\infty).(p)\,,$
and we used the fact that the Fourier transform diagonalises $D\,.$
To use Fubini to rearrange these integrals, we need to show that the integrand 
is integrable. % and this will also produce estimates to prove part~(ii).
We need to separate the low $p$ from the high $p$ behaviour in the last integral.
For  the low $p$ behaviour, consider the integral
\[
\int_0^1 dp\int dx\int dy\,f(x)\,g(y)\,{e^{-p}\over 1-e^{-p}}
\,\sin p(z+y-x)\;.
\]
Rearrange the integrand to $f(x)\,g(y)\,{e^{-p}\over 1-e^{-p}}\,p(z+y-x)
\,{\sin p(z+y-x)\over p(z+y-x)}$ and observe that $p(z+y-x)\in S=\R+i[0,1]\,$
since $p\in[0,1]\,.$
Now $H(z):=\sin(z)\big/z$ is analytic in $S,$ so $|H(z)|$ takes its maximum
on the boundary. On $\R,$ $|H(x)|\leq 1,$ and for $\R+i$ we have
\[
\big|H(x+i)\big|^2=\left|{\sin(x+i)\over x+i}\right|^2
={\cosh^21-\cos^2x\over x^2+1}\leq\cosh^2 1
\]
and thus for the integrand
\begin{equation}
\label{LowIngrand}
\left|f(x)\,g(y)\,{e^{-p}\over 1-e^{-p}}\,p(z+y-x)
\,{\sin p(z+y-x)\over p(z+y-x)}\right|
\leq \left|f(x)\,g(y)\right|\,{e^{-p}\over 1-e^{-p}}\,p|z+y-x|\cosh 1
\end{equation}
which is clearly integrable because $f$ and $g$ are Schwartz functions
so take care of the linear factor in $x$ and $y\,.$
For the high $p$ behaviour, consider the remaining part of the
integral~(\ref{GIntegr}), i.e.
\[
\int_1^\infty dp\int dx\int dy\,f(x)\,g(y)\,{e^{-p}\over 1-e^{-p}}
\,\sin p(z+y-x)\;.
% =\int_1^\infty dp\,{e^{-p}\over 1-e^{-p}}\Big[e^{ipz}\ol\wh{f}(p).\,
% \wh{g}(p)-e^{-ipz}\wh{f}(p)\,\ol\wh{g}(p).\Big]
\]
Now for $z=t+is\in S,$ $t\in\R,$ $s\in[0,1]$ we have
\begin{eqnarray*}
\left|f(x)\,g(y)\,{e^{-p}\over 1-e^{-p}}
\,\sin p(z+y-x)\right| &\leq&
\hlf\left|f(x)\,g(y)\right|\,{e^{-p}\over 1-e^{-p}}
\,\left|e^{i p(z+y-x)}-e^{-i p(z+y-x)}\right|\\[1mm]
\leq \hlf\left|f(x)\,g(y)\right|\,{e^{-p}\over 1-e^{-p}}
\,\left|e^{-ps}+e^{ps}\right| &\leq& \left|f(x)\,g(y)\right|\,{
e^{p(s-1)}\over 1-e^{-p}}
% \leq \left|f(x)\,g(y)\right|\,{1\over 1-e}\,.
\end{eqnarray*}
This is integrable for $s\in[0,1),$ but not for $s=1\,.$
However, we will see below that $G$ is continuous on $S,$
and this will be enough.
Thus the integrand in (\ref{GIntegr}) is integrable
for $z\in\R+i[0,1),$ and we can apply
the Fubini theorem to rearrange the order of integrals,
and we get:
\begin{equation}
\label{GFubini}
G(z)=(g,F(D)f) +2i\int dx\int dy\,f(x)\,g(y)\,\int_0^\infty dp\,{e^{-p}\over 1-e^{-p}}
\,\sin p(z+y-x)\,.
\end{equation}
To prove (i), let $z=0,$ so $G(0)=\theta(f,g)$ and $F(p)=\chi\s[0,\infty).(p)$
hence $P:=F(D)=2\pi\times\hbox{projection}$ onto positive spectrum of $D\,.$ So
\begin{eqnarray*}
G(0)=\theta(f,g)&=& 
(g,Pf) +2i\int dx\int dy\,f(x)\,g(y)\,\int_0^\infty dp\,{e^{-p}\over 1-e^{-p}}
\,\sin p(y-x) \\[1mm]
&=&(g,Pf)-2i\int dx\int dy\,f(x)\,g(y)\,(y-x)\,\int_0^\infty dp\,
\ln( 1-e^{-p})\,\cos p(y-x)\\[1mm]
&=&\big(g,(P+T)f)
\end{eqnarray*}
through an integration by parts. Consider the operator
\[
\big(Tf\big)(x):=2i\int dy\,f(y)\,(x-y)\,\int_0^\infty dp\,
\ln( 1-e^{-p})\,\cos p(x-y)=:\int dy\,f(y)\,K(x-y)
\]
which is obviously a kernel operator with kernel $K.$ Due to the factor
${(x-y)}$ in $K,$ $T$ is unbounded. Note however that 
\[
\left|{\partial^n\over\partial t^n}\,\ln(1-e^{-p})\,\cos pt\right|
\leq\big|p^n\ln(1-e^{-p})\big|
\]
which is integrable in $p.$ [To see this, note that for $0<p\leq1\,,$
$\big|p^n\ln(1-e^{-p})\big|\leq\big|\ln(1-e^{-p})\big|
\leq\big|\ln p\big|+\big|\ln({1-e^{-p}\over p})\big|$
which is integrable, and for $e^p>2$ we have
$\big|p^n\ln(1-e^{-p})\big|=p^n\big(e^{-p}+e^{-2p}/2+e^{-3p}/3+\cdots\big)
\leq p^n\,e^{-p}\big(1+e^{-p}/2+e^{-2p}/3+\cdots\big)
\leq p^n\,e^{-p}\sum_{k=0}^\infty(\hlf)^k\leq 2 p^n\,e^{-p}$
which is integrable.]
Thus by dominated convergence the function
$t\to{\int_0^\infty dp\,
\ln( 1-e^{-p})\,\cos pt}$ is
 is smooth. Thus the kernel $K$ of $T$ is smooth. If $J$ is a compact interval
then $P_JTP_J$ has kernel $\chi\s J.(x)K(x-y)\chi\s J.(y)$ which is smooth
and bounded on $J\,.$ Thus $P_JTP_J$ is trace--class by Theorem~1 p128
in Lang~\cite{La}. We also have an explicit proof that 
$P_JTP_J$ is trace--class below in the proof of Theorem~\ref{LocBd}.
Selfadjointness now follows from the fact 
that $\theta(f,f)$ is real by its formula. This proves (i).

To prove (ii), Note that we already proved above that $G(z)$ is well-defined
for $z\in\R+i[0,1)\,.$ 
To prove that it is well defined on all of $S,$ it is only necessary to prove
integrability for the high $p$ part of the integral. For this
\begin{eqnarray}
&&\Big|2i\int_1^\infty dp\int dx\int dy\,f(x)\,g(y)\,{e^{-p}\over 1-e^{-p}}
\,\sin p(z+y-x)\Big|\nonumber \\[1mm]
&=&\Big|\int_1^\infty dp\,{e^{-p}\over 1-e^{-p}}
\Big[e^{ipz}\wh{f}(p)\,\ol\wh{g}(p).-e^{-ipz}\ol\wh{f}(p).\,\wh{g}(p)
\Big]\Big|\nonumber \\[1mm]
&\leq&\int_1^\infty dp\,{e^{-p}\over 1-e^{-p}}
\big(e^{-ps}+e^{ps}\big)\,\big|\wh{f}(p)\,\wh{g}(p)\big|\nonumber \\[1mm]
\label{HighpIneq}
&\leq&\int_1^\infty dp\,{2\over 1-e^{-p}}
\big|\wh{f}(p)\,\wh{g}(p)\big|\leq{2\over 1-e^{-1}}\|f\|\,\|g\|
\end{eqnarray}
where $z=t+is\in S,$
and so $G(z)$ is well-defined for $z\in S\,.$ 

To establish the stated inequality for $z=s+it\in S,$ consider
Equation~(\ref{GFubini}). Now $|F(p)|=e^{-sp}\chi\s[0,\infty).(p)$
implies $\|F\|=1,$ so $\|F(D\|=1$ and hence 
${\big|(g,F(D)f,
)\big|}\leq{\|f\|\,\|g\|}\,.$ 
The high $p$ part of the integral in (\ref{GFubini}) has an estimate
(\ref{HighpIneq}), so for the low $p$ integral we have for its integrand
the inequality (\ref{LowIngrand}) above, so that
\begin{eqnarray*}
&&\!\!\!\!\Big|\int_0^1 dp\int dx\int dy\,f(x)\,g(y)\,{e^{-p}\over 1-e^{-p}}
\,\sin p(z+y-x)\Big|  \\[1mm]
&\leq&\cosh1\cdot\int dx\int dy\,\big|f(x)\,g(y)\big|\,\big(1+|t|+|x-y|\big)\,
\int_0^1 dp\,{p\,e^{-p}\over 1-e^{-p}}   \\[1mm]
&=& C+|t|E
\end{eqnarray*}
for some finite constants $C$ and $E.$
Combining this with (\ref{HighpIneq}) and the estimate for 
${\big|(g,\,F(D)f)\big|},$ we obtain 
${\big|G(t+is)\big|}\leq A+B|t|$ for constants $A,\,B$ as desired.
It remains to prove that $G$ is continuous on $S$ and analytic in its interior.
Now
\[
\Big|{d\over dz}\,f(x)\,g(y)\,{e^{-p}\over 1-e^{-p}}\,\sin p(z+y-x)\Big|
\leq\big|f(x)\,g(y)\big|\,{p\,e^{-p}\over 1-e^{-p}}\,\hlf(e^{-sp}+e^{sp})
\]
which is $L^1$ for all $s\in(0,1)\,.$ Thus the last integral in
(\ref{GFubini}) is analytic on the interior of $S\,.$ That
${(g,F(D)f)}$ is analytic in $z$ follows from spectral theory,
hence $G$ is analytic on the interior of $S\,.$
For continuity on $S,$ we already have that ${(g,F(D)f)}$ is
continuous in $z,$ and by inequalities (\ref{HighpIneq}) and
(\ref{LowIngrand}) we can get $L^1$ estimates to ensure that 
$G$ is continuous on $S.$

\subsection*{Proof of Theorem~\ref{QFCAR}}

We now show that
 quasi--free functional $\psi$ with two point functional $\theta$
is a graded KMS--functional on 
$\ClifS\,.$
Its domain $\dom\psi=\hbox{*-alg}\set c(f), {f\in\al S.(\R)}.$
is clearly a unital dense *-algebra of $\ClifS$
which is invariant w.r.t. both the grading $\gamma$ and the
time evolution $\alpha_t\,,$ so part (i) of 
Definition~\ref{KMSfunDef} is satisfied.
For the KMS-condition (ii), it suffices to check it for the
monomials $c(f_1)\cdots c(f_k)\,.$ Let $A=c(f_1)\cdots c(f_k)$
and $B=c(g_1)\cdots c(g_m)\,$ where $k+m=2n\,.$ Then
from (\ref{QFodd}) and (\ref{QFeven}) we get
 \begin{eqnarray}
F\s A,B.(t)&:=&\psi\big(A\,\alpha_t(B)\big)=
\psi(c(f_1)\cdots c(f_{k})c(T_tg_1)\cdots c(T_tg_m)\nonumber\\[1mm]
\label{QFtheta}
&=&
(-1)^{\left({n\atop 2}\right)}\sum_P(-1)^P\prod_{j=1}^n
\theta\big(h\s P(j).,\,h\s P(n+j).\big)
 \end{eqnarray}
where $h_1=f_1,\ldots,h_k=f_k,\,h_{k+1}=T_tg_1,\ldots,h_{2n}=T_tg_m$
and $T_tf:=f_t$ is translation by $t\,.$ Since $P(j)<P(n+j)$ always,
the terms ${\theta\big(h\s P(j).,\,h\s P(n+j).\big)}$ can only be one of 
the types
\[
\theta\big(f_i,T_tg_j\big)\qquad\hbox{or}\qquad
\theta\big(T_tf_i,T_tg_j\big)\qquad\hbox{if $k<n,$ or}\qquad
\theta\big(f_i,\,g_j\big)\qquad\hbox{if $k>n\,.$}
\]
Since by the formula for $\theta$ we have $\theta\big(T_tf,T_tg\big)=
\theta\big(f,\,g\big)$ the last two types are the same and constant in $t.$
For the first type, we get by definition functions 
$G(t)={\theta\big(f,T_tg\big)}$ as in Theorem~\ref{ThetaProp}(ii),
which we therefore know extend analytically to the strip $S\,.$
Thus since Equation~(\ref{QFtheta}) expresses $F\s A,B.(t)$ as a polynomial
of constant functions and functions of the form $G\,,$ it follows that
$F\s A,B.(t)$ extends to a continuous function on $S$ which is analytic
on its interior.
Now 
 \begin{eqnarray}
G(t+i)&=& 
\lim_{\varepsilon\to 0^+}\Big( \int_{-\infty}^{-\varepsilon}
+\int_{\varepsilon}^\infty\Big){e^{ip(t+i)}\over 1-e^{-p}}\,
\,{\widehat{f}(p)}\,\ol\hat{g}(p).\,dp \nonumber\\[1mm]
&=& \lim_{\varepsilon\to 0^+}\Big( \int_{-\infty}^{-\varepsilon}
+\int_{\varepsilon}^\infty\Big){e^{ipt}\over e^{p}-1}\,
\,{\widehat{f}(p)}\,\ol\hat{g}(p).\,dp \nonumber\\[1mm]
&=& \lim_{\varepsilon\to 0^+}\Big( \int_{-\infty}^{-\varepsilon}
+\int_{\varepsilon}^\infty\Big){e^{-ipt}\over e^{-p}-1}\,
\,{\widehat{f}(-p)}\,\ol\hat{g}(-p).\,dp \nonumber\\[1mm]
&=&- \lim_{\varepsilon\to 0^+}\Big( \int_{-\infty}^{-\varepsilon}
+\int_{\varepsilon}^\infty\Big){e^{-ipt}\over 1- e^{-p}}\,
\,\ol{\widehat{f}(p)}.\,\hat{g}(p)\,dp \nonumber\\[1mm]
\label{GKMStheta}
&=&-\theta(T_tg,f)=\psi\big(\alpha_t(c(g))\gamma(c(f))\big)
 \end{eqnarray}
which is the graded KMS-condition for $F\s c(f),c(g).(t)=
\psi\big(c(f)\alpha_t(c(g))\big)\,.$
The terms $\theta(T_tg,f)$ are exactly the ones which occur in the 
corresponding expression~(\ref{QFtheta}) for 
${\psi\big(\alpha_t(B)\,\gamma(A)\big)}$ 
so the graded KMS-condition for $F\s A,B.$ follows from the
one for $G,$ Equation~(\ref{GKMStheta}).

It remains to prove the growth condition~(iii)
of Definition~\ref{KMSfunDef}. We already have 
\[
\big|G(t+is)\big|\leq a+b|t|\quad\hbox{for}\;t\in\R,\;
s\in[0,1]
\]
by Theorem~\ref{ThetaProp}(ii). So from formula~(\ref{GKMStheta})
we get that for $t+is\in S:$
\[
\big|F\s A,B.(t+is)\big|\leq
(a_1+b_1|t|)\cdots(a_n+b_n|t|)\leq C\,(1+|t|)^n
\]
for suitable constants $a_i,\,b_i$ and $C\,.$
Thus $\psi$ is a graded KMS-functional.

\subsection*{Proof of Theorem~\ref{KMSfSUSY}}

By construction $\dom\varphi$ contains the *-algebra generated by
all $c(f),\;f\in\al S.(\R)$ as well as $\al R.(\al S.(\R),\sigma)$
and so it will certainly contain the *-algebra generated by $\un$ and
all $c(f),\;\rlf\,,$ which is $\al A._0\,.$ So (i) is trivially true.

Next, for (ii),  we need to prove the SUSY-invariance of $\varphi\,,$
and for this, we need  the following lemma.
\begin{lem}
\label{PhijR} 
For all $g,\;f_i\in\al S.(\R)\backslash 0$
and $\lambda_i\in\R\backslash 0$ we have
% \[
% \varphi\big(j(f_0)\,R(\lambda_1,f_1)\cdots R(\lambda_n,f_n)\big)
% =\sum_{k=1}^n s(f_k,f_0)\cdot   %\varphi\big(j(f_0)\,j(f_k)\big)\cdot
% \varphi\left(R(\lambda_1,f_1)\cdots R(\lambda_k,f_k)^2
% \cdots R(\lambda_n,f_n)\right)
% \]
\begin{eqnarray*}
&&\!\!\!\!\!\!\!\!\!\!
\varphi\Big(R(\lambda_1,f_1)\cdots R(\lambda_n,f_n)
\,j(g)\,R(\lambda_{n+1},f_{n+1})\cdots R(\lambda_m,f_m)\Big) \\[1mm]
&=&\sum_{k=1}^n s(f_k,g)\,
\varphi\Big(R(\lambda_1,f_1)\cdots R(\lambda_k,f_k)^2\cdots 
 R(\lambda_n,f_n)
\,R(\lambda_{n+1},f_{n+1})\cdots R(\lambda_m,f_m)\Big) \\[1mm]
&&+\sum_{k=n+1}^m
s(g,f_k)\;
\varphi\Big(R(\lambda_1,f_1)\cdots R(\lambda_n,f_n)
\,R(\lambda_{n+1},f_{n+1})\cdots R(\lambda_k,f_k)^2\cdots R(\lambda_m,f_m)\Big)
\end{eqnarray*}
where $\varphi$ is a strongly regular state on $\al R.(\al S.(\R),\sigma)$
so these expressions make sense on $\al E._0\,.$
\end{lem}
\begin{beweis}
Recall by Theorem~\ref{KMSresolv} that $\varphi$ is a
 quasi--free state on $\CCR$ defined by
$\varphi(\delta_f):=\exp[-s(f,f)/2]\,,$ $f\in\al S.(\R,\R)$ where
$s$ is given in Equation~(\ref{KMSCCR}). Since the maps
$t,r\to s(rf,tg)$ are smooth, we can apply 
Proposition~\ref{MasterLemma} to $\varphi$ w.r.t. the maps $t\to tf\,.$
From the two relations
${d\over dt}R(\lambda,tf)=j(f)R(\lambda,tf)^2$ and
$\lim\limits_{t\to 0}\lambda i\pi_\varphi(R(\lambda,tf))\psi=\psi,$
we get
\begin{eqnarray*}
&&\!\!\!\!\!\!\!\!\!\!
\varphi\Big(R(\lambda_1,f_1)\cdots R(\lambda_n,f_n)
\,j(g)\,R(\lambda_{n+1},f_{n+1})\cdots R(\lambda_m,f_m)\Big) \\[1mm]
&=&-\mu^2\,\lim_{t\to 0}{\partial\over \partial t}\,
\varphi\Big(R(\lambda_1,f_1)\cdots R(\lambda_n,f_n)
\,R(\mu,tg)\,R(\lambda_{n+1},f_{n+1})\cdots R(\lambda_m,f_m)\Big) \\[1mm]
&=& -\mu^2\,\lim_{t\to 0}\bigg\{
-\sum_{k=1}^n{d\over dt}\,s(f_k,tg)\,
{\partial^2\over \partial \mu\partial\lambda_k}\,
\varphi\Big(R(\lambda_1,f_1)\cdots R(\lambda_n,f_n)
\,R(\mu,tg)\,R(\lambda_{n+1},f_{n+1})\cdots R(\lambda_m,f_m)\Big) \\[1mm]
&&-\hlf\,{d\over d t}\,
s(tg,tg)\;{\partial^2\over \partial \mu^2}\,
\varphi\Big(R(\lambda_1,f_1)\cdots R(\lambda_n,f_n)
\,R(\mu,tg)\,R(\lambda_{n+1},f_{n+1})\cdots R(\lambda_m,f_m)\Big) \\[1mm]
&&-\sum_{k=n+1}^m{d\over dt}\,
s(tg,f_k)\;{\partial^2\over \partial \mu\partial\lambda_k}\,
\varphi\Big(R(\lambda_1,f_1)\cdots R(\lambda_n,f_n)
\,R(\mu,tg)\,R(\lambda_{n+1},f_{n+1})\cdots R(\lambda_m,f_m)\Big) 
\bigg\}  \\[1mm]
&&\qquad\qquad\qquad\hbox{(By Proposition~\ref{MasterLemma})} \\[1mm]
&=&\sum_{k=1}^n s(f_k,g)\,
\varphi\Big(R(\lambda_1,f_1)\cdots R(\lambda_k,f_k)^2\cdots 
 R(\lambda_n,f_n)
\,R(\lambda_{n+1},f_{n+1})\cdots R(\lambda_m,f_m)\Big) \\[1mm]
&&+\sum_{k=n+1}^m
s(g,f_k)\;
\varphi\Big(R(\lambda_1,f_1)\cdots R(\lambda_n,f_n)
\,R(\lambda_{n+1},f_{n+1})\cdots R(\lambda_k,f_k)^2\cdots R(\lambda_m,f_m)\Big) 
\end{eqnarray*}
where we used ${d\over d\lambda}\rlf=-i\rlf^2\,.$
\end{beweis}
Next we need to show that $\varphi\circ\delta$ is zero on $\al D._S\,,$
i.e. that it vanishes on all the monomials:
\[
\zeta(f_1)\cdots\zeta(f_n)R(\lambda_1,g_1)\cdots R(\lambda_m,g_m)
=c(f_1)\cdots c(f_n)R(1,f_1)\cdots R(1,f_n)R(\lambda_1,g_1)\cdots R(\lambda_m,g_m)\,.
\]
Recall that $\delta$ is a restriction to $\al D._S$ of a graded derivation
on $\al E.$ defined by
\[
{\delta}(j(f)) = i c(f^\prime), \qquad
 {\delta}(R(\lambda,f)) = i c(f^\prime) R(\lambda,f)^2, \qquad\hbox{and}\qquad
 {\delta}(c(f)) = j(f)\,.
 \]
 So we calculate:
 \begin{eqnarray*}
&&\!\!\!\!\!\varphi\circ\delta\left(c(f_1)\cdots c(f_n)R(1,f_1)\cdots R(1,f_n)
R(\lambda_1,g_1)\cdots R(\lambda_m,g_m)\right) \\[1mm]
&=& \sum_{k=1}^n(-1)^{k+1}\varphi\Big(c(f_1)\cdots\v k.\cdots c(f_n)
\,j(f_k)\,R(1,f_1)\cdots R(1,f_n)
R(\lambda_1,g_1)\cdots R(\lambda_m,g_m)\Big) \\[1mm]
&&\qquad+i(-1)^n\sum_{\ell=1}^n\varphi\left(c(f_1)\cdots c(f_n)c(f'_\ell)R(1,f_1)\cdots
R(1,f_\ell)^2\cdots R(1,f_n)
R(\lambda_1,g_1)\cdots R(\lambda_m,g_m)\right) \\[1mm]
&&\qquad+i(-1)^n\sum_{p=1}^m\varphi\left(c(f_1)\cdots c(f_n)c(g'_p)R(1,f_1)\cdots
\cdots R(1,f_n)R(\lambda_1,g_1)\cdots 
R(\lambda_p,g_p)^2\cdots R(\lambda_m,g_m)\right) \\[1mm]
&=& \sum_{k=1}^n(-1)^{k+1}\varphi\big(c(f_1)\cdots\v k.\cdots c(f_n)\big)
\bigg\{  \\[1mm]
&&\qquad\quad
\sum_{r=1}^n    s(f_k,f_r)\,  %\varphi\big(j(f_k)\,j(f_r)\big)\,
\varphi\left(R(1,f_1)\cdots
R(1,f_r)^2\cdots R(1,f_n)R(\lambda_1,g_1) 
\cdots R(\lambda_m,g_m)\right) \\[1mm]  
&&\qquad\quad+
\sum_{t=1}^m s(f_k,g_t)\,  %\varphi\big(j(f_k)\,j(g_t)\big)\,
\varphi\left(R(1,f_1)\cdots
R(1,f_n)R(\lambda_1,g_1)\cdots 
R(\lambda_t,g_t)^2\cdots R(\lambda_m,g_m)\right)\bigg\} \\[1mm]
&&+i(-1)^n\sum_{\ell=1}^n\varphi\left(c(f_1)\cdots c(f_n)c(f'_\ell)\right)
\,\varphi\left(R(1,f_1)\cdots
R(1,f_\ell)^2\cdots R(1,f_n)
R(\lambda_1,g_1)\cdots R(\lambda_m,g_m)\right) \\[1mm]
&&+i(-1)^n\sum_{p=1}^m\varphi\left(c(f_1)\cdots c(f_n)c(g'_p)\right)
\,\varphi\left(R(1,f_1)
\cdots R(1,f_n)R(\lambda_1,g_1)\cdots 
R(\lambda_p,g_p)^2\cdots R(\lambda_m,g_m)\right) 
 \end{eqnarray*}
 where we made use of the lemma~\ref{PhijR}. Note that as $\varphi$ is quasifree,
 $n$ must be odd for the last expression to be nonzero, and also:
 \[
 \varphi\big(c(f_1)\cdots c(f_n)\big)=
 \sum_{k=1}^{n-1}(-1)^{k+1}\,\varphi\big(c(f_k)\,c(f_n)\big)\,
 \varphi\big(c(f_1)\cdots\v k.\cdots c(f_{n-1})\big)\,.
 \]
So we get:
\begin{eqnarray*}
&&\!\!\!\!\!\!\!
\varphi\circ\delta\left(c(f_1)\cdots c(f_n)R(1,f_1)\cdots R(1,f_n)
R(\lambda_1,g_1)\cdots R(\lambda_m,g_m)\right) \\[1mm]
&=& \sum_{k=1}^n(-1)^{k+1}\varphi\big(c(f_1)\cdots\v k.\cdots c(f_n)\big)
\bigg\{  
\sum_{r=1}^n\Big[s(f_k,f_r)  % \varphi\big(j(f_k)\,j(f_r)\big)
-i\varphi\big(c(f_k)c(f'_r)\big)\Big]  \\[1mm]
&&\qquad\qquad\qquad\times\varphi\left(R(1,f_1)\cdots
R(1,f_r)^2\cdots R(1,f_n)R(\lambda_1,g_1) 
\cdots R(\lambda_m,g_m)\right) \\[1mm] 
&&\qquad+\sum_{p=1}^m\Big[ s(f_k,g_p)  %\varphi\big(j(f_k)\,j(g_p)\big)
-i\varphi\big(c(f_k)\,c(g'_p)\big)\Big]  \\[1mm]
&&\qquad\qquad\qquad\times\varphi\left(R(1,f_1)\cdots
R(1,f_n)R(\lambda_1,g_1)\cdots 
R(\lambda_p,g_p)^2\cdots R(\lambda_m,g_m)\right)\bigg\} \;.
\end{eqnarray*}
However for the two-point functions we have:
\begin{eqnarray*}
\varphi\big(c(f)\,c(g')\big)&=&
\lim_{\varepsilon\to 0^+}\Big( \int_{-\infty}^{-\varepsilon}
+\int_{\varepsilon}^\infty\Big){-ip\over 1-e^{-p}}\,
\,{\widehat{f}(p)}\,\ol\hat{g}(p).\,dp \\[1mm]
&=& -i\int_{-\infty}^\infty{p\over 1-e^{-p}}\,{\widehat{f}(p)}\,\ol\hat{g}(p).\,dp
=-is(f,g)       % \varphi\big(j(f)\,j(g)\big)
\end{eqnarray*}
and so we get $\varphi\circ\delta=0$ as desired.

Finally, for (iii) we need to prove that
\begin{equation}
\label{PhiDT}
 \varphi\big(B
M_A\overline{\delta}_0(A)C\big)=-i{d\over dt}\,
\varphi\big(BM_A\,\alpha_t(A)C\big)\Big|_0
\end{equation}
for all $A\in\al D._S$ and $B,C\in\al A._0\,.$
Since $\alpha_t$ and $\ol\delta._0$ do not mix 
$\ClifS$
 and $\al R.(\al S.(\R),\sigma)$ and $\varphi$ has a 
product structure, it suffices to verify
(\ref{PhiDT}) on the CAR and CCR parts separately. First, on 
the Clifford algebra we have $\ol\delta._0(c(f))=ic(f'),$ 
and by the derivative property we only need to check
(\ref{PhiDT}) for $A=c(f)\,.$ However, $\varphi$ is quasifree
so it suffices to check for the two-point functions that
\[
{d\over dt}\,\varphi\big(c(g)\,\alpha_t(c(f))\big)\Big|_0
=i\varphi\big(c(g)\ol\delta._0(c(f))\big)=-\varphi\big(
c(g)\,c(f')\big)
\]
for all $f,\,g\in\al S.(\R)\,.$
The differentiability of $G(t)=\varphi\big(c(g)\,\alpha_t(c(f))\big)$
was proven above in Theorem~\ref{ThetaProp}. So:
\begin{eqnarray*}
&&\!\!\!\!\!\!\!\!\!\!\!
{d\over dt}\,\varphi\big(c(f)\,\alpha_t(c(g))\big)\Big|_0
={d\over dt}\,\lim_{\varepsilon\to 0^+}\Big( \int_{-\infty}^{-\varepsilon}
+\int_{\varepsilon}^\infty\Big){e^{ipt}\over 1-e^{-p}}\,
\,{\widehat{f}(p)}\,\ol\hat{g}(p).\,dp\,\Big|_0 \\[1mm]
&=& \lim_{\varepsilon\to 0^+}\Big( \int_{-\infty}^{-\varepsilon}
+\int_{\varepsilon}^\infty\Big)
{ip\over 1-e^{-p}}\,
\,{\widehat{f}(p)}\,\ol\hat{g}(p).\,dp \\[1mm]
&=& -\varphi\big(c(f)\,c(g')\big)
\end{eqnarray*}
as required. Next, we need to check (\ref{PhiDT}) on the resolvent
algebra. By the derivative property, it suffices to do this for
$A=\rlf\,,$ and by linearity for the remaining terms being
monomials of resolvents. That is, we need to prove that
\begin{eqnarray}
&& \!\!\!\!\!\!\!\!\!\!\!  {d\over dt}\,
\varphi\Big(R(\lambda_1,f_1)\cdots R(\lambda_n,f_n)
R(\mu,T_tg)\,R(\lambda_{n+1},f_{n+1})\cdots
R(\lambda_m,f_m)\Big)\Big|_0
\nonumber \\[1mm]
\label{RPhiDT}
&=& i\,\varphi\Big(R(\lambda_1,f_1)\cdots R(\lambda_n,f_n)
\ol\delta._0\big(R(\mu,g)\big)\,R(\lambda_{n+1},f_{n+1})\cdots
R(\lambda_m,f_m)\Big)
\end{eqnarray}
where $\ol\delta._0\big(R(\mu,g)\big)=i\,R(\mu,g)\,j(g')\,R(\mu,g)\,.$
This is an expression of the form of Proposition~\ref{MasterLemma},
so to apply this, we need to check that the functions
$t\to s(f,T_tg)$ are smooth (note that $s(T_tf,T_tg)=s(f,g))\,,$ and this is 
an easy verification. In fact, $s(\cdot,\cdot)$ is clearly a distribution
in each entry as it is an expectation value $(f,Ag)$ where 
$A$ is multiplication by a smooth function which is 
polynomially bounded.
% \begin{eqnarray*}
% % && \!\!\!\!\!\!\!\!\!\!\! 
% \lim_{t\to 0}\varphi\big(f,{T_tg-g\over t}-g'\big)
% &=& \lim_{t\to 0}
% \int{p\over 1-e^{-p}}\ol\widehat{f}(p).\,\widehat{g}(p)\,
% \Big({e^{-itp}-1\over t}-(-ip)\Big)\,dp  \\[1mm]
% &=&0\qquad\quad\hbox{by dominated convergence, using} \\[1mm]
% % &&\!\!\!\!\!\!\!\!\!\!\! 
% \Big|{e^{-itp}-1\over -itp}-1\Big|\leq\Big|{\sin(tp/2)\over tp/2}
% \Big|+1\leq 2 &&\hbox{and that} % \\[1mm]
% % &&\!\!\!\!\!\!\!\!\!\!\!
% \qquad{ip^2\over 1-e^{-p}}\ol\widehat{f}(p).\,\widehat{g}(p)
% \qquad\quad\hbox{is integrable.}
% \end{eqnarray*}
Applying Proposition~\ref{MasterLemma} we get:
\begin{eqnarray*}
&& \!\!\!\!\!\!\!\!\!\!\!  {d\over dt}\,
\varphi\Big(R(\lambda_1,f_1)\cdots R(\lambda_n,f_n)
R(\mu,T_tg)\,R(\lambda_{n+1},f_{n+1})\cdots
R(\lambda_m,f_m)\Big)\Big|_{t=0}
 \\[1mm]
&=& -  \sum_{k=1}^n
  \frac{d }{dt}s(f_k, T_tg)  
\, \frac{\partial^2}{\partial \mu\partial \lambda_k}
% \frac{\partial}{\partial \lambda_k}
\, \varphi\Big(R(\lambda_1, f_1)) \cdots 
% \\[1mm] &&\qquad\qquad\cdots 
R(\lambda_n, f_n)
R(\mu,T_tg)\,R(\lambda_{n+1},f_{n+1})\cdots
R(\lambda_m,f_m)\Big)\Big|_{t=0}
\\[1mm]
&&\qquad-\hlf\frac{d }{dt}s( T_tg,T_tg) 
\, \frac{\partial^2}{\partial \mu^2}
\, \varphi\Big(R(\lambda_1, f_1) \cdots 
% \\[1mm] &&\qquad\qquad\cdots 
 R(\lambda_n, f_n)
R(\mu,T_tg)\,R(\lambda_{n+1},f_{n+1})\cdots
R(\lambda_m,f_m)\Big)\Big|_{t=0}
\\[1mm] 
&& -  \sum_{k=n+1}^m
  \frac{d }{dt}s(T_tg,f_k)
\, \frac{\partial^2}{\partial \mu\partial \lambda_k}
% \frac{\partial}{\partial \lambda_k}
\, \varphi\Big(R(\lambda_1, f_1) \cdots 
% \\[1mm] &&\qquad\qquad\cdots 
 R(\lambda_n, f_n)
R(\mu,T_tg)\,R(\lambda_{n+1},f_{n+1})\cdots
R(\lambda_m,f_m)\Big)\Big|_{t=0}
\\[1mm]
&=& -  \sum_{k=1}^n
 s(f_k,g')  
\, \varphi\Big(R(\lambda_1, f_1)) \cdots R(\lambda_k,f_k)^2\cdots
R(\lambda_n, f_n)
R(\mu,g)^2\,R(\lambda_{n+1},f_{n+1})\cdots
R(\lambda_m,f_m)\Big)
\\[1mm]
&& -  \sum_{k=n+1}^m
  s(g',f_k)
\, \varphi\Big(R(\lambda_1, f_1) \cdots 
 R(\lambda_n, f_n)\,
R(\mu,g)^2\,R(\lambda_{n+1},f_{n+1})\cdots
 R(\lambda_k,f_k)^2\cdots R(\lambda_m,f_m)\Big)
\end{eqnarray*}
where we used ${d\over dt}\,s(f,T_tg)\big|_0={d\over dt}
\int{p\over 1-e^{-p}}\,\wh{f}\,e^{ip}\ol\wh{g}.\,dp\big|_0
=\int{ip^2\over 1-e^{-p}}\wh{f}\,\ol\wh{g}.\,dp
=-s(f,g'),$  as well as $s(T_tf,T_tg)=s(f,g)\,.$
On the other hand, for the right hand side of Equation~(\ref{RPhiDT})
we have from Lemma~\ref{PhijR}, that: 
\begin{eqnarray*}
&& \!\!\!\!\!\!\!\!\!\!\! 
i\,\varphi\Big(R(\lambda_1,f_1)\cdots R(\lambda_n,f_n)
\ol\delta._0\big(R(\mu,g)\big)\,R(\lambda_{n+1},f_{n+1})\cdots
R(\lambda_m,f_m)\Big) \\[1mm]
&=& -\varphi\Big(R(\lambda_1,f_1)\cdots R(\lambda_n,f_n)
R(\mu,g)\,j(g')\,R(\mu,g)\,R(\lambda_{n+1},f_{n+1})\cdots
R(\lambda_m,f_m)\Big) \\[1mm]
&=&-\sum_{k=1}^n s(f_k,g')\,
\varphi\Big(R(\lambda_1,f_1)\cdots R(\lambda_k,f_k)^2\cdots 
 R(\lambda_n,f_n)\,R(\mu,g)^2
\,R(\lambda_{n+1},f_{n+1})\cdots R(\lambda_m,f_m)\Big) \\[1mm]
&&-\sum_{k=n+1}^m
s(g',f_k)\;
\varphi\Big(R(\lambda_1,f_1)\cdots R(\lambda_n,f_n)\,R(\mu,g)^2
\,R(\lambda_{n+1},f_{n+1})\cdots R(\lambda_k,f_k)^2\cdots R(\lambda_m,f_m)\Big)
\\[1mm]
&& -s(g,g')\,
\varphi\Big(R(\lambda_1,f_1)\cdots
 R(\lambda_n,f_n)\,R(\mu,g)^3
\,R(\lambda_{n+1},f_{n+1})\cdots R(\lambda_m,f_m)\Big) \\[1mm]
&& -s(g',g)\;
\varphi\Big(R(\lambda_1,f_1)\cdots R(\lambda_n,f_n)\,R(\mu,g)^3
\,R(\lambda_{n+1},f_{n+1})\cdots  R(\lambda_m,f_m)\Big)\,.
\end{eqnarray*}
However, $s(g,g')=-s(g',g)$ and so the last two terms cancel
and hence we have proven~(\ref{RPhiDT}).

\subsection*{Proof of Theorem~\ref{LocBd}}

Recall that 
\begin{eqnarray*}
\al A._0(J)&:=&\hbox{*-alg}\set{c(f),\;\rlf},{\supp f\subseteq J,\;f\in\al S.(\R),\;
\lambda\in\R\backslash0}.=\al C.(J)\otimes\al R._0(J) \\[1mm]
\hbox{where}\qquad \al C.(J)&:=&\hbox{*--alg}\set c(f),{\supp f\subseteq J,\;f\in\al S.(\R)}.\\[1mm]
\hbox{and}\qquad \al R._0(J)&:=&\hbox{*--alg}\set\rlf,{\supp f\subseteq J,\;f\in\al S.(\R),\;
\lambda\in\R\backslash0}.\,.
\end{eqnarray*}
Since $\varphi=\psi\otimes\omega$ is a product functional, and $\omega$
is a state, we have that $\big\|\varphi\restriction\al A._0(J)\big\|
=\big\|\psi\restriction\al C.(J)\big\|\,,$ and so this is what 
we need to estimate. 
Without loss of generality we may assume $J$ to be a closed interval,
and also symmetrical about the origin (since $\varphi$ is invariant w.r.t.
translations).
Recall from Theorem~\ref{ThetaProp}
that $\psi(c(f)c(g))=\theta(f,g)=(g,\,(P+T)f)$ where $P$ is a projection
(after a normalisation)
and $P_JTP_J$ is trace-class and selfadjoint
 for all compact intervals $J\,,$ and so this is the
case for $c(f),\,c(g)\in\al C.(J)\,.$ 
%
% Now recall that for
% a quasifree functional with $\psi(c(f)c(g))=(g,\,(\un+S)f)$
% and $S^*=S,$
% it will actually be a state iff $\|S\|\leq 1\,$ (cf.~\cite{MRC}).
% Thus in our case where $S:=P-\un+T,$ we have to control contributions of 
% $S$ corresponding to spectral values in the complement of 
% ${[-1,1]}\,.$ 
% 
 We will need to use the isomorphism of the Clifford algebra 
 $\ol{\al C.(J)}.$ with a self--dual 
 CAR--algebra explicitly. First observe that
 $\ol{\al C.(J)}.={\rm Cliff}\big(L^2(J,\R)\big)$ by continuity of $c(f)\,.$
 Since $J$ is symmetrical about the origin, we can define
 $\Gamma:L^2(J,\R)\to L^2(J,\R)$ by $(\wh{\Gamma f})(p):=\wh{f}(-p)\,.$
 Then $(\Gamma f)(x)=f(-x)$ and $\Gamma P=(\un-P)\Gamma\,,$
 and  $\Gamma T\Gamma=-T$      by the explicit formula for $T\,.$
 Define for $f\in  L^2(J,\R)$
 \[
 \Phi(f):=\f 1,\sqrt{2}.\,\Big(c(Pf)-ic(\Gamma Pf)+c(P\Gamma f)
 +ic\big((\un-P)f\big)\Big)
 \]
 and observe that $\Phi(\Gamma f)=\Phi(f)^*$ and
 $\{\Phi(f),\Phi(g)\}=(\Gamma f,g)\un,$ which establishes the isomorphism.
 By requiring complex linearity for $\Phi(f),$ we get
 $\Phi(f)+i\Phi(g)=:\Phi(f+ig)$ hence get in fact isomorphism 
 of $\ol{\al C.(J)}.$ with ${\rm CAR}(L^2(J,\C))\,.$ Note that by
 $\Phi(\Gamma f)=\Phi(f)^*$ the involution $\Gamma$ has to extend
 to $L^2(J,\C)=:\al K.$ in a conjugate linear way.
The image of $\al C.(J)$ under the isomorphism, is the dense
*-algebra  ${\rm CAR}_0$ generated by all
$\Phi(f)\,,$ $f\in \al K.\,,$ and this is the domain of
$\psi\,.$

With respect to the decomposition $\al K.=P\al K.\oplus(\un-P)\al K.
\ni f\oplus g$ we decompose
$T=\Bigl({A\atop C}{B\atop D}\Bigr)$
so by $T=T^*$ we get $A=A^*,\;D=D^*$ and $B=C^*\,,$
and these operators are trace--class because $T$ is.
From the relation $\Gamma T\Gamma=-T$ we then find that 
$A=-D$ and $B^*=-B\,,$ hence 
$T=\Bigl({A\atop -B}{B\atop -A}\Bigr)\,.$
Since $T$ preserves the original real space $L^2(J,\R)$ we have
$\ol Af.=A\,\ol f.$ and $\ol Bf.=B\,\ol f.\,.$
 Then the two-point function of $\psi$ on ${\rm CAR}_0$
 is for $f\oplus g,\;h\oplus k\in L^2(J,\R)\;:$
\begin{eqnarray*}
&&\!\!\!\!\!\psi\big(\Phi(f\oplus g)\,\Phi(h\oplus k)\big)  \\[1mm]
&=&\hlf\,\psi\Big(\big(c(f\oplus0)-i\,c(0\oplus f)+c(g\oplus0)+i\,
c(0\oplus g)\big)\cdot\big(c(h\oplus0)-i\,c(0\oplus h)+c(k\oplus0)+i\,
c(0\oplus k)\big)\Big) \\[1mm]
&=&\hlf\,\big[(h,f)+(k,f)+(h,g)+(k,g)\big]
+(h,Af)-i(k,Bf)+i(h,Bg)+(k,Ag) \\[1mm]
&=&\Big(\Gamma(h\oplus k),\,\Big[\hlf\,
\left(\scriptstyle{\matrix{I & I \cr I & I \cr}}\right)+
\left(\scriptstyle{\matrix{-iB & A \cr A & iB \cr}}\right)\Big](f\oplus g)\Big)\;.
\end{eqnarray*}
This expression is complex linear in both entries, so we can extend it by linearity
to $\al K.$ to get for all $f,\,g\in\al K.$ that
\begin{eqnarray}
\label{RQstart}
\psi\big(\Phi(f)\,\Phi(g)\big)&=&\big(\Gamma(g),\,\big(
R+Q\big)f\big)\qquad\hbox{where:} \\[1mm]
R&:=&\hlf\,
\left(\scriptstyle{\matrix{I & I \cr I & I \cr}}\right)\qquad\hbox{and}\qquad
Q:=
\left(\scriptstyle{\matrix{-iB & A \cr A & iB \cr}}\right) \nonumber
\end{eqnarray}
Define an operator $S$ by $\hlf +S:=R+Q\,,$ then $S$ is a bounded selfadjoint
operator  which satisfies $\Gamma S\Gamma =S\,.$
 As $S_0:=\hlf\Bigl({0\atop I}{I\atop 0}\Bigr)$ has eigenvalues $\pm\hlf$ and $Q$ is trace class,
 $S=S_0+Q$ has discrete spectrum with the only possible accumulation
 points $\pm\hlf$ (cf. Theorem~9.6~\cite{Weid}). We now need:
% Recall that spectral theory for selfadjoint
% operators is well-defined on real Hilbert spaces (cf. 7.25, p199~\cite{Weid}).
% Then we have:
\begin{lem}
\label{RomeWork}
\begin{itemize}
\item[(i)] $E([-\hlf,\hlf]^c)\cdot(2|S|-\un)$ on $L^2(J)$ is trace class
where $E$ is the spectral resolution of $S\,.$
Moreover we have for the trace--norms $\|\cdot\|_1$ that
\[
\left\|E([-\hlf,\hlf]^c)\cdot(2|S|-\un)\right\|_1\leq
b\left\|P_JTP_J\right\|_1\,
\]
for a positive constant $b$ (independent of $J$).
\item[(ii)]
Let $\{e_j\,|\,j\in\Mb{J}\}$ be an orthonormal system of eigenvectors
of $S$ corresponding to the eigenvalues $s_j\in[-\hlf,\hlf]^c\,,$
and exhausting these eigenspaces.
For each $j \in \Mb{J},$ let
${\cal C}_j$ be the two--dimensional abelian *-algebras
 generated by
$\Phi(e_j)^*\Phi(e_j)$, and let $\al C._0:=\hbox{*--alg}\set\Phi(f),
{f\in E\big([-\hlf,\hlf]\big)
\al K.}.$. Then
\begin{equation}
\label{psiProd}
\big\|\psi\restriction\al C.(J)\big\|
 = \prod_{j \in \{0\} \cup \Mb{J}} \| \psi_j \|
=\prod_{j \in  \Mb{J}}2|s_j| <\infty
\end{equation}
where $\psi_j$ denotes the
restriction of $\psi$ to $ {\cal C}_j\,.$
\end{itemize}
\end{lem}
\begin{beweis}
(i) Let Let $E_0(\,\cdot\,)$ be the spectral resolution corresponding to 
$S_0\,.$
For any $s\in(0,\hlf)$ which is not in the spectrum $\sigma(S)$ of $S\,,$ 
% Picking any $0 < s < \mbox{\footnotesize $\frac{1}{2}$}$
% which does not belong to the spectrum of $S$ and taking
since $S$ and $ S_0$ are bounded, we obtain by
spectral calculus that
\begin{eqnarray}
E((s,\infty)) - E_0((s,\infty)) &=&
(2 \pi i)^{-1} \, \int_{C} \! dz \, ((z-S)^{-1}  
- (z-S_0)^{-1})\nonumber\\[1mm]
\label{DiffRes}
&=&(2 \pi i)^{-1} \, \int_{C} \! dz \, (z-S_0)^{-1} Q (z-S)^{-1}
\end{eqnarray}
where $C$ is a suitable closed path in $\C\,,$ e.g. a large anticlockwise 
simple contour with $\sigma(S)\cap [s,\infty)$ and 
$\sigma(S_0)\cap [s,\infty)$ in its interior, and crossing the real axis only
at $s$ and some $t>s\,.$ % In the last step we used:
% \[
% (z-S)^{-1} - (z-S_0)^{-1} =  (z-S_0)^{-1} Q (z-S)^{-1}
% \]
So from (\ref{DiffRes})
we conclude that $E((s,\infty)) - E_0((s,\infty))$ is a
trace class operator. Taking into account that
$ E_0((s,\infty)) (2S_0 -1) = 0$ we obtain
\begin{equation}
\label{EStrace}
E((s,\infty)) (2S -1) = \big(E((s,\infty)) - E_0((s,\infty))\big) (2S -1)
+ E_0((s,\infty)) \, 2Q,
\end{equation}
showing that $E((s,\infty)) (2S -1)$ and hence {\it a fortiori}
$E((\mbox{\footnotesize $\frac{1}{2}$}, \infty))\, (2S -1)$ is 
trace class. A similar argument establishes that 
$E((-\infty,-\hlf)) (2S +1)$ is trace class, and hence
 $E([-\hlf,\hlf]^c)\cdot(2|S|-\un)$  is trace class.
From Equation~(\ref{EStrace}) we get that
\begin{eqnarray}
\left\|E((s,\infty)) (2S -1)\right\|_1 &\leq&
\left\|\big(E((s,\infty)) - E_0((s,\infty))\big) (2S -1)\right\|_1
+\left\|E_0((s,\infty)) \, 2Q\right\|_1 \nonumber\\[1mm]
\label{Etraceineq}
&\leq&\left\|\big(E((s,\infty)) - E_0((s,\infty))\big) (2S -1)\right\|_1
+2\|Q\|_1
\end{eqnarray}
since all the terms in (\ref{EStrace}) are trace class, and
$\|AQ\|_1\leq\|A\|\cdot\|Q\|_1$ for $A$ bounded. 
Let $P_0$ be the projection onto the eigenspace of $S_0$
with eigenvalue $\hlf$
(since
  $S_0=\hlf\Bigl({0\atop I}{I\atop 0}\Bigr)$  this is just the space
of even functions w.r.t. the decomposition associated with $P\,).$
Then by substituting
\[
(z-S_0)^{-1}=(z-\hlf)^{-1}P_0+(z+\hlf)^{-1}(\un-P_0)
\qquad\hbox{and}\qquad(z-S)^{-1}(2S-\un)=(2z-1)(z-S)^{-1}-2\un
\]
into  (\ref{DiffRes})$\times(2S-\un)$ we get that
\begin{eqnarray*}
&&\!\!\!\!\!\!\!
\big[E((s,\infty)) - E_0((s,\infty))\big](2S-\un) \\[1mm]
&=& \f 1,2\pi i. \Big[P_0Q\,\int_{C} \! dz\,(z-\hlf)^{-1}\,
\Big(\f 2z-1,z-S.-2\Big)
+(\un-P_0)Q\,\int_{C} \! dz\,(z+\hlf)^{-1}\,\Big(\f 2z-1,z-S.-2\Big)
\Big] \\[1mm]
&=&\f 1,2\pi i.P_0Q\,\int_{C} \! dz\,2 \,(z-S)^{-1}-2P_0Q
+\f 1,2\pi i.(\un-P_0)Q\,\int_{C} \! dz\,
{\Big(\f 2z-1,z+1/2.\Big)\over z-S} \\[1mm]
&=&P_0\,Q\,2E((s,\infty))-2P_0\,Q+(\un-P_0)Q\,f(S)
\end{eqnarray*}
where $f(z):=\f 2z-1,z+1/2.\,\chi\s H.(z)$ and $H:=\set z\in\C,
{{\rm Re}(z)> s}.\,.$ 
Now $\|f(S)\|\leq\|f\restriction(s,\infty)\|_\infty=2\,,$ and so
\[
\Big\|\big[E((s,\infty)) - E_0((s,\infty))\big](2S-\un)\Big\|_1 
\leq6\|Q\|_1\leq a\|P_JTP_J\|_1
\]
for a constant $a>0\,,$ where we obtain the last inequality
$\|Q\|_1\leq\hbox{const.}\|P_JTP_J\|_1$ from the decomposition
in (\ref{RQstart}) from which $PQP=-iB=-iP(P_JTP_J)(\un-P),$
$(\un-P)QP=P(P_JTP_J)P$ etc. Thus by (\ref{Etraceineq}) we get
\[
\|E((\hlf,\infty))(2S-1)\|_1=\|E(\hlf,\infty)E(s,\infty)(2S-1)\|_1
\leq\|E(s,\infty)(2S-1)\|_1\leq a\|P_JTP_J\|_1\,.
\]
A similar argument establishes that $\|E(-\infty,-\hlf)(2S+1)\|_1
\leq a\|P_JTP_J\|_1$ and hence that 
$
\left\|E([-\hlf,\hlf]^c)\cdot(2|S|-\un)\right\|_1\leq
b\left\|P_JTP_J\right\|_1\,
$
for a positive constant $b\,.$
\chop
(ii)
Recall that $S$ has a purely discrete spectrum in $[-\hlf,\hlf]^c\,,$
so choose an orthonormal system
$\set{ e_j\in\al K.},j\in\J\subseteq\N.$ 
of eigenvectors of $S,$  corresponding to eigenvalues
$s_j\in[-\hlf,\hlf]^c$ and exhausting these eigenspaces
(some $s_j$ will coincide for higher multiplicities).
Let $E_j$ be the one--dimensional orthogonal projection onto $e_j$,
$j \in \J$,
and let $\mbox{\boldmath $\lambda$} = (\lambda_1, \lambda_2, \dots ) 
\in \bigoplus_{j \in \J} \, \{-1,1\}$. We define on ${\cal K}$
the unitaries
$$V(\mbox{\boldmath $\lambda$}) := E([-\hlf,\hlf]) + \sum_{j \in \J}
\lambda_j E_j\,.$$
Since $V(\mbox{\boldmath $\lambda$})$ commutes with $\Gamma$,
these unitaries induce an action $\gamma:\bigoplus_{j \in \J} \, \{-1,1\}
 \to\aut{\rm CAR}(\al K.)$ given by
$$
\gamma_{\mbox{\boldmath $\scriptstyle\lambda$}} (\Phi(f)) :=
\Phi(V(\mbox{\boldmath $\lambda$}) f).
$$
Since $V(\mbox{\boldmath $\lambda$})^2=\un\,,$ we can decompose 
${\rm CAR}(\al K.)$ into odd and even parts w.r.t. 
each $\gamma_{\mbox{\boldmath $\scriptstyle\lambda$}}\,.$
Moreover, since $V(\mbox{\boldmath $\lambda$})$ commutes with $S$
we have that
$
\psi \circ \gamma_{\mbox{\boldmath $\scriptstyle\lambda$}} = \psi
$
and so $\psi$ must vanish on the odd part of ${\rm CAR}_0$
with respect to each $\gamma_{\mbox{\boldmath $\scriptstyle\lambda$}}\,.$
Let $\al C.\subset{\rm CAR}_0$
be the *--algebra generated by
$\set{\Phi( e_j)},j\in\J.\cup\set{\Phi(f)},{f\in E\big([-\hlf,\hlf]\big)
\al K.}.\,,$ then $\al C.$ is mapped to itself by all 
$\gamma_{\mbox{\boldmath $\scriptstyle\lambda$}}\,.$
Since the two-point functional $\theta$ is bounded on $L^2(J),$
it suffices to
 calculate the norm of $\psi$ on $\al C.,$ and in fact
on the intersection of all the even parts of $\al C.$
with respect to $\gamma_{\mbox{\boldmath $\scriptstyle\lambda$}}\,,$
and we denote this *-algebra by $\varepsilon(\al C.)\,.$
It is produced by a projection $\varepsilon$ which   we can 
consider as the projection defined on $\al C.$
by averaging over the action of $\gamma_{\mbox{\boldmath $\scriptstyle\lambda$}}$
on $\al C..$ Since for each $A\in\al C.$ only a finite number
of $j\hbox{'s}$ are involved, these averages will
again be in the *-algebra $\al C..$
Now we only need to consider  monomials 
in the $\Phi( e_j)$ and $\Phi( e_j)^*$ which are 
even in each index $j\,.$
In a given monomial $\Phi( e_{j_1})\cdots\Phi( e_{j_n})\in\varepsilon(\al C.)$
if we collect all (even number of) terms with the same $j$ together,
we can then simplify it with the relations
$2 \, \Phi(e_j)^2 = \langle \Gamma e_j | e_j \rangle 1 $ and
$\big[\Phi(e_j)^*\Phi(e_j)\big]^2=\Phi(e_j)^*\Phi(e_j)-\f 1,4.\big|(\Gamma e_j,
e_j)\big|^2\un\,.$ Thus  $\varepsilon(\al C.)$ is generated by
the two-dimensional abelian *-algebras $\al C._j:=\hbox{*--alg}\{
\Phi(e_j)^*\Phi(e_j)\}$
 and  $\al C._0:=\hbox{*--alg}\set\Phi(f),{f\in E\big([-\hlf,\hlf]
\big)\al K.}.\,.$ Since for $i\not=j$ we have
$$
\varepsilon\left(\big[\Phi(e_i)^*\Phi(e_i),\,\Phi(e_j)^*\Phi(e_j)\big]\right)
=\varepsilon\left(\ol\big(\Gamma e_i,e_j\big).\Phi(e_j)\Phi(e_i)+
\big(\Gamma e_i,e_j\big)\Phi(e_j)^*\Phi(e_i)^*\right)=0
$$
it follows that all the $\al C._i$ commute, and in fact we have
the (incomplete) tensor product decomposition
$$
\varepsilon({\cal C}) = \bigotimes_{j \in \{0\} \cup \J} \, {\cal C}_j.
$$
Moreover, $\psi$ is a product functional on this tensor product.
Hence its norm, if it exists, is given by
$$\| \psi \| = \prod_{j \in \{0\} \cup \J} \| \psi_j \|,$$
where $\psi_j$ denotes the
restriction of $\psi$ to $ {\cal C}_j$. Now $\psi_0$ is by
construction a state on ${\cal C}_0$
because $\hlf+S$ is positive on ${E\big([-\hlf,\hlf]
\big)\al K.},$ hence the two-point function is positive
and so $\| \psi_0 \| = 1$.
Since for $a,b \in \C$
$$
\psi(a1+b\,\Phi(e_j)^* \Phi(e_j)) = a + b \, (\mbox{\footnotesize
  $\frac{1}{2}$} + s_j)
$$
and
$$
\| a1+b\,\Phi(e_j)^* \Phi(e_j) \| = \max\{|a|,|a+b|\}
$$
one obtains $ \| \psi_j \| = 2 |s_j| $, $j \in \J$.
Thus since $2|s_j|>1,$ 
a necessary and sufficient condition for the existence of
$\| \psi \|$ is $\sum_{j \in \J} (2 |s_j| -1) < \infty$.
However, this is guaranteed by part (i)
\end{beweis}
Using this lemma, we can now prove:
\begin{lem}
\label{eT1norm}
We have $$\big\|\varphi\restriction\al A._0(J)\big\|
\leq\exp\big(b\|P_JTP_J\|_1\big)$$ where $b$ is a positive constant
(independent of $J$) and $\|\cdot\|_1$ denotes the trace--norm.
\end{lem}
\begin{beweis}
recall from Equation~(\ref{psiProd}) that
\[
\big\|\varphi\restriction\al A._0(J)\big\|
 = \prod_{j \in \{0\} \cup \Mb{J}} \| \psi_j \|
=\prod_{j \in  \Mb{J}}2|s_j|=\prod_{j \in  \Mb{J}}(1+t_j) <\infty
\]
where $t_j:=2|s_j|-1\,.$ Now $\ln(1+x)\leq x$ for $x\geq 0\,,$ so
\[
\ln\prod_{j =1}^N(1+t_j)=\sum_{j=1}^N\ln(1+t_j)\leq\sum_{j=1}^Nt_j
\qquad\hbox{hence}\qquad\prod_{j \in  \Mb{J}}(1+t_j)\leq
\exp\Big(\sum_{j\in\J}t_j\Big)
\]
Now $\sum\limits_{j\in\J}t_j=\sum\limits_{j\in\J}(2|s_j|-1)=\big\|
E([-\hlf,\hlf]^c)\cdot(2|S|-\un)\big\|_1
\leq
b\left\|P_JTP_J\right\|_1$
for a constant $b>0$ by Lemma~\ref{RomeWork}(i).
 Combining these claims prove the lemma
\end{beweis}
To conclude the proof of the theorem, we need
to estimate $\left\|P_JTP_J\right\|_1\,.$
Recall from Theorem \ref{ThetaProp} that
\begin{eqnarray}
\big(P_JTP_Jf\big)(x)&:=&2i\,\chi\s J.(x)\int dy\,f(y)\,(x-y)\,\int_0^\infty dp\,
\ln( 1-e^{-p})\,\cos p(x-y)\,\chi\s J.(y) \nonumber \\[1mm]
&=&
i\,\chi\s J.(x)\int dy\,f(y)\,(x-y)\,\int_{-\infty}^\infty dp\,
\ln( 1-e^{-|p|})\,e^{ip(x-y)}\,\chi\s J.(y) \nonumber \\[1mm]
\label{PTPcomm}
&=&{\rm const.}\Big(\chi\s J.(X)\,\Big[X,\,\ln(1-e^{-|P|})\Big]\,
\chi\s J.(X)\,f\Big)(x)
\end{eqnarray}
where $X$ is the multiplication operator $(Xf)(x)=xf(x),$ and $P$ is as
usual $i{d\over dx}\,,$ and the constant incorporates the $2\pi$
factors from the Fourier transforms. Now $D:=-\ln(1-e^{-|P|})$ is a positive
operator, so write trivially 
$ \chi\s J.(X)\,X\,D\,\chi\s J.(X)=\big(\chi\s J.(X)\,X\,D^{1/2}\big)
\big(D^{1/2}\,\chi\s J.(X)\big)\,,$
then we show that both factors are Hilbert--Schmidt. Now
\begin{eqnarray*}
\big(\chi\s J.(X)\,X\,D^{1/2}f\big)(x)&=&\int dy\,K(x,y)\qquad
\qquad\hbox{where:}\\[1mm]
K(x,y)&=&\chi\s J.(x)\,x\,\int dp\,\big[-\ln( 1-e^{-|p|})\big]^{1/2}\,
e^{ip(x-y)} \\[1mm]
\hbox{so}\qquad
\big\|\chi\s J.(X)\,X\,D^{1/2}\big\|_2^2&=&
\int dx\,dy\; \big|K(x,y)\big|^2   %\\[1mm]
=\int_Jdx\, x^2\,\int dp\,\big[-\ln( 1-e^{-|p|})\big] \\[1mm]
&\leq& {\rm const.}|J|^3\,,
\end{eqnarray*}
using the integrability of $\ln( 1-e^{-|p|})\,.$ Likewise, we get
\[
\big\|(D^{1/2}\,\chi\s J.(X)\big\|_2^2=
\int_Jdx\,\int dp\,\big[-\ln( 1-e^{-|p|})\big]\leq {\rm const.}|J|\,.
\]
Thus $\chi\s J.(X)\,X\,D\,\chi\s J.(X)$ is trace class, and as
$P_JTP_J$ is, so is ${\chi\s J.(X)\,D\,X\,\chi\s J.(X)}\,.$
For their trace norms we find
\[
\big\|\chi\s J.(X)\,X\,D\,\chi\s J.(X)\big\|_1
\leq\big\|\chi\s J.(X)\,X\,D^{1/2}\big\|_2\cdot
\big\|(D^{1/2}\,\chi\s J.(X)\big\|_2
\leq{\rm const.}|J|^2
\]
and likewise $\big\|\chi\s J.(X)\,D\,X\,
\chi\s J.(X)\big\|_1\leq{\rm const.}|J|^2\,.$
Thus by (\ref{PTPcomm}) we get that
\[
\big\|P_JTP_J\big\|_1\leq
{\rm const.}\big\|\chi\s J.(X)\,X\,D\,\chi\s J.(X)\big\|_1
+{\rm const.}\big\|\chi\s J.(X)\,D\,X\,
\chi\s J.(X)\big\|_1\leq K|J|^2
\]
for a constant $K$ (independent of $J$).
Now from Lemma~\ref{eT1norm} we get the claim of
the theorem, i.e. that
\[
\big\|\varphi\restriction\al A._0(J)\big\|
\leq\exp\big(K\,|J|^2\big)\,.
\]

\subsection*{Proof of Theorem~\ref{CoBd}}

% Let $a_i\in\al D._c$ where $1=0,\ldots,\,n\,.$ Thus there is some 
Fix a compact interval $J=[-k,\,k]$ and let $a_i\in\al A._0(J)$ for
all $i\,,$ then as $\alpha_t(a_i)\in\al A._0(J+t)\,$ we have
$\alpha\s t_0.(a_0)\cdots\alpha\s t_n.(a_n)\in \al A._0([-M,M])$
where $M:=k+\sup\limits_i|t_i|\,.$ Thus by Theorem~\ref{LocBd}
we get
\begin{eqnarray*}
&&\!\!\!\!\!\!\!\!
\Big|\varphi\big(\alpha\s t_0.(a_0)\cdots\alpha\s t_n.(a_n)\big)\Big|
\leq e^{4KM^2}\,\|a_0\|\cdots\|a_n\|\;. \\[1mm]
\hbox{Now}\qquad
&&M^2= k^2+\sup_i t_i^2+2k\,\sup_i|t_i|
\leq k^2+\sum_it_i^2+2k\,\sup_i(1+t_i^2) \\[1mm]
&&\leq k^2+\sum_it_i^2+2k\Big(1+\sum_it_i^2)\Big)
=k^2+2k+(1+2k)\,\sum_it_i^2 \\[1mm]
\hbox{hence:}\qquad
&&\!\!\!\!\!\!\!\!
\Big|\varphi\big(\alpha\s t_0.(a_0)\cdots\alpha\s t_n.(a_n)\big)\Big|
\leq A\,\exp\Big(B\,\sum_it_i^2\Big)\,\|a_0\|\cdots\|a_n\|\;.
\end{eqnarray*}
for suitable constants $A$ and $B$ depending only on $k$
(but not on $n$).
Now let $t_0= 0,\ab\;t_1=s_1,\ab\;t_2=s_1+s_2,\ldots,\,t_n=
s_1+\cdots+s_n\,,$ and define for all $s_i\in\R$:
\[
F\s a_o,\ldots a_n.(s_1,\ldots,\,s_n):=
\exp\Big(-B\hbox{$\sum\limits_{k=1}^n$}(s_1+\cdots+s_k)^2\Big)\cdot
\varphi\Big(a\s 0.\,\alpha\s s_1.(a_1)\cdots\alpha\s s_1+\cdots+s_n.(a_n)\Big)\,.
\]
Then we have $\big|F\s a_o,\ldots a_n.(s_1,\ldots,\,s_n)\big|
\leq A\,\|a_0\|\cdots\|a_n\|\;,$ and by the KMS--property of $\varphi,$
the function $F\s a_o,\ldots a_n.$ can be analytically continued in each variable
$s_j$ into the strip $S_j:=\{z_j\in\C^n\,\big|\,{\rm Im}\,z_j\in[0,1]\},$ 
keeping the other variables real.
This produces functions $F^{(j)}\s a_o,\ldots a_n.$ analytic in the flat tubes
$T^j:=\R^{n-1}\times S_j\,,$ and hence by using the 
Flat Tube Theorem~\ref{FTT} inductively, we obtain an analytic continuation
of $F\s a_o,\ldots a_n.$ into the tube $\al T._n:=\R^n+i\Sigma_n$ where
$\Sigma_n:=
\set{\bf s}\in\R^n,0\leq s_i\;\forall\,i,\;\;s_1+\cdots+s_n\leq1.\,,$
coinciding with all $F^{(j)}\s a_o,\ldots a_n.$ on $T^j\,.$
We want to obtain a bound for this analytic function $F\,.$ 
We start by finding bounds for the $F^{(j)}\,.$ 
Let $G(s_1,\ldots,\,s_n):={\varphi\Big(a\s 0.\,\alpha\s s_1.(a_1)
\cdots\alpha\s s_1+\cdots+s_n.(a_n)\Big)}$ which has analytic extensions to each
$T^j\,,$ and by the definition of KMS--functionals we know that
$\big|G(s_1,\ldots,\,s_j+ir_j,\ldots,\,s_n)\big|\leq C(1+|s_j|)^N$
where $C$ and $N$ are independent of $s_j$ and $r_j\in[0,1]\,.$ Now
\begin{eqnarray}
&&\!\!\!\!\!\!
F\s a_o,\ldots a_n.({s_1,\ldots,\,s_j+ir_j,\ldots,\,s_n})= \nonumber\\[1mm]
\label{Festimate}
&&G({ s_1,\ldots,\,s_j+ir_j,\ldots,\,s_n})\,\exp\big[Br_j^2(n+1-j)-
B\hbox{$\sum\limits_{k=1}^n$}(s_1+\cdots+s_k)^2
+i\theta\big]
\end{eqnarray}
where $\theta$ is real. Thus from the exponential damping factor in 
$s_j$ we conclude that $F^{(j)}\s a_o,\ldots a_n.$ is bounded.
By the maximum modulus principle (applied after first mapping $S_j$ to a unit disk
by the Schwartz mapping principle), the bound of $|F^{(j)}\s a_o,\ldots a_n.|$
is attained on the boundary of $S_j$ (this also follows from the 
Phragmen Lindel\"of theorem, cf. p138 in~\cite{Conw}). So on the
real part of the boundary 
of $S_j$ we have already from above that
$
\big|F^{(j)}\s a_o,\ldots a_n.
(s_1,\ldots,\,s_n)\big|
\leq A\,\|a_0\|\cdots\|a_n\|
$
and by the KMS--condition and translation invariance of $\varphi$
we have on the other part
\begin{eqnarray*}
&&\big|G(s_1,\ldots,\,s_j+i,\ldots,\,s_n)\big|=
\Big|\varphi\Big(\alpha\s s_1+\cdots s_j.(a_j)\cdots \alpha\s s_1+\cdots+ s_n.(a_n)
a\s 0.\,\alpha\s s_1.(a_1)\cdots\alpha\s s_1+\cdots+s_{j-1}.
(a_{j-1})\Big)\Big| \\[1mm]
&&\qquad\qquad\qquad
\leq A\,\exp\Big(B\hbox{$\sum\limits_{k=1}^n$}(s_1+\cdots+s_k)^2\Big)\cdot
\|a_0\|\cdots\|a_n\|\qquad\hbox{hence by (\ref{Festimate}):} \\[1mm]
&&\big|F^{(j)}\s a_o,\ldots a_n.(s_1,\ldots,s_j+i,\ldots,\,s_n)\big|
\leq A\,e^{Bn}\,\|a_0\|\cdots\|a_n\|\qquad\hbox{and thus
as $e^{Bn}>1\,,$} \\[1mm]
&&\big|F^{(j)}\s a_o,\ldots a_n.(s_1,\ldots,z_j,\ldots,\,s_n)\big|\leq
A\,e^{Bn}\,\|a_0\|\cdots\|a_n\|=:C\qquad\hbox{for all $z_j\in S_j\,.$}
\end{eqnarray*}
Now define $H_\alpha(z_1,\ldots,\,z_n):=[F\s a_o,\ldots a_n.(z_1,\ldots,\,z_n)
-e^{i\alpha}C]^{-1}$ where $\alpha\in[0,2\pi]\,$ for $(z_1,\ldots,\,z_n)
\in \al T._n\,.$
Then by the estimates above for $|F^{(j)}\s a_o,\ldots a_n.|\,,$
each map $z_j\to H_\alpha(s_1,\ldots,z_j,\ldots,\,s_n)$ is analytic 
on the strip $S_j\,,$ and thus by the 
Flat Tube Theorem~\ref{FTT}, $H_\alpha$ has a unique extension 
as an analytic function to $\al T._n\,,$ and hence cannot have any singularities 
in  $\al T._n\,,$ i.e. $F\s a_o,\ldots a_n.(z_1,\ldots,\,z_n)
\not=e^{i\alpha}C$ for all $\alpha\,.$
By continuity of $F\,,$ the image set $F\s a_o,\ldots a_n.(\al T._n)$
must be connected. By assumption, this set has some points inside the
circle $|z|=C\,,$ hence the entire image set is inside the circle
$|z|=C\,,$ i.e. 
\[
\big|F\s a_o,\ldots a_n.(z_1,\ldots,\,z_n)\big|\leq
A\,e^{Bn}\,\|a_0\|\cdots\|a_n\|
\qquad\forall\;(z_1,\ldots,\,z_n)\in\al T._n\quad
\hbox{and}\quad
a_i\in\al A._0(J)\,.
\]
Consider now the Chern character formula~(\ref{ChCa}):
\begin{eqnarray*}
\tau_n(a_0,\ldots,a_n)&:=& i^{\epsilon_n}\int_{\sigma_n}
\varphi\Big(a_0\,\alpha_{is_1}\big(\delta\gamma(a_1)\big)\,
\alpha_{is_2}\big(\delta(a_2)\big)\,\alpha_{is_3}\big(\delta\gamma(a_3)\big)
\cdots   \\[1mm]
& &\qquad\qquad\cdots\alpha_{is_n}\big(\delta\gamma^n(a_n)\big)\Big)\,
ds_1\cdots ds_n\,,\qquad a_i\in\al D._c\,, \\[1mm]
&=& i^{\epsilon_n}
\int_{\Sigma_n}\varphi\big(b_0\,\alpha_{ir_1}(b_1)\cdots\alpha_{ir_1+\cdots+ ir_n}(b_n)\big)
\,dr_1\cdots dr_n
\end{eqnarray*}
where we made
a change of variables $s_1=r_1,\;s_2=r_1+r_2,\ldots,s_n=r_1+\cdots+r_n$
and substitutions $a_0=b_0,$ $b_1=\delta\gamma(a_1),\ldots, b_n=\delta\gamma^n(a_n)$
as in Section~\ref{JLOsection}, making use of the Flat Tube Theorem.
(Note that $b_i\in\al A._0(J)\;\forall\,i$ for some $J\,.)$
In fact, from the uniqueness part of the extensions to $\al T._n$
we have that on  $\al T._n$
\[
\varphi\big(b_0\,\alpha_{z_1}(b_1)\cdots\alpha_{z_1+\cdots+ z_n}(b_n)\big) 
=\exp\big[B\sum_{k=1}^n(z_1+\cdots+ z_n)^2\big]\cdot
F\s b_o,\ldots b_n.(z_1,\ldots,\,z_n)
\]
and so for $(z_1,\ldots,\,z_n)=i(r_1,\ldots,\,r_n)\in i\Sigma_n$ we have
\begin{eqnarray*}
\Big|\varphi\big(b_0\,\alpha_{ir_1}(b_1)\cdots\alpha_{ir_1+\cdots+ ir_n}(b_n)\big)
\Big| &\leq& \exp\Big[-B\sum_{k=1}^n(r_1+\cdots+ r_n)^2+Bn\Big]
\,A\,\|b_0\|\cdots\|b_n\|  \\[1mm]
\hbox{hence:}\qquad
\big|\tau_n(a_0,\ldots,a_n)\big| &\leq&
{A\over n!}\,e^{Bn}\,\|b_0\|\cdots\|b_n\|\leq{A\over n!}\,e^{Bn}\,
\|a_0\|_*\cdots\|a_n\|_* 
\end{eqnarray*}
where we used first, that the volume of $|\Sigma_n|=1/n!\,,$ and second, that
$\|b_j\|\leq\|a_j\|_*$ because $b_j=\delta\gamma(a_j)=-\gamma\delta(a_j)$
for $j>0\,.$ 
Thus $\|\tau_n\|_*\leq A\,e^{Bn}\big/n!$ and hence it is clear that
$\lim\limits_{n\to\infty}n^{1/2}\|\tau\|_*^{1/n}\leq
e^B\,\lim\limits_{n\to\infty}n^{1/2}(A/n!)^{1/n}=0$
by Stirling's formula, which concludes the proof.

\subsection*{Proof of Theorem~\ref{TCycCoc}}

By Theorem~\ref{CoBd} we already have the entireness condition for 
$\widetilde{\tau}$ so it is only necessary to prove the cocycle
condition for $a_i\in\al D._c\,:$
\[
(b{\tau}_{n-1})(a_0,\ldots,a_n)=(B{\tau}_{n+1})(a_0,\ldots,a_n)
\,,\qquad n=1,3,5,,\ldots
\]
with $b$ and $B$ given by Equations (\ref{Defb}) and (\ref{DefB}).
We will roughly follow the technique used in \cite{JLOK}, but due to the different
analytic properties of our model, we will need to go explicitly through the steps. 
In order to manipulate the expressions involved with Equation~(\ref{CocEq}),
we need the results in the following Lemma.
\begin{lem}
\label{CocLem}
Let $b_i\in\al A._0\,,$ then:
\begin{itemize}
\item[(i)]
$\varphi\Big(b_0\,\alpha\s{is_1}.(b_1)\cdots\alpha\s{is_n}.(b_n)\Big)
=\varphi\Big(\gamma(b_n)\,\alpha\s{i(1-s_n)}.(b_0)\,\alpha\s{i(1-s_n+s_1)}.(b_1)
\cdots\alpha\s{i(1-s_n+s_{n-1})}.(b_{n-1})\Big)$
for all $(s_1,\ldots,\,s_n)\in\sigma_n\;.$
\item[(ii)]
$\displaystyle{\int_{\sigma_n}\varphi\Big(b_0\,\alpha\s{is_1}.(b_1)
\cdots\alpha\s{is_n}.(b_n)\Big)\,ds_1\cdots ds_n=
\int_{\sigma_n}\varphi\Big(\gamma(b_n)\,\alpha\s{is_1}.(b_0)
\cdots\alpha\s{is_n}.(b_{n-1})\Big)}\,ds_1\cdots ds_n$
\item[(iii)]
The functions $(t_1,\ldots,\,t_n)\to
\varphi\Big(b_0\,\alpha\s{t_1}.(b_1)\cdots\delta\big(\alpha\s t_k.(b_k)\big)
\cdots\alpha\s{t_n}.(b_n)\Big)$ and\chop
$(t_1,\ldots,\,t_n)\to
\varphi\Big(b_0\,\alpha\s{t_1}.(b_1)\cdots\gamma\big(\alpha\s t_k.(b_k)\big)
\cdots\alpha\s{s_n}.(b_n)\Big)$
both have analytic continuations to $\R^n+i\sigma_n\,,$ and for these we have\chop
$\varphi\Big(b_0\,\alpha\s{is_1}.(b_1)\cdots\delta\big(\alpha\s is_k.(b_k)\big)
\cdots\alpha\s{is_n}.(b_n)\Big)
=\varphi\Big(b_0\,\alpha\s{is_1}.(b_1)\cdots\alpha\s is_k.\big(\delta(b_k)\big)
\cdots\alpha\s{is_n}.(b_n)\Big)$ and \chop
$\varphi\Big(b_0\,\alpha\s{is_1}.(b_1)\cdots\gamma\big(\alpha\s is_k.(b_k)\big)
\cdots\alpha\s{is_n}.(b_n)\Big)
=\varphi\Big(b_0\,\alpha\s{is_1}.(b_1)\cdots\alpha\s is_k.\big(\gamma(b_k)\big)
\cdots\alpha\s{is_n}.(b_n)\Big)\,.$
\item[(iv)] For $j=2,\ldots,\,n$ we have:
\begin{eqnarray}
&&\!\!\!\!
\int_{\sigma_{n+1}}{\partial\over\partial s_j}
\varphi\Big(b_0\,\alpha\s{is_1}.(b_1)
\cdots\alpha\s{is_{n+1}}.(b_{n+1})\Big)\,ds_1\cdots ds_{n+1} 
\nonumber\\[1mm]
&&=\int_{\sigma_n}\Big[\varphi\Big(b_0\,\alpha\s{is_1}.(b_1)
\cdots\alpha\s is_j.(b_jb_{j+1})\cdots\alpha\s{is_n}.(b_{n+1})\Big)
\nonumber\\[1mm]
\label{ByParts1}
&&\qquad\qquad-
\varphi\Big(b_0\,\alpha\s{is_1}.(b_1)
\cdots\alpha\s is_{j-1}.(b_{j-1}b_j)\cdots\alpha\s{is_n}.(b_{n+1})\Big)\Big]
\,ds_1\cdots ds_n \\[1mm]
&&\!\!\!\!
\int_{\sigma_{n+1}}{\partial\over\partial s_1}
\varphi\Big(b_0\,\alpha\s{is_1}.(b_1)
\cdots\alpha\s{is_{n+1}}.(b_{n+1})\Big)\,ds_1\cdots ds_{n+1} 
\nonumber\\[1mm]
 \label{ByParts2}
&&=\int_{\sigma_n}\Big[
% \nonumber\\[1mm]
% \label{ByParts2}
% &&\qquad\qquad+
\varphi\Big(b_0\,\alpha\s{is_1}.(b_1b_2)
\,\alpha\s is_{2}.(b_3)\cdots\alpha\s{is_n}.(b_{n+1})\Big)
-\varphi\Big(b_0\,b_1\,\alpha\s{is_1}.(b_2)
\cdots\alpha\s{is_n}.(b_{n+1})\Big)\Big]
\,ds_1\cdots ds_n\qquad\qquad{} \\[1mm]
&&\!\!\!\!
\int_{\sigma_{n+1}}{\partial\over\partial s_{n+1}}
\varphi\Big(b_0\,\alpha\s{is_1}.(b_1)
\cdots\alpha\s{is_{n+1}}.(b_{n+1})\Big)\,ds_1\cdots ds_{n+1} 
\nonumber\\[1mm]
 \label{ByParts3}
&&=\int_{\sigma_n}\Big[\varphi\Big(\gamma(b_{n+1})\,b_0\,\alpha\s{is_1}.(b_1)
\cdots\alpha\s{is_n}.(b_{n})\Big)
% \nonumber\\[1mm]
% \label{ByParts3}
% &&\qquad\qquad-
-\varphi\Big(b_0\,\alpha\s{is_1}.(b_1)
\cdots\alpha\s{is_n}.(b_nb_{n+1})\Big)\Big]
\,ds_1\cdots ds_n 
\end{eqnarray}
\end{itemize}
\end{lem}
\begin{beweis}
(i) Recall that the left hand side is defined by the analytic extension of the function
${F\s t_1\cdots,t_n.(b_0,\cdots,\,b_n)}:= {\varphi\big(b_0\,\alpha\s t_1.(b_1)\cdots
\alpha\s t_n.(b_n)\big)}$ to the tube $\R^n+i\sigma_n$ by the KMS--condition and
Flat Tube theorem, so
$\varphi\Big(b_0\,\alpha\s{is_1}.(b_1)\cdots\alpha\s{is_n}.(b_n)\Big)
:={F\s is_1\cdots,is_n.(b_0,\cdots,\,b_n)}\,.$
By the invariance ${\varphi\circ\alpha_t}=\varphi$ we have 
\begin{eqnarray*}
F\s t_1\cdots,t_n.(b_0,\cdots,\,b_n)&=&\varphi\big(\alpha_t(b_0)\,\alpha\s t+t_1.(b_1)
\cdots\alpha\s t+t_n.(b_n)\big)\\[1mm]
&=&F\s t,\,t+t_1\cdots,\,t+t_n.(\un,\,b_0,\cdots,\,b_n)=
F\s t+t_1\cdots,\,t+t_n.(\alpha_t(b_0),\,b_1,\cdots,\,b_n)\,.
\end{eqnarray*}
The latter function has an analytic continuation in the variables 
${(t+t_1,\ldots,\,t+t_n)}$ to $\R^n+i\sigma_n$ and from the former function
it also has an analytic extension in $t$ to the strip $\R+i[0,1]\,.$
Thus by the flat tube theorem we get a unique analytic extension to 
all of $\R^{n+1}+i\sigma\s n+1.\,.$ Put $t_j=is_j$ where ${\bf s}\in\sigma_n$
and $t=i(1-s_n)\,,$ then
\[
F\s is_1\cdots,is_n.(b_0,\cdots,\,b_n)
=F\s i(1-s_n),\,i(1-s_n+s_1),\cdots,\,i(1-s_n+s_{n-1}),\,i.
(\un,\,b_0,\cdots,\,b_n)\;
\]
which is justified because we have that the variables 
\[
{\bf r}=(r_1,\cdots,r_n):=
(1-s_n,\,1-s_n+s_1,\ldots,\,1-s_n+s_{n-1})\in\sigma_n\,.
\]
Now the function ${F\s ir_1,\cdots,\,ir_n,\,i.
(\un,\,b_0,\cdots,\,b_n)}$ is obtained from the analytic extension of 
${F\s t_1,\cdots,\,t_n,\,v.
(\un,\,b_0,\cdots,\,b_n)}\,,$ $t_i,\;v\in\R$ to $\R^{n+1}+i\sigma\s n+1.\,.$
By the KMS--condition:
\begin{eqnarray*}
&&\!\!\!\!\!
F\s t_1,\cdots,\,t_n,\,i.(\un,\,b_0,\cdots,\,b_n) =
\varphi\Big(b_n\,\gamma\big(\alpha\s{t_1}.(b_0)\cdots\alpha\s{t_n}.(b_{n-1})\big)\Big)
\\[1mm]
&& = \varphi\Big(\gamma(b_n)\,\alpha\s{t_1}.(b_0)\cdots\alpha\s{t_n}.(b_{n-1})\Big)
=F\s t_1,\cdots,\,t_n.(\gamma(b_n),\,b_0,\cdots,\,b_{n-1})
\end{eqnarray*}
Thus by uniqueness of the analytic continuations we have
\[
F\s is_1\cdots,is_n.(b_0,\cdots,\,b_n)=F\s ir_1,\cdots,\,ir_n,\,i.
(\un,\,b_0,\cdots,\,b_n)=F\s ir_1,\cdots,\,ir_n.(\gamma(b_n),\,b_0,\cdots,\,b_{n-1})
\]
which is the statement (i) of the lemma.\chop
(ii) By part (i) we have:
\begin{eqnarray*}
&&\!\!\!\!\!
\int_{\sigma_n}\varphi\Big(b_0\,\alpha\s{is_1}.(b_1)\cdots\alpha\s{is_n}.(b_n)\Big)
\,ds_1\cdots ds_n \\[1mm]
&&=\int_{\sigma_n}\varphi\Big(\gamma(b_n)\,\alpha\s{i(1-s_n)}.(b_0)\,
\alpha\s{i(1-s_n+s_1)}.(b_1)
\cdots\alpha\s{i(1-s_n+s_{n-1})}.(b_{n-1})\Big) \,ds_1\cdots ds_n \\[1mm]
&&=\int_{\sigma_n}\varphi\Big(\gamma(b_n)\,\alpha\s{ir_1}.(b_0)\,
\cdots\alpha\s{ir_n}).(b_{n-1})\Big)\,dr_1\cdots dr_n 
\end{eqnarray*}
making use of the change of variables ${\bf s}\to{\bf r}$ above (with 
Jacobian $=1),$ and the fact that
${\bf r}\in\sigma_n$ iff ${\bf s}\in\sigma_n\,.$\chop
(iii)
Since
$\varphi\Big(b_0\,\alpha\s{t_1}.(b_1)\cdots\delta\big(\alpha\s t_k.(b_k)\big)
\cdots\alpha\s{t_n}.(b_n)\Big)=\varphi\Big(b_0\,\alpha\s{t_1}.(b_1)\cdots
 \alpha\s t_k.\big(\delta(b_k)\big)
\cdots\alpha\s{t_n}.(b_n)\Big)$ and the latter obviously has an analytic
extension to $\R^n+\sigma_n$ the claim follows. Likewise for the other one.
\chop
(iv) For $2\leq j\leq n$ we have
\begin{eqnarray*}
&&\!\!\!\!\!
\int_{s_{j-1}}^{s_{j+1}}{\partial\over\partial s_j}
\varphi\Big(b_0\,\alpha\s{is_1}.(b_1)\cdots\alpha\s{is_{n+1}}.(b_{n+1})\Big)
\,ds_j \\[1mm]
&&=\varphi\Big(b_0\,\alpha\s{is_1}.(b_1)\cdots\alpha\s{is_{j-1}}.(b_{j-1})
\,\alpha\s{is_{j+1}}.(b_{j})\,\alpha\s{is_{j+1}}.(b_{j+1})
\cdots\alpha\s{is_{n+1}}.(b_{n+1})\Big) \\[1mm]
&&\quad-\varphi\Big(b_0\,\alpha\s{is_1}.(b_1)\cdots\alpha\s{is_{j-1}}.(b_{j-1})
\,\alpha\s{is_{j-1}}.(b_{j})\,\alpha\s{is_{j+1}}.(b_{j+1})
\cdots\alpha\s{is_{n+1}}.(b_{n+1})\Big) \\[1mm]
&&=\varphi\Big(b_0\,\alpha\s{is_1}.(b_1)\cdots\alpha\s{is_{j-1}}.(b_{j-1})
\,\alpha\s{is_{j+1}}.(b_{j}b_{j+1})
\cdots\alpha\s{is_{n+1}}.(b_{n+1})\Big) \\[1mm]
&&\quad-\varphi\Big(b_0\,\alpha\s{is_1}.(b_1)\cdots\alpha\s{is_{j-1}}.(b_{j-1}
b_{j})\,\alpha\s{is_{j+1}}.(b_{j+1})
\cdots\alpha\s{is_{n+1}}.(b_{n+1})\Big)
\end{eqnarray*}
from which equation~(\ref{ByParts1}) follows by a change of label of the 
integration variables.
For equation~(\ref{ByParts2}) we substitute $j=1,\;s\s j-1.=0$ into the last equation.
For equation~(\ref{ByParts3}) we substitute $j=n+1,\;s\s j+1.=1$ into the last equation
to get
\begin{eqnarray*}
&&\!\!\!\!\!
\int_{\sigma_{n+1}}{\partial\over\partial s_{n+1}}
\varphi\Big(b_0\,\alpha\s{is_1}.(b_1)
\cdots\alpha\s{is_{n+1}}.(b_{n+1})\Big)\,ds_1\cdots ds_{n+1} \\[1mm]
&&=\int_{\sigma_{n}}\Big[\varphi\Big(b_0\,\alpha\s{is_1}.(b_1)\cdots
\alpha\s{is_{n}}.(b_{n})\,\alpha\s{i}.(b_{n+1})\Big)
-\varphi\Big(b_0\,\alpha\s{is_1}.(b_1)\cdots
\alpha\s{is_{n}}.(b_{n}b_{n+1})\Big)\Big]\,ds_1\cdots ds_{n}  \\[1mm]
&&=\int_{\sigma_n}\Big[\varphi\Big(\gamma(b_{n+1})\,b_0\,\alpha\s{is_1}.(b_1)
\cdots\alpha\s{is_n}.(b_{n})\Big)
-\varphi\Big(b_0\,\alpha\s{is_1}.(b_1)
\cdots\alpha\s{is_n}.(b_nb_{n+1})\Big)\Big]
\,ds_1\cdots ds_n 
\end{eqnarray*}
making use of part (i) for the KMS--condition.
\end{beweis}
Let us begin with the right hand side of our desired Equation~(\ref{CocEq}).
From the definition~(\ref{DefB}) we have for $a_i\in\al D._c\,$ via $\delta(\un)=0$
that:
\begin{eqnarray*}
&&\!\!\!\!\!
\big(B{\tau}_{n+1}\big)(a_0,\ldots,\,a_n)
=i^{\epsilon_{n+1}}\int_{\sigma_{n+1}}ds_1\cdots ds_{n+1}\,\Big[
\varphi\Big(\alpha\s is_1.(\delta\gamma a_0)\cdots\alpha\s is_{n+1}.
(\delta\gamma^{n+1} a_n)\Big) \\[1mm]
&&\quad +\sum_{j=1}^n(-1)^{nj}\,\varphi\Big(\alpha\s is_1.(\delta
\gamma^2 a_{n+1-j})\cdots
\alpha\s is_j.(\delta\gamma^{j+1} a_n)\,\alpha\s is_{j+1}.(\delta\gamma^{j+1} a_0)
\cdots\alpha\s is_{n+1}.(\delta\gamma^{n+1} a_{n-j})\Big)\Big]\,.
\end{eqnarray*}
We can now use Lemma~\ref{CocLem}(ii) in all the
terms on the right hand side to bring the factor with $a_0$ to the front:
\begin{eqnarray*}
&&\!\!\!\!\!
\big(B{\tau}_{n+1}\big)(a_0,\ldots,\,a_n)
=i^{\epsilon_{n+1}}\int_{\sigma_{n+1}}ds_1\cdots ds_{n+1}\,\Big[
\varphi\Big(\delta\gamma( a_0)\,\alpha\s is_1.(\delta\gamma^2 a_1)
\cdots \\[1mm]
&&\quad\cdots\alpha\s is_{n}.
(\delta\gamma^{n+1} a_n)\,\alpha\s is_{n+1}.(\un)\Big) 
 +\sum_{j=1}^n(-1)^{nj}\,\varphi\Big(\delta\gamma^{j+1} (a_0)\,
 \alpha\s is_1.(\delta\gamma^{j+2} a_{1})\cdots \\[1mm]
&&\qquad\cdots
\alpha\s is_{n-j}.(\delta\gamma^{n+1} a_{n-j})\,\alpha\s is_{n-j+1}.(\un)\,
\alpha\s is_{n-j+2}.(\gamma\delta\gamma^2 a_{n+1-j})
\cdots\alpha\s is_{n+1}.(\gamma\delta\gamma^{j+1} a_{n})\Big)\Big] \\[1mm]
&&\hbox{now substitute $\varphi\to\varphi\circ\gamma^{j+1}$ in the last term, and use
$\gamma\circ\delta=-\delta\circ\gamma$ and Lemma~\ref{CocLem}(iii):}\\[1mm]
&&=i^{\epsilon_{n+1}}\int_{\sigma_{n+1}}ds_1\cdots ds_{n+1}\,\Big[(-1)^{n+1}
\varphi\Big(\delta( a_0)\,\alpha\s is_1.(\delta\gamma a_1)
\cdots\alpha\s is_{n}.
(\delta\gamma^{n} a_n)\,\alpha\s is_{n+1}.(\un)\Big) \\[1mm]
&&\quad +\sum_{j=1}^n(-1)^{nj}\,\varphi\Big((-1)^{j+1}\delta(a_0)\,(-1)^{j+1}
 \alpha\s is_1.(\delta\gamma a_{1})\cdots \\[1mm]
&&\qquad\cdots(-1)^{j+1}
\alpha\s is_{n-j}.(\delta\gamma^{n-j} a_{n-j})\,\alpha\s is_{n-j+1}.(\un)\,
(-1)^{j}\alpha\s is_{n-j+2}.(\delta\gamma^j a_{n+1-j})
\cdots(-1)^j\alpha\s is_{n+1}.(\delta\gamma a_{n})\Big)\Big] \\[1mm]
&&=i^{\epsilon_{n+1}}(-1)^{n+1}\int_{\sigma_{n+1}}ds_1\cdots ds_{n+1}\,\Big[
\varphi\Big(\delta( a_0)\,\alpha\s is_1.(\delta\gamma a_1)
\cdots\alpha\s is_{n}.
(\delta\gamma^{n} a_n)\,\alpha\s is_{n+1}.(\un)\Big) \\[1mm]
&&\quad +\sum_{j=1}^n\varphi\Big(\delta(a_0)\,
 \alpha\s is_1.(\delta\gamma a_{1})\cdots \\[1mm]
&&\qquad\cdots
\alpha\s is_{n-j}.(\delta\gamma^{n-j} a_{n-j})\,\alpha\s is_{n-j+1}.(\un)\,
\alpha\s is_{n-j+2}.(\delta\gamma^j a_{n+1-j})
\cdots\alpha\s is_{n+1}.(\delta\gamma a_{n})\Big)\Big]\,.
\end{eqnarray*}
Now recall that
 $\wt\tau:=(\tau_0,0,-\tau_2,0,\tau_4,\ldots)\in \al C.(\al D._c),$
and hence we may assume that $n$ is odd in the preceding expression
(if $n$ is even, $B\wt\tau_{n+1}=0\,).$ Thus
\begin{eqnarray*}
&&\!\!\!\!\!
\big(B{\tau}_{n+1}\big)(a_0,\ldots,\,a_n)
=\int_{\sigma_{n+1}}ds_1\cdots ds_{n+1}\,\Big[
\sum_{j=0}^n\varphi\Big(\delta(a_0)\,
 \alpha\s is_1.(\delta\gamma a_{1})\cdots \\[1mm]
&&\qquad\cdots
\alpha\s is_{n-j}.(\delta\gamma^{n-j} a_{n-j})\,\alpha\s is_{n-j+1}.(\un)\,
\alpha\s is_{n-j+2}.(\delta\gamma^{n+1-j} a_{n+1-j})
\cdots\alpha\s is_{n+1}.(\delta\gamma^n a_{n})\Big)\Big]\,.
\end{eqnarray*}
Since $\alpha\s is_{n-j+1}.(\un)=\un\,,$ we can do the integrals w.r.t.
$s_{n-j+1},$ and so using $0\leq s_1\leq s_2\leq\cdots\leq s_{n+1}\leq 1$
and a relabelling of variables, we get
\begin{eqnarray}
&&\!\!\!\!\!
\big(B{\tau}_{n+1}\big)(a_0,\ldots,\,a_n)
=\int_{\sigma_{n}}ds_1\cdots ds_{n}\,\Big[
\sum_{j=1}^{n-1}(s\s n-j+1.-s\s n-j.)\,\varphi\Big(\delta(a_0)\,
 \alpha\s is_1.(\delta\gamma a_{1})\cdots \nonumber\\[1mm]
&&\qquad\qquad\qquad\cdots
\alpha\s is_{n-j}.(\delta\gamma^{n-j} a_{n-j})\,
\alpha\s is_{n-j+1}.(\delta\gamma^{n+1-j} a_{n+1-j})
\cdots\alpha\s is_{n}.(\delta\gamma^n a_{n})\Big) \nonumber \\[1mm]
&&\qquad\qquad\qquad\qquad\qquad
+(s_1+1-s_n)\,\varphi\Big(\delta(a_0)\,\alpha\s is_1.(\delta\gamma a_{1})\cdots
\alpha\s is_{n}.(\delta\gamma^n a_{n})\Big)
\Big] \nonumber \\[1mm]
\label{BtFinal}
&&=\int_{\sigma_{n}}ds_1\cdots ds_{n}\,
\varphi\Big(\delta(a_0)\,\alpha\s is_1.(\delta\gamma a_{1})\cdots
\alpha\s is_{n}.(\delta\gamma^n a_{n})\Big)\,.
\end{eqnarray}
Next, we turn our attention to the left hand side of our desired Equation~(\ref{CocEq}).
Observe first that we have 
\begin{eqnarray*}
\tau_n\big(\gamma a_0,\ldots,\,\gamma a_n\big)
&=&(-1)^n\tau_n\big( a_0,\ldots,\, a_n\big)\qquad\quad \hbox{because:}  \\[1mm]
\varphi\left(\gamma(a_0)\,\alpha\s is_1.(\delta\gamma(\gamma a_1))\cdots
\alpha\s is_n.(\delta\gamma^n(\gamma a_1))\right) 
&=& 
(-1)^n\,\varphi\left(a_0\,\alpha\s is_1.(\delta\gamma a_1)\cdots
\alpha\s is_n.(\delta\gamma^na_n)\right)
\end{eqnarray*}
since $\delta\circ\gamma=-\gamma\circ\delta,$ $\varphi\circ\gamma=\varphi$
and by Lemma~\ref{CocLem}(iii). Thus $\wt{\tau}\circ\gamma=\wt{\tau},$
and so we have $\wt{a}=\gamma a$ in definition~(\ref{Defb}). An application of
 definition~(\ref{Defb}) to the left hand side of Equation~(\ref{CocEq}) yields:
\begin{eqnarray}
&&\big(b\,\tau_{n-1}\big)(a_0,\ldots,\,a_n)=
i^{\epsilon_{n-1}}\int_{\sigma_{n-1}}ds_1\cdots ds_{n-1}\Big[
\sum_{j=0}^{n-1}(-1)^j\varphi\Big(a_0\,\alpha\s is_1.(\delta\gamma a_1)
\cdots 
\alpha\s is_j.(\delta\gamma^j(a_j\,a_{j+1}))\cdots \nonumber \\[1mm]
\label{LHSCocEq}
&&\qquad\qquad\cdots\alpha\s is_{n-1}.(\delta\gamma^{n-1}a_n)\Big)
+(-1)^{n}\,\varphi\left((\gamma a_n)\,a_0\,
\alpha\s is_1.(\delta\gamma a_1)\cdots \alpha\s is_{n-1}.(\delta\gamma^{n-1}a_{n-1})
\right)\Big]\,.
\end{eqnarray}
We examine the terms in this sum more closely:
\begin{eqnarray*}
\hbox{$j=0$:} &&
\int_{\sigma_{n-1}}ds_1\cdots ds_{n-1}\,\varphi\Big(a_0\,a_1\,\alpha\s is_1.(\delta\gamma a_2)
\cdots \alpha\s is_{n-1}.(\delta\gamma^{n-1}a_n)\Big)   \\[1mm]
\hbox{$j=1$:} &&
-\int_{\sigma_{n-1}}ds_1\cdots ds_{n-1}\,\varphi\Big(a_0\,\alpha\s is_1.(\delta\gamma
(a_1a_2))\,\alpha\s is_2.(\delta a_3)\cdots \alpha\s is_{n-1}.(\delta
\gamma^{n-1}a_n)\Big)   \\[1mm]
&&=-\int_{\sigma_{n-1}}ds_1\cdots ds_{n-1}\,\varphi\Big(a_0\,\alpha\s is_1.
\big((\delta\gamma(a_1)\,\gamma a_2+a_1\,\delta\gamma a_2\big)\,\alpha\s is_2.
(\delta a_3)\cdots \alpha\s is_{n-1}.(\delta
\gamma^{n-1}a_n)\Big)   \\[1mm]
&&=-\int_{\sigma_{n-1}}ds_1\cdots ds_{n-1}\,\Big[\varphi\Big(a_0\,\alpha\s is_1.
\big((\delta\gamma(a_1)\,\gamma a_2\big)\,\alpha\s is_2.
(\delta a_3)\cdots \alpha\s is_{n-1}.(\delta
\gamma^{n-1}a_n)\Big)   \\[1mm]
&&\qquad\qquad +\; \varphi\Big(a_0\,a_1\,\alpha\s is_1.
(\delta\gamma a_2)\,\alpha\s is_2.
(\delta a_3)\cdots \alpha\s is_{n-1}.(\delta
\gamma^{n-1}a_n)\Big)\Big]   \\[1mm]
&&-\int_{\sigma_{n}}ds_1\cdots ds_{n}\,{\partial\over\partial s_1}
\,\varphi\Big(a_0\,\alpha\s is_1.(a_1)\,\alpha\s is_2.(\delta\gamma a_2)\,
\alpha\s is_3.(\delta a_3)\cdots\alpha\s is_{n}.(\delta\gamma^{n+1}a_n)\Big)
\end{eqnarray*}
where we made use of Equation~(\ref{ByParts2}) in the last step.
Notice that we get a cancellation between the middle term and the 
$j=0$ term in the sum. For $1<j\leq n-1$ we have the terms:
\begin{eqnarray*}
&&(-1)^j\int_{\sigma_{n-1}}ds_1\cdots ds_{n-1}\,\varphi\Big(a_0\,\alpha\s is_1.
(\delta\gamma a_1)\cdots\alpha\s is_j.\big(\delta\gamma^j(a_ja_{j+1})\big)\cdots
\alpha\s is_{n-1}.(\delta\gamma^{n-1}a_n)\Big) \\[1mm]
&&\!\!\!\!\!\!\!\!
=(-1)^j\int_{\sigma_{n-1}}ds_1\cdots ds_{n-1}\,\varphi\Big(a_0\,\alpha\s is_1.
(\delta\gamma a_1)\cdots\alpha\s is_j.\big((\delta\gamma^ja_j)\gamma^ja_{j+1}
+\gamma^{j+1}(a_j)\delta\gamma^ja_{j+1}\big)\cdots
\alpha\s is_{n-1}.(\delta\gamma^{n-1}a_n)\Big) \\[1mm]
&&\quad=(-1)^j\int_{\sigma_{n-1}}ds_1\cdots ds_{n-1}\,\Big[
\varphi\Big(a_0\,\alpha\s is_1.
(\delta\gamma a_1)\cdots\alpha\s is_j.\big((\delta\gamma^ja_j)\gamma^ja_{j+1}\big)
\cdots\alpha\s is_{n-1}.(\delta\gamma^{n-1}a_n)\Big) \\[1mm]
&&\qquad\qquad +\; 
\varphi\Big(a_0\,\alpha\s is_1.(\delta\gamma a_1)\cdots\alpha\s is_{j-1}.
\big((\delta\gamma^{j-1}a_{j-1})\gamma^{j+1}a_j\big)
\cdots\alpha\s is_{n-1}.(\delta\gamma^{n-1}a_n)\Big)\Big] \\[1mm]
&&+\;(-1)^j\int_{\sigma_{n}}ds_1\cdots ds_{n}\,{\partial\over\partial s_j}
\,\varphi\Big(a_0\,\alpha\s is_1.(\delta\gamma a_1)\cdots
\alpha\s is_j.(\delta\gamma^{j+1}a_j)
\cdots\alpha\s is_{n}.(\delta\gamma^{n-1}a_n)\Big)
\end{eqnarray*}
where we made use of Equation~(\ref{ByParts1}). Thus we get for 
Equation~(\ref{LHSCocEq}), taking into account cancellations between subsequent terms 
in the sum, that
\begin{eqnarray*}
&&\big(b\,\tau_{n-1}\big)(a_0,\ldots,\,a_n)=i^{\epsilon_{n-1}}
(-1)^{n-1}\int_{\sigma_{n-1}}ds_1\cdots ds_{n-1}\,
\varphi\Big(a_0\,\alpha\s is_1.(\delta\gamma a_1)\cdots
\alpha\s is_{n-1}.\big((\delta\gamma^{n-1}a_{n-1})\,\gamma^{n-1}a_n\big)\Big)
 \\[1mm]
&&\quad +\;i^{\epsilon_{n-1}}
\sum_{j=1}^{n-1}(-1)^j\int_{\sigma_{n}}ds_1\cdots ds_{n}\,{\partial\over\partial s_j}
\,\varphi\Big(a_0\,\alpha\s is_1.(\delta\gamma a_1)\cdots
\alpha\s is_j.(\delta\gamma^{j+1}a_j)
\cdots\alpha\s is_{n}.(\delta\gamma^{n-1}a_n)\Big) \\[1mm]
&&\qquad +\;i^{\epsilon_{n-1}}
(-1)^n\int_{\sigma_{n-1}}ds_1\cdots ds_{n-1}\,
\varphi\left((\gamma a_n)\,a_0\,
\alpha\s is_1.(\delta\gamma a_1)\cdots \alpha\s is_{n-1}.(\delta\gamma^{n-1}a_{n-1})
\right)  \\[1mm]
&&=\sum_{j=1}^n(-1)^j\int_{\sigma_{n}}ds_1\cdots ds_{n}\,{\partial\over\partial s_j}
\,\varphi\Big(a_0\,\alpha\s is_1.(\delta\gamma a_1)\cdots
\alpha\s is_j.(\delta\gamma^{j+1}a_j)
\cdots\alpha\s is_{n}.(\delta\gamma^{n-1}a_n)\Big)
\end{eqnarray*}
where we made use of Equation~(\ref{ByParts3}) and used the fact that since
$\wt{\tau}={(\tau_0,\,0,\,-\tau_2,\,0,\tau_4,\ldots)},$ we may take $n$ 
to be odd. Then
\begin{eqnarray}
&&\big(b\,\tau_{n-1}\big)(a_0,\ldots,\,a_n) \nonumber\\[1mm]
&&=\sum_{j=1}^n(-1)^{j+n}\int_{\sigma_{n}}ds_1\cdots ds_{n}\,{\partial\over\partial s_j}
\,\varphi\Big((\gamma a_0)\,\alpha\s is_1.(\delta a_1)\cdots
\alpha\s is_j.(\delta\gamma^{j}a_j)
\cdots\alpha\s is_{n}.(\delta\gamma^{n}a_n)\Big)  \nonumber\\[1mm]
\label{btauMid}
&&\!\!\!\!\!\!\!\!
=\int_{\sigma_{n}}ds_1\cdots ds_{n}\,\sum_{j=1}^n{\partial\over\partial s_j}
\,\varphi\Big((\gamma a_0)\,\alpha\s is_1.(\gamma\delta\gamma a_1)\cdots
\alpha\s is_{j-1}.(\gamma\delta\gamma^{j-1}a_{j-1})\,\alpha\s is_j.(\delta\gamma^{j}a_j)
\cdots\alpha\s is_{n}.(\delta\gamma^{n}a_n)\Big)\;.\qquad\quad{}
\end{eqnarray}
To make further progress, we need the following lemma.
\begin{lem}
\label{phideltas}
Let $a_i\in\al D._S$ and $(s_1,\ldots,\,s_n)\in\sigma_n,$ then
\begin{eqnarray*}
&&\quad\varphi\Big(\delta(a_0)\,\alpha\s is_1.(\delta a_1)\cdots\alpha\s is_{n}.(\delta
a_n)\Big) \\[1mm]
&&=\sum_{j=1}^n{\partial\over\partial s_j}\,
\varphi\Big((\gamma a_0)\,\alpha\s is_1.(\gamma\delta a_1)\cdots
\alpha\s is_{j-1}.(\gamma\delta a_{j-1})\,\alpha\s is_j.(a_j)\,
\alpha\s is_{j+1}.(\delta a_{j+1})\cdots
\alpha\s is_{n}.(\delta a_n)\Big)\;.
\end{eqnarray*}
\end{lem}
\begin{beweis}
A close examination of the proof of Theorem~\ref{KMSfSUSY} shows that we actually
proved that 
\[
{d\over dt}\,
\varphi\big(B\,\alpha_t(A)C\big)\Big|_0=
 i\,\varphi\big(B
\overline{\delta}_0(A)C\big)=i\,\varphi\big(B
\overline{\delta}^2(A)C\big)
\]
for all $A\in\al D._S$ and $B,\; C\in\al A._0$ where the right hand side
makes sense because $\varphi$ is strongly regular on 
${\al R.\big(\al S.(\R),\,\sigma\big)}$
hence is well defined on $\al E._0\,.$
Now from the graded product rule for $\overline{\delta}$ on $\al E._0$ we get
\[
\overline{\delta}(b_0)\,B_1\cdots B_n=
\overline{\delta}(b_0\,B_1\cdots B_n)-\sum_{j=1}^n
\gamma(b_0\,B_1\cdots B_{j-1})\,\overline{\delta}(B_j)\,
B_{j+1}\cdots B_n
\]
for $b_0\in\al D._S,$ $B_1,\ldots,\,B_n\in\al A._0\,.$
Let $B_i=\overline{\delta}(b_i)$ for $b_i\in\al D._S,$ then
\[
\overline{\delta}(b_0)\,\overline{\delta}(b_1)\cdots \overline{\delta}(b_n)=
\overline{\delta}\big(b_0\,\overline{\delta}(b_1)\cdots \overline{\delta}(b_n)
\big)-\sum_{j=1}^n
\gamma\big(b_0\,\overline{\delta}(b_1)\cdots \overline{\delta}(b_{j-1})\big)\,
\overline{\delta}^2(b_j)\,\overline{\delta}(b_{j+1})\cdots \overline{\delta}(b_n)\,.
\]
Hence, using $\varphi\circ\overline{\delta}=0$ we get:
\begin{eqnarray*}
\varphi\Big(\overline{\delta}(b_0)\,\overline{\delta}(b_1)\cdots \overline{\delta}(b_n)\Big)
&=&
-\sum_{j=1}^n\varphi\Big(\gamma\big(b_0\,\overline{\delta}(b_1)\cdots
 \overline{\delta}(b_{j-1})\big)\,
\overline{\delta}^2(b_j)\,\overline{\delta}(b_{j+1})\cdots \overline{\delta}(b_n)\Big)
\\[1mm]
&=&i{d\over dt}\,\sum_{j=1}^n\varphi\Big(\gamma\big(b_0\,\overline{\delta}(b_1)\cdots
 \overline{\delta}(b_{j-1})\big)\,\alpha_t(b_j)\,\overline{\delta}(b_{j+1})
\cdots \overline{\delta}(b_n)\Big)\Big|_0\,.
\end{eqnarray*}
Now make the replacements $b_0\to a_0,$ $b_i\to\alpha\s t_i.(a_i)\,,$ $i=1,\ldots,\, n$
for $a_i\in\al D._S$
and use the fact that $\alpha_t\circ\overline{\delta}=\overline{\delta}\circ\alpha_t$
to find that:
\begin{eqnarray*}
&&\!\!\!\!\!\!\!
\varphi\Big(\delta(a_0)\,\alpha\s t_1.(\delta\,a_1)
\cdots \alpha\s t_n.(\delta\,a_n)\Big) \\[1mm]
&&\qquad\quad
=i\sum_{j=1}^n{\partial\over\partial t_j}\,
\varphi\Big(\gamma(a_0)\,\alpha\s t_1.(\gamma\delta\,a_1)\cdots
\alpha\s t_{j-1}.(\gamma\delta\,a_{j-1})\,\alpha\s t_j.(a_j)\,
\alpha\s t_{j+1}.(\delta\,a_{j+1})\cdots\alpha\s t_n.(\delta\,a_n)\Big)
\end{eqnarray*}
where we replaced $\overline{\delta}$ by $\delta$ because it is now evaluated on
$\al D._S$ only. Now by the KMS--condition, analyticity, flat tube theorem
and a complex linear change of variables, we
find as in Section~\ref{JLOsection} that the functions  
\begin{eqnarray*}  
&& (t_1,\ldots,\,t_n) \to \varphi\Big(\delta(a_0)\,\alpha\s t_1.(\delta\,a_1)
\cdots \alpha\s t_n.(\delta\,a_n)\Big) \\[1mm]
&& (t_1,\ldots,\,t_n) \to \varphi\Big(\gamma(a_0)\,\alpha\s t_1.(\gamma\delta\,a_1)\cdots
\alpha\s t_{j-1}.(\gamma\delta\,a_{j-1})\,\alpha\s t_j.(a_j)\,
\alpha\s t_{j+1}.(\delta\,a_{j+1})\cdots\alpha\s t_n.(\delta\,a_n)\Big) 
\end{eqnarray*}
extend analytically to the flat tube $\al T._n:=\R^n+i\sigma^n$ such
that
\begin{eqnarray*}
&&\!\!\!\!\!\!\!
\varphi\Big(\delta(a_0)\,\alpha\s z_1.(\delta\,a_1)
\cdots \alpha\s z_n.(\delta\,a_n)\Big) \\[1mm]
&&\qquad\quad
=i\sum_{j=1}^n{\partial\over\partial z_j}\,
\varphi\Big(\gamma(a_0)\,\alpha\s z_1.(\gamma\delta\,a_1)\cdots
\alpha\s z_{j-1}.(\gamma\delta\,a_{j-1})\,\alpha\s z_j.(a_j)\,
\alpha\s z_{j+1}.(\delta\,a_{j+1})\cdots\alpha\s z_n.(\delta\,a_n)\Big)\;.
\end{eqnarray*}
In the case that $z_k=is_k$ where $(s_1,\ldots,\,s_n)\in\sigma^n$
we can use $\partial\big/\partial z_k=-i\,\partial\big/\partial s_k$
to obtain from the last equation the statement of the Lemma.
\end{beweis}
Application of the Lemma to Equation~(\ref{btauMid}) then produces
\begin{eqnarray*}
&&\big(b\,\tau_{n-1}\big)(a_0,\ldots,\,a_n)
=\int_{\sigma_{n}}ds_1\cdots ds_{n}\,
\varphi\Big(\delta(a_0)\,\alpha\s is_1.(\delta\gamma a_1)\cdots
\alpha\s is_{n}.(\delta\gamma^n a_n)\Big)\\[1mm]
&&=\big(B\tau_{n+1}\big)(a_0,\ldots,\,a_n)
\end{eqnarray*}
by Equation~(\ref{BtFinal}) and hence $\wt{\tau}$ is a cyclic cocycle.

\section*{Acknowledgements.}

We gratefully acknowledge discussions with Arthur Jaffe, Roberto Longo
and Hajime Moriya on various aspects of supersymmetry.
DB wishes to thank the Department of Mathematics of the 
University of New South Wales and HG wishes to thank the 
Institute for Theoretical Physics of the University of G\"ottingen
for hospitality and financial support which facilitated this research.
The work was also supported in part by the FRG grant PS01583.

\bigskip

\providecommand{\bysame}{\leavevmode\hbox to3em{\hrulefill}\thinspace}

\end{document}